\pdfoutput=1

\documentclass[11pt,twoside,a4paper,cmspaper,final,collab]{cms-tdr}
\def\svnVersion{a8a4ad4}\def\svnDate{2019/06/22}\def\cmsCernNoTag{CERN-EP-2018-329}\def\cmsCernDate{\today}\def\cmsMessage{"Published in Physical Review D as \href{http://dx.doi.org/10.1103/PhysRevD.99.112003}{\doi{10.1103/PhysRevD.99.112003}.}"}

\begin{document}\cmsNoteHeader{HIG-18-002}

\hyphenation{had-ron-i-za-tion}
\hyphenation{cal-or-i-me-ter}
\hyphenation{de-vices}
\RCS$HeadURL$
\RCS$Id$
\newlength\cmsFigWidth
\ifthenelse{\boolean{cms@external}}{\setlength\cmsFigWidth{0.35\textwidth}}{\setlength\cmsFigWidth{0.5\textwidth}}
\ifthenelse{\boolean{cms@external}}{\providecommand{\CL}{C.L.\xspace}}{\providecommand{\CL}{CL\xspace}}
\newlength\cmsTabSkip\setlength\cmsTabSkip{1.2ex}
\newcommand{\Eq}[1]{Eq.~(\ref{#1})}
\newcommand{\Eqs}[3]{Eqs.~(\ref{#1}) #3 (\ref{#2})}
\newcommand{\Eqss}[4]{Eqs.~(\ref{#1}), (\ref{#2}), #4~(\ref{#3})}
\newcommand\sss{}
\newcommand{\cPe}{\Pe\xspace}
\newcommand{\cPm}{\PGm\xspace}
\newcommand{\ff}{\ensuremath{\cmsSymbolFace{ff}}\xspace}
\newcommand{\ffbar}{\ensuremath{\cmsSymbolFace{f}\cmsSymbolFace{\overline{f}}}\xspace}
\newcommand{\ROOT}{\textsc{ROOT}\xspace}
\newcommand{\jhugen}{\textsc{JHUGen}\xspace}
\newcommand{\minlo}{\textsc{MINLO}\xspace}
\newcommand{\mela}{\textsc{mela}\xspace}
\newcommand{\hnnlo}{\textsc{hnnlo}\xspace}
\newcommand{\F}{\ensuremath{\cmsSymbolFace{F}}\xspace}
\newcommand{\V}{\ensuremath{\cmsSymbolFace{V}}\xspace}
\newcommand{\VV}{\ensuremath{\cmsSymbolFace{VV}}\xspace}
\newcommand{\WW}{\ensuremath{\PW\PW}\xspace}
\newcommand{\ZZ}{\ensuremath{\PZ\PZ}\xspace}
\newcommand{\VVV}{\ensuremath{\cmsSymbolFace{VVV}}\xspace}
\newcommand{\HVV}{\ensuremath{\PH\V\V}\xspace}
\newcommand{\HWW}{\ensuremath{\PH\PW\PW}\xspace}
\newcommand{\VBF}{\ensuremath{\mathrm{VBF}}\xspace}
\newcommand{\VH}{\ensuremath{\V\PH}\xspace}
\newcommand{\WH}{\ensuremath{\PW\PH}\xspace}
\newcommand{\ZH}{\ensuremath{\PZ\PH}\xspace}
\newcommand{\tqH}{\ensuremath{\PQt\cPq\PH}\xspace}
\newcommand{\ttH}{\ensuremath{\PQt\PAQt\PH}\xspace}
\newcommand{\bbH}{\ensuremath{\PQb\PAQb\PH}\xspace}
\newcommand{\ggH}{\ensuremath{\Pg\Pg\PH}\xspace}
\newcommand{\ZX}{\ensuremath{\PZ+\X}\xspace}
\newcommand{\X}{\ensuremath{\cmsSymbolFace{X}}\xspace}
\newcommand{\Hboson}{\ensuremath{\PH} boson\xspace}
\newcommand{\Hell}{\ensuremath{\PH\to4\ell}\xspace}
\newcommand{\qq}{\ensuremath{\Pq\Pq}\xspace}
\newcommand{\usedLumiD}{35.9\fbinv\xspace}
\newcommand{\usedLumiE}{41.5\fbinv\xspace}
\newcommand{\GH}{\ensuremath{\Gamma_\PH}\xspace}
\newcommand{\Gref}{\ensuremath{\Gamma_0}\xspace}
\newcommand{\GHSM}{\ensuremath{\Gamma_\PH^{\mathrm{SM}}}\xspace}
\newcommand{\mH}{\ensuremath{m_{\PH}}\xspace}
\newcommand{\mell}{\ensuremath{m_{4\ell}}\xspace}
\newcommand{\bbar}{\ensuremath{\PQb\PAQb}\xspace}
\newcommand{\glufu}{\ensuremath{\Pg\Pg}\xspace}
\newcommand{\muF}{\ensuremath{\mu_{\F}}\xspace}
\newcommand{\muV}{\ensuremath{\mu_{\V}}\xspace}
\newcommand{\muVVH}{\ensuremath{\mu_{\cmsSymbolFace{vv}\PH}}\xspace}
\newcommand{\LC}[1]{\ensuremath{\Lambda_{#1}}\xspace}
\newcommand{\LZGs}{\ensuremath{\LC{1}^{\PZ\gamma}}\xspace}
\newcommand{\AC}[1]{\ensuremath{a_{#1}}\xspace}
\newcommand{\ai}{\ensuremath{\AC{i}}\xspace}
\newcommand{\fLC}[1]{\ensuremath{f_{\Lambda #1}}\xspace}
\newcommand{\fAC}[1]{\ensuremath{f_{a #1}}\xspace}
\newcommand{\fLZGs}{\ensuremath{\fLC{1}^{\PZ\gamma}}\xspace}
\newcommand{\fai}{\fAC{i}}
\newcommand{\sLC}[1]{\ensuremath{\tilde{\sigma}_{\Lambda #1}}\xspace}
\newcommand{\sAC}[1]{\ensuremath{\sigma_{#1}}\xspace}
\newcommand{\sLZGs}{\ensuremath{\sLC{1}^{\PZ\gamma}}\xspace}
\newcommand{\sai}{\sAC{i}}
\newcommand{\pLC}[1]{\ensuremath{\phi_{\Lambda #1}}\xspace}
\newcommand{\pAC}[1]{\ensuremath{\phi_{a #1}}\xspace}
\newcommand{\pLZGs}{\ensuremath{\pLC{1}^{\PZ\gamma}}\xspace}
\newcommand{\pai}{\pAC{i}}
\newcommand{\cospLC}[1]{\ensuremath{\cos\left(\pLC{#1}\right)}\xspace}
\newcommand{\cospAC}[1]{\ensuremath{\cos\left(\pAC{#1}\right)}\xspace}
\newcommand{\cospLZGs}{\ensuremath{\cos\left(\pLZGs\right)}\xspace}
\newcommand{\cospai}{\cospAC{i}}
\newcommand{\fcospLC}[1]{\ensuremath{\fLC{#1}\cospLC{#1}}\xspace}
\newcommand{\fcospAC}[1]{\ensuremath{\fAC{#1}\cospAC{#1}}\xspace}
\newcommand{\fcospLZGs}{\ensuremath{\fLZGs\cospLZGs}\xspace}
\newcommand{\fcospai}{\fcospAC{i}}
\newcommand{\Offshell}{Off-shell\xspace}
\newcommand{\offshell}{off-shell\xspace}
\newcommand{\onshell}{on-shell\xspace}
\newcommand{\Dbkg}{\ensuremath{{\mathcal{D}}_{\text{bkg}}}\xspace}
\newcommand{\Dai}{\ensuremath{{\mathcal{D}}_{ai}}\xspace}
\newcommand{\Dint}{\ensuremath{{\mathcal{D}}_{\mathrm{int}}}\xspace}
\newcommand{\Dgg}{\ensuremath{{\mathcal{D}}_{\glufu}}\xspace}
\newcommand{\Dbsi}{\ensuremath{{\mathcal{D}}_{\mathrm{bsi}}}\xspace}
\newcommand{\Dcp}{\ensuremath{{\mathcal{D}}_{C\!P}}\xspace}
\newcommand{\Dbkgkin}{\ensuremath{{\mathcal{D}}^{\text{kin}}_{\text{bkg}}}\xspace}
\newcommand{\DbkgVBFdec}{\ensuremath{{\mathcal{D}}^{{\VBF}+\text{dec}}_{\text{bkg}}}\xspace}
\newcommand{\DbkgVHdec}{\ensuremath{{\mathcal{D}}^{{\VH}+\text{dec}}_{\text{bkg}}}\xspace}
\newcommand{\DjjVBF}{\ensuremath{{\mathcal{D}}^{\VBF}_{\text{2jet}}}\xspace}
\newcommand{\DjjZH}{\ensuremath{{\mathcal{D}}^{\ZH}_{\text{2jet}}}\xspace}
\newcommand{\DjjWH}{\ensuremath{{\mathcal{D}}^{\WH}_{\text{2jet}}}\xspace}
\newcommand{\DjjVH}{\ensuremath{{\mathcal{D}}^{\VH}_{\text{2jet}}}\xspace}
\newcommand{\mlplm}{\ensuremath{m_{\ell^{+}\ell^{-}}}\xspace}
\newcommand{\vv}{\ensuremath{\cmsSymbolFace{vv}}\xspace}

\cmsNoteHeader{HIG-18-002}
\title{Measurements of the Higgs boson width and anomalous \HVV couplings from \onshell and \offshell production in the four-lepton final state}
\date{\today}

\abstract{
Studies of \onshell and \offshell Higgs boson production in the four-lepton final state are presented, using data
from the CMS experiment at the LHC that correspond to an integrated luminosity of 80.2\fbinv at
a center-of-mass energy of 13\TeV. Joint constraints are set on the Higgs boson total width and parameters that
express its anomalous couplings to two electroweak vector bosons. These results are combined with those
obtained from the data collected at center-of-mass energies of 7 and 8\TeV, corresponding to integrated
luminosities of 5.1 and 19.7\fbinv, respectively. Kinematic information from the decay particles
and the associated jets are combined using matrix element techniques to identify the production mechanism
and to increase sensitivity to the Higgs boson couplings in both production and decay. 
The constraints on anomalous \HVV couplings are found to be consistent 
with the standard model expectation in both the on-shell and off-shell regions.
Under the assumption of a coupling structure similar to that in the standard model, the Higgs boson width is constrained
to be $3.2^{+2.8}_{-2.2}$\MeV while the expected constraint based on simulation is $4.1^{+5.0}_{-4.0}$\MeV.
The constraints on the width remain similar with the inclusion of the tested anomalous \HVV interactions.
}

\author{CMS Collaboration}

\hypersetup{
pdfauthor={CMS Collaboration},
pdftitle={Measurements of the Higgs boson width and anomalous HVV couplings from on-shell and off-shell production in the four-lepton final state},
pdfsubject={CMS},
pdfkeywords={Higgs, width, spin, parity, CP, anomalous couplings}
}

\maketitle

\section{Introduction}
\label{sec:Introduction}

The standard model (SM) of particle physics postulates the existence of a Higgs field responsible for the
generation of the masses of fundamental particles. The excitation of this
field is known as the Higgs boson ($\PH$)~\cite{StandardModel67_1, Englert:1964et,Higgs:1964ia,
Higgs:1964pj,Guralnik:1964eu,StandardModel67_2,StandardModel67_3}.
The observation of an \Hboson with a mass of around 125\GeV by the ATLAS and CMS
Collaborations~\cite{Aad:2012tfa,Chatrchyan:2012xdj,Chatrchyan:2013lba} is consistent with
the expectations of the SM, but further tests of the properties of this particle,
such as its width and the structure of its couplings to the known SM particles, are needed
to determine its nature.

The CMS and ATLAS experiments have set constraints of $\GH < 13$\MeV at $95\%$ confidence level (\CL)
on the \Hboson total width~\cite{Khachatryan:2014iha,Aad:2015xua,Khachatryan:2015mma,Khachatryan:2016ctc,Aaboud:2018puo}
using the \offshell production method~\cite{Caola:2013yja,Kauer:2012hd,Campbell:2013una},
which relies on the relative measurement of \offshell and \onshell production.
The upper bound on \GH was set considering the gluon fusion and electroweak (EW) production mechanisms in the analysis.
The precision on \GH from \onshell measurements of the width of the resonance peak
alone is approximately 1\GeV~\cite{Khachatryan:2014jba,Aad:2014aba,Sirunyan:2017exp},
which is significantly worse than the result from the \offshell method.
The constraint on the \Hboson lifetime is equivalent to a lower bound on the width and was
derived from the flight distance in the CMS detector as $\GH > 3.5\times10^{-9}$\MeV at $95\%$~\CL~\cite{Khachatryan:2015mma}.
The SM expectation of the width of the \Hboson is around $4$\MeV~\cite{deFlorian:2016spz}.

The CMS~\cite{Chatrchyan:2012jja,Chatrchyan:2013mxa,Khachatryan:2014kca,Khachatryan:2015mma,Khachatryan:2016tnr,Sirunyan:2017tqd}
and ATLAS~\cite{Aad:2013xqa,Aad:2015mxa,Aad:2016nal,Aaboud:2017oem,Aaboud:2017vzb,Aaboud:2018xdt}
experiments have set constraints on the spin-parity properties and anomalous couplings of the \Hboson,
finding its quantum numbers to be consistent with $J^{PC}=0^{++}$, but allowing small anomalous couplings
to two EW gauge bosons (anomalous \HVV couplings).
\Offshell signal production may be enhanced in the presence of these anomalous
\HVV couplings~\cite{Khachatryan:2014iha, Gainer:2014hha, Englert:2014aca, Ghezzi:2014qpa,Khachatryan:2015mma,deFlorian:2016spz}.
As a result, the measurement of \GH using the \offshell technique may be affected by these deviations of the
\Hboson couplings from the SM expectations.
An attempt to measure \GH using the \offshell technique while including anomalous \HVV interactions has been made by
the CMS experiment~\cite{Khachatryan:2015mma}. In that previous study, constraints are placed on \GH and the \onshell
cross-section fraction \fLC{Q} that expresses an anomalous coupling contribution sensitive to the invariant mass of the \Hboson,
using a realistic treatment of interference between the \Hboson signal and the continuum background.
Extending the application of the \offshell technique to a wider range of anomalous \HVV contributions,
studied previously using \onshell \Hboson production~\cite{Sirunyan:2017tqd}, is the goal of this paper.

The presented investigation on the \Hboson width targets both gluon fusion and EW production mechanisms
and tests the effects of possible anomalous \HVV couplings in either production or decay. Nevertheless,
it still relies on the knowledge of coupling ratios between the \offshell and \onshell production, the
dominance of the top quark loop in the gluon fusion production mechanism, and the absence of new particle
contributions in the loop. A violation of the last assumption by itself would be a manifestation of physics
beyond the SM (BSM), which may become evident if the measured width deviates from the SM expectation.
The measured width may also deviate from the SM expectation if the \Hboson has new BSM decay channels
or the known channels have non-SM rates. Therefore, the measurement of the width complements the search
for \Hboson decay to invisible or undetected particles, and the measurement of the \Hboson couplings to the
known SM particles.

The data sample used in this analysis corresponds to integrated luminosities of $\usedLumiD$ collected in
2016 and $\usedLumiE$ collected in 2017 during Run~2 of the CERN LHC at a center-of-mass energy of 13\TeV.
These results are combined with results obtained earlier from the data collected at center-of-mass energies
of 7\TeV (in 2011), 8\TeV (in 2012),  and 13\TeV (in 2015), corresponding to integrated luminosities of
5.1,  19.7, and 2.7\fbinv, respectively~\cite{Khachatryan:2014kca, Sirunyan:2017tqd}.
The increase in either energy and integrated luminosity leads to substantial improvement in the precision
of the width measurement using the \offshell technique, either under the assumption of SM couplings
or with BSM effects.

This analysis follows closely the general \Hell (leptons $\ell=\cPe$ or $\cPm$) selection and reconstruction
documented in Ref.~\cite{Sirunyan:2017exp} using the data collected in 2016, and the \onshell study of
anomalous \HVV couplings with the combined 2015 and 2016 data set in Ref.~\cite{Sirunyan:2017tqd}.
Many of the technical details of the search for a scalar resonance $\X\to\ZZ$ at high mass in Run~2 data,
documented in Ref.~\cite{Sirunyan:2018qlb}, are also shared in the analyses presented here.
The rest of the paper is organized as follows.
The phenomenology of anomalous \HVV interactions is discussed in Sec.~\ref{sec:Pheno}.
The CMS detector, reconstruction techniques, and Monte Carlo (MC) simulation methods are introduced in Sec.~\ref{sec:CMS}.
The addition of the 2017 data to that used in Refs.~\cite{Sirunyan:2017exp,Sirunyan:2017tqd},
and the relevant differences in the detector and reconstruction techniques are also discussed in this section.
The details of the analysis are discussed in Secs.~\ref{sec:AnalysisStrategyIntro} and~\ref{sec:offshell},
and the results are presented in Sec.~\ref{sec:results}. We provide a summary of these results in Sec.~\ref{sec:Summary}.

\section{Phenomenology of anomalous \texorpdfstring{\HVV}{HVV} interactions}
\label{sec:Pheno}

The constraints on \GH are set using the \offshell production method,
which considers the \PH boson production relationship between the \onshell
($105 < \mell< 140$\GeV) and \offshell ($\mell > 220$\GeV) regions.
Denoting each production mechanism with
$\vv \to \PH \to \V\V\to 4\ell$ for the \Hboson coupling to either strong ($\vv=\Pg\Pg$)
or EW ($\vv=\WW,\ZZ,\PZ\gamma,\gamma\gamma$) vector bosons in its production,
the \onshell and \offshell \Hboson signal yields are related by~\cite{Caola:2013yja}
\begin{equation}
\label{eq:resonant}
\sigma^\text{\onshell}_{\vv \to \Hell} \propto \muVVH
\quad\text{and}\quad
\sigma^\text{\offshell}_{\vv \to \Hell} \propto \muVVH \ \GH,
\end{equation}
where \muVVH is defined as the \onshell signal strength,
the ratio of the observed number of \onshell four-lepton events relative to the SM expectation.
This ratio is interpreted as either \muF for \Hboson production via gluon fusion (\ggH) or in
association with a \ttbar (\ttH) or \bbar pair (\bbH), or \muV for \Hboson production via vector
boson fusion (\VBF) or in association with an EW vector boson $\PW$ or $\PZ$ (\VH).
There is sizable interference between the \Hboson signal and the continuum background
in the \offshell region~\cite{Kauer:2012hd}, contrary to \onshell production,
and this formalism scales the interference contribution with $\sqrt{\muVVH \ \GH}$.

This analysis is based on a phenomenological
framework~\cite{Nelson:1986ki,Soni:1993jc,Plehn:2001nj,Choi:2002jk,Buszello:2002uu,Hankele:2006ma,Accomando:2006ga,
Godbole:2007cn,Hagiwara:2009wt,Gao:2010qx,DeRujula:2010ys,Christensen:2010pf,Bolognesi:2012mm,Ellis:2012xd,
Chen:2012jy,Artoisenet:2013puc,Anderson:2013afp,Chen:2013waa,Dolan:2014upa,Gonzalez-Alonso:2014eva,Greljo:2015sla,
Gritsan:2016hjl,deFlorian:2016spz}
that describes the anomalous couplings of a Higgs-like boson
to two gauge bosons, such as $\WW, \ZZ, \PZ\gamma, \gamma\gamma$, and \glufu.
These couplings appear in either the production of the \Hboson or its decay,
regardless of the \mell region in which the \Hboson is produced.
The relationship in \Eq{eq:resonant} is therefore meant to imply concurrent variations
in $\vv\PH$ couplings in both \onshell and \offshell regions.
The coupling of the \Hboson to two gluons is assumed
to be as in the SM, via quark loops with Yukawa couplings to quarks, where the contribution from the top-quark is dominant.
This assumption is valid as long as the production is dominated
by the top-quark loop and no new particles contribute to this loop.
The Yukawa couplings also appear in direct interactions with fermion-antifermion pairs,
such as in \ttH and \bbH productions. These interactions are of less importance in this study,
since they are highly suppressed at high \offshell mass, but they are included in the analysis
of the \onshell \Hboson production with similar assumptions as in the case of production via gluon fusion.
Variation of the \HVV couplings, in either the \VBF or \VH~productions, or the \Hell decay,
are allowed to depend on anomalous coupling contributions.

In the following, we assume that the \Hboson couples to two gauge bosons \VV, such as
\WW, \ZZ, $\PZ\gamma$ or $\gamma\gamma$, which in turn couple to fermions,
either four leptons in \Hboson decay, or quarks or leptons in its production or in the decay
of associated EW bosons. It is assumed that the \Hboson does not couple to fermions through
a new heavy state, generating a so-called contact interaction~\cite{Gonzalez-Alonso:2014eva,Greljo:2015sla}.
However, the inclusion of amplitude terms pertaining to contact interactions is equivalent to the
anomalous \HVV couplings already tested~\cite{Khachatryan:2014kca} under the assumption
of flavor universality in $\V\ffbar$ couplings.
Both approaches test three general tensor structures allowed by Lorentz symmetry,
with form factors $F_i(q_1^2,q_2^2)$ in front of each term,
where $q_1$ and $q_2$ are the four-momenta of the two difermion states, such as $(\EE)$ and $(\MM)$ in the
$\PH \to \EE\MM$ decay, and equivalent states in production. We also fix all lepton and quark couplings to vector bosons
according to SM expectations. Relaxing this requirement would make it equivalent to flavor nonuniversal couplings of the contact terms,
but would also introduce too many unconstrained parameters, which cannot be tested with the present data sample.
Only the lowest order operators, or lowest order terms in the $(q_j^2/\Lambda^2)$ form-factor expansion, are tested,
where $\Lambda$ is the energy scale of new physics.

The signal scattering amplitude describing the interaction between a spin-zero \Hboson and two spin-one
gauge bosons $\V\V$ is written as~\cite{Anderson:2013afp}
\ifthenelse{\boolean{cms@external}}{
\begin{multline}
A\sim\left[ a_{1}^{\VV}
- \frac{\kappa_1^{\VV}q_{1}^2 + \kappa_2^{\VV} q_{2}^{2}}{\left(\Lambda_{1}^{\VV} \right)^{2}}
- \frac{\kappa_3^{\VV}(q_{1} + q_{2})^{2}}{\left(\Lambda_{Q}^{\VV} \right)^{2}}
\right]\\
\times
m_{\sss\V1}^2 \epsilon_{\sss\V1}^* \epsilon_{\sss\V2}^*\\
+ a_{2}^{\VV}  f_{\mu \nu}^{*(1)}f^{*(2)\,\mu\nu}
+ a_{3}^{\VV}   f^{*(1)}_{\mu \nu} {\tilde f}^{*(2)\,\mu\nu}.
\label{eq:formfact-fullampl-spin0}
\end{multline}
}{
\begin{equation}
A\sim\left[ a_{1}^{\VV}
- \frac{\kappa_1^{\VV}q_{1}^2 + \kappa_2^{\VV} q_{2}^{2}}{\left(\Lambda_{1}^{\VV} \right)^{2}}
- \frac{\kappa_3^{\VV}(q_{1} + q_{2})^{2}}{\left(\Lambda_{Q}^{\VV} \right)^{2}}
\right]
m_{\sss\V1}^2 \epsilon_{\sss\V1}^* \epsilon_{\sss\V2}^*
+ a_{2}^{\VV}  f_{\mu \nu}^{*(1)}f^{*(2)\,\mu\nu}
+ a_{3}^{\VV}   f^{*(1)}_{\mu \nu} {\tilde f}^{*(2)\,\mu\nu}.
\label{eq:formfact-fullampl-spin0}
\end{equation}
}
In this expression of the scattering amplitude, $\epsilon_{i}$ is the polarization vector of gauge boson
$\V_{i}$, $f^{(i){\mu \nu}} = \epsilon_{i}^{\mu}q_{i}^{\nu} - \epsilon_{i}^\nu q_{i}^{\mu}$ is a scalar tensor
constructed from this polarization vector and the momentum of the gauge boson, and
${\tilde f}^{(i)}_{\mu \nu} = \frac{1}{2} \epsilon_{\mu\nu\rho\sigma} f^{(i)\,\rho\sigma}$ is the pseudoscalar
tensor counterpart. When at least one of the gauge bosons \V is massive, $m_{\sss\V1}$ is the pole mass
of that gauge boson. The scales of BSM physics are denoted with $\Lambda_{1}$ and $\Lambda_{Q}$, so
$\ai^{\VV}$, or $1/\Lambda_{1}$ and $1/\Lambda_{Q}$, become the coupling-strength modifiers
of the relevant \HVV amplitudes, where $\ai^{\VV}$ may in general be any complex number,
and $\left| \kappa_{1,2,3}^{\VV} \right|=0$ or~$1$ are complex numbers.
Under the assumption that the couplings are constant and real, the above formulation is equivalent
to an effective Lagrangian notation. Therefore, in this paper, the real coupling constants are tested.
The above approach allows a sufficiently general test of the \Hell
kinematics in decay and equivalent kinematics in production, as discussed below, including production
and decay of virtual intermediate photons. If deviations from the SM are detected,
a more detailed study of $F_i(q_1^2,q_2^2)$ could be performed, eventually providing a measurement of the
double-differential cross section for each tensor structure tested.

In the above, the only leading tree-level contributions are $\AC{1}^{\ZZ}\ne 0$ and $a_{1}^{\WW} \ne 0$,
and in the following we assume the custodial symmetry $\AC{1}^{\ZZ}=\AC{1}^{\WW}$. The rest of the couplings
are considered anomalous contributions, which are either tiny contributions arising in the SM due to loop effects
or new BSM contributions. The SM loop contributions are not accessible experimentally with the available data.
Among anomalous contributions, considerations of symmetry and gauge invariance require
$\kappa_1^{\ZZ}=\kappa_2^{\ZZ}=-\exp({i\pLC{1}^{\ZZ}})$,
$\kappa_1^{\gamma\gamma}=\kappa_2^{\gamma\gamma}=0$,
$\kappa_1^{\Pg\Pg}=\kappa_2^{\Pg\Pg}=0$,
$\kappa_1^{\PZ\gamma}=0$,
and $\kappa_2^{\PZ\gamma}=-\exp({i\phi^{\PZ\gamma}_{\Lambda{1}}})$.
While not strictly required, the same symmetry is considered in the $\WW$ case
$\kappa_1^{\WW}=\kappa_2^{\WW}=-\exp({i\pLC{1}^{\WW}})$.

Neither $\PH\PZ\gamma$ nor $\PH\gamma\gamma$ couplings produce a sizable \offshell enhancement,
since there is no interplay between the vector bosons or the \Hboson going \offshell,
and there is no \offshell threshold for these couplings. Therefore, \offshell treatment for these
couplings can be ignored. While the $a^{\PZ\gamma}_{2,3}$ and $a^{\gamma\gamma}_{2,3}$ terms
are tested in the Run~1 analysis~\cite{Khachatryan:2014kca}, the precision of those constraints
is still not competitive with the \onshell photon measurements in $\PH\to \PZ\gamma$ and $\gamma\gamma$.
Therefore, we omit those measurements in this paper.
The \LZGs coupling, on the other hand, can only be observed with \offshell photons decaying to a pair of fermions, 
so it is considered in the \onshell analysis. The \LC{Q} term depends
only on the invariant mass of the \Hboson, so its contribution is not distinguishable from the SM in the \onshell
region and is only testable through the \offshell region. Tight constraints are already set on this parameter
in the Run~1 analysis~\cite{Khachatryan:2015mma}, so it is also not considered in this paper.

In the following, the \ZZ labels for the \ZZ interactions are omitted, and we use a generic notation \ai
to denote $\AC{3}$, $\AC{2}$, $1/\LC{1}$, and $1/\LZGs$, which are the four couplings tested in
this paper as listed in Table~\ref{tab:xsec_ratio}. Furthermore, the \WW measurements
are integrated into the \ZZ measurements assuming $a_i^{\ZZ}=a_i^{\WW}$.
The \HWW contributions appear in the \VBF and \WH productions.
This assumption does not affect the kinematic analysis of events because there is very little difference
in kinematic distributions in events initiated by either \WW or \ZZ fusion.
However, this assumption may affect the interpretation of the results should a different relationship
between $a_i^{\ZZ}$ and $a_i^{\WW}$ be assumed. Therefore, such a scenario is discussed in more
detail below by introducing the parameter $r_{ai}$, following Ref.~\cite{Khachatryan:2014kca}, as
\begin{equation}
r_{ai} = \frac{a_i^{\WW} / a_1^{\WW}  }{  a_i / a_1}.
\label{eq:ratio_ww_zz}
\end{equation}
Including the parameter $r_{ai}$ in the probability parametrization despite the lack of sensitivity
of the data would introduce complexity without a comparable gain in physics content.
We proceed with the analysis assuming $r_{ai} = 1$, but point out below how results could be
reinterpreted should a different value be assumed.

Most systematic uncertainties cancel when taking ratios to the total cross section, so measurements
of \ai relative to the dominant SM-like contribution \AC{1} are the preferred approach.
For this purpose, the effective fractional \ZZ cross sections \fai and phases \pai are defined as
\begin{equation}\begin{aligned}
\fai &= \frac{\abs{\ai}^2 \sai}{\sum_{j=1,2,3\ldots}{\abs{\AC{j}}^2 \sAC{j}}},\\
 \pai &= \text{arg}\left(\frac{\ai}{\AC{1}}\right),
\label{eq:fa_definitions}
\end{aligned}\end{equation}
where $\sigma_{i}$ is the cross section for the process corresponding to $a_{i} = 1$, $a_{j \neq i} = 0$,
while $\tilde{\sigma}_{\Lambda{1}}$ is the effective cross section for the process corresponding to
$\Lambda_{1} =1\TeV$, given in units of $\text{fb}\,\TeV^4$.
The cross-section ratios are quoted in Table~\ref{tab:xsec_ratio}.
The $\ai/\AC{1}$ ratios can be obtained from the ratio $\fai/\fAC{1}$, the cross-section ratios, and the phase \pai as
\begin{equation}
\frac{\ai}{\AC{1}}= \sqrt{\frac{\fai}{\fAC{1}} \ \frac{\sAC{1}}{\sai}} \, \re^{i \pai}.
\label{eq:fa_a_correspondence}
\end{equation}
The effective fractions \fai are bounded between 0 and 1 and do not depend on the coupling convention.
In most cases, uncertainties on these measurements scale with integrated luminosity as $1/\sqrt{\mathcal{L}}$
until effects of interference become important.
Furthermore, the values of \fai have a simple interpretation as the fractional size of the BSM contribution for
the $\PH\to2\Pe2\PGm$ decay.  For example, $\fai=0$ indicates a pure SM-like \Hboson,
$\fai=1$ gives a pure BSM particle, and $\fai=0.5$ means that the two couplings contribute equally
to the $\PH\to2\Pe2\PGm$ process.

\begin{table*}[!bht]
\centering
\topcaption{
List of the anomalous \HVV couplings considered in the measurements assuming a spin-zero \Hboson.
The definition of the effective fractions $\fai$ is discussed in the text and the translation constants
are the cross-section ratios corresponding to the processes
$\PH\to2\Pe2\mu$ with the \Hboson mass $m_{\sss\PH}=125\GeV$ and calculated
using \textsc{JHUGen}~\cite{Gao:2010qx,Bolognesi:2012mm,Anderson:2013afp}.
}
\label{tab:xsec_ratio}
\begin{scotch}{llll}
  \multicolumn{1}{c}{Anomalous}& \multicolumn{1}{c}{Coupling} & \multicolumn{1}{c}{Effective} & \multicolumn{1}{c}{Translation} \\
  \multicolumn{1}{c}{Coupling}    & \multicolumn{1}{c}{Phase} & \multicolumn{1}{c}{Fraction}   & \multicolumn{1}{c}{Constant} \\
\hline
 \AC{3} & \pAC{3} & \fAC{3} & $\sAC{1}/\sAC{3}= 6.53$ \\
 \AC{2} & \pAC{2} & \fAC{2} & $\sAC{1}/\sAC{2}= 2.77$ \\
 \LC{1} & \pLC{1} & \fLC{1} & $\sAC{1}/\sLC{1}= 1.47\times 10^{4}\TeV^{-4}$ \\
 \LZGs & \pLZGs & \fLZGs & $\sAC{1}/\sLZGs = 5.80\times 10^{3}\TeV^{-4}$ \\
\end{scotch}
\end{table*}

As mentioned above in application to \Eq{eq:ratio_ww_zz}, the measurement of \fai is performed
under the $r_{ai} = 1$ assumption. Let us denote this to be an effective $f_{ai}^\text{eff}$.
Without such an assumption,  there is a certain dependence of \fai on $r_{ai}$ and $f_{ai}^\text{eff}$,
such that $\fai=f_{ai}^\text{eff}$ for $r_{ai} = 1$. This dependence is different for different processes,
such as \VBF production or \Hell decay, where the latter case is in fact independent of
$r_{ai}$ because the \HWW coupling does not affect this decay process. In the former case, let us consider
the relative contributions of \WW and \ZZ fusion \onshell. For example, the ratio of \VBF cross sections
driven by \WW and \ZZ fusion is $\sigma_{1}^{\WW}/\sigma_{1}^{\ZZ} = 2.59$ for the SM tree-level
couplings under custodial symmetry $a_1^{\WW}=a_1^{\ZZ}$ at 13\TeV $\Pp\Pp$ collision energy.
The same ratio for the $CP$-odd couplings is $\sigma_{3}^{\WW}/\sigma_{3}^{\ZZ} = 3.15$,
where $\sigma_{3}^{\VV}$ are calculated for $a_3^{\WW}=a_3^{\ZZ}$.
The dependence of \fai on $r_{ai}$ and $f_{ai}^\text{eff}$, as measured in the \VBF process, becomes
\ifthenelse{\boolean{cms@external}}{
\begin{multline}
\fai = \Bigl[ 1+(1/f_{ai}^\text{eff}-1)\\
\times (\sigma_{i}^{\ZZ} + r_{ai}^2 \sigma_{i}^{\WW} )/(\sigma_{i}^{\ZZ} + \sigma_{i}^{\WW} ) \Bigr]^{-1},
\label{eq:fai_conversion}
\end{multline}
}{
\begin{equation}
\fai = \left[ 1+(1/f_{ai}^\text{eff}-1) (\sigma_{i}^{\ZZ} + r_{ai}^2 \sigma_{i}^{\WW} )/(\sigma_{i}^{\ZZ} + \sigma_{i}^{\WW} ) \right]^{-1},
\label{eq:fai_conversion}
\end{equation}
}
where custodial symmetry $a_1^{\WW}=a_1^{\ZZ}$ is assumed and
the effects of interference between \WW and \ZZ fusion are negligible and are therefore ignored.

All of the above discussion, including \Eq{eq:formfact-fullampl-spin0}, describes the production of a resonance
via gluon fusion, \VBF with associated jets, or associated production with an EW vector boson, \VH.
These mechanisms, along with the \ttH and \bbH production, are considered
in the analysis of the spin-zero hypothesis of the \Hboson, where the gluon fusion production is expected to dominate.
It is possible to study \HVV interactions using the kinematics of particles produced in
association with the \Hboson, such as \VBF jets or vector boson daughters in \VH~production, as we show below.
More details can be found in, \eg, Ref.~\cite{Anderson:2013afp} and the experimental application in
Refs.~\cite{Khachatryan:2016tnr, Sirunyan:2017tqd}.
While the $q_{i}^2$ range in the \HVV process does not exceed approximately 100\GeV because of
the kinematic bound, no such bound exists in the associated production, so consideration of more restricted
$q_{i}^2$ ranges might be required~\cite{Anderson:2013afp}. However, we only consider that
the $q_{i}^2$ range is not restricted in the allowed phase space.

\section{The CMS detector, simulation, and reconstruction}
\label{sec:CMS}

The \Hell decay candidates are reconstructed in the CMS detector~\cite{Chatrchyan:2008zzk}.
The CMS detector is comprised of a silicon pixel and strip tracker, a lead tungstate crystal electromagnetic
calorimeter (ECAL), and a brass/scintillator hadron calorimeter, each composed of a barrel and two end cap sections,
all within a superconducting solenoid of 6\unit{m} internal diameter, providing a magnetic field of 3.8\unit{T}.
Extensive forward calorimetry complements the coverage provided by the barrel and end cap detectors.
Outside the solenoid are the gas-ionization detectors for muon measurements, which are embedded
in the steel flux-return yoke. A detailed description of the CMS detector, together with a definition of the
coordinate system used and the relevant kinematic variables, can be found in Ref.~\cite{Chatrchyan:2008zzk}.

The \textsc{JHUGen}~7.0.2~\cite{Gao:2010qx,Bolognesi:2012mm,Anderson:2013afp,Gritsan:2016hjl}
Monte Carlo (MC) program is used to simulate anomalous couplings in the \Hboson production and
$\PH\to \ZZ$ / $\PZ\gamma^*$ / $\gamma^*\gamma^*\to4\ell$ decay. The gluon fusion production is simulated with
the \POWHEG~2~\cite{Frixione:2007vw,Bagnaschi:2011tu,Nason:2009ai,Luisoni:2013kna,Hartanto:2015uka}
event generator at next-to-leading order (NLO) in QCD, and simulation with the \minlo~\cite{Hamilton:2012np}
program at NLO in QCD is used for evaluation of systematic uncertainties related to modeling of two associated
jets. The kinematics of events produced in gluon fusion with two associated jets are also modified by anomalous
$\PH\Pg\Pg$ couplings. These effects are studied using \jhugen, and it is found that the kinematic distributions
relevant for this analysis are not affected significantly.

The production of the \Hboson through \VBF, in association with a $\PW$ or $\PZ$ boson, or with a \ttbar pair,
is simulated using both \jhugen at LO in QCD and \POWHEG at NLO in QCD.
Production in association with a \bbar pair is simulated only at LO in QCD via \jhugen.
In the \VBF, \VH, and \ttH production modes, the \jhugen and \POWHEG simulations are explicitly compared
after parton showering in the SM case, and no significant differences are found in kinematic observables.
Therefore, the \jhugen simulation is adopted to describe kinematics in the \VBF, \VH, and \ttH production
modes with anomalous couplings in the \onshell region, with expected yields taken from the \POWHEG
simulation. The \POWHEG program is used to simulate wide resonances at masses ranging from
115\GeV to 3\TeV, produced in gluon fusion, \VBF, or \VH. The events from the \POWHEG simulation
are later reweighted using the package for the matrix element likelihood approach
(\mela)~\cite{Chatrchyan:2012xdj,Gao:2010qx,Bolognesi:2012mm,Anderson:2013afp,Gritsan:2016hjl}
to model \offshell \Hboson production distributions, as discussed below.

The $\Pg\Pg\to\ZZ/\PZ\gamma^*\to 4\ell$ background process is simulated with
\MCFM~7.0.1~\cite{MCFM,Campbell:2011bn,Campbell:2013una,Campbell:2015vwa}.
The vector boson scattering and triple-gauge-boson (\VVV) backgrounds are obtained by reweighting the \POWHEG
simulation with the matrix elements provided by the \mela package using the \MCFM and \jhugen matrix elements,
and the reweighted simulation is checked against the predictions of the \textsc{phantom}~1.3~\cite{Ballestrero:2007} simulation.
Both the \MCFM and \textsc{phantom} generators allow one to model the \Hboson signal, background, and their interference
in the \offshell production. However, they do not allow modeling of the anomalous interactions considered in this analysis.
Therefore, a combined program has been developed for both gluon fusion and \VBF with triple-gauge-boson production
based on the modeling of signal and background scattering amplitudes from \MCFM and anomalous contributions in the
signal scattering amplitude from \jhugen. This program is included within the \jhugen and \mela packages, as detailed
in Ref.~\cite{deFlorian:2016spz}. A large number of MC events with anomalous couplings in the signal and their
interference with background have been generated with these packages. The simulated events also include alternative
weights to model various anomalous couplings in the signal.

In the gluon fusion process, the factorization and renormalization scales are chosen to be running as $\mell/2$.
In order to include higher-order QCD corrections, LO, NLO, and next-to-NLO (NNLO) signal cross-section calculations
are performed using the \MCFM and \hnnlo~2 programs~\cite{Catani:2007vq,Grazzini:2008tf,Grazzini:2013mca}
for a wide range of masses using a narrow width approximation. The ratios between the NNLO and LO values (NNLO K factors)
are used to reweight~\cite{deFlorian:2016spz} the \mell distributions from the \MCFM and \jhugen simulation at LO in QCD, and
a uniform factor of 1.10 across all of the \mell range is applied to normalize the cross section of the \Hboson production via gluon fusion
to the predictions for $\mell\approx125$\GeV at next-to-NNLO ($N^3\mathrm{LO}$) in QCD~\cite{deFlorian:2016spz}.
The simulated \mell shapes or yields obtained from the \POWHEG simulation of the gluon fusion process are corrected based
on the above reweighted distributions. While the NNLO K factor calculation is directly applicable to the signal contribution,
it is approximate for the background and its interference with the signal. The NLO calculation with some
approximations~\cite{Caola:2015psa,Melnikov:2015laa,Campbell:2016ivq,Caola:2016trd} is available for the background
and interference. Comparison with this calculation shows that while there is some increase of the NLO K factor for the
interference close to the \ZZ threshold, the NLO K factors for the background and interference are consistent with the
signal within approximately 10\% in the mass range $\mell > 220$\GeV relevant for this analysis.
We therefore multiply the background and interference contributions by the same NNLO K factor and uniform
$N^3\mathrm{LO}$ correction, both calculated for signal and including associated uncertainties, and introduce
an additional unit factor with a 10\% uncertainty for the background and the square root of this factor for the interference.

The \mela package contains a library of matrix elements from \jhugen and \MCFM for the signal,
and \MCFM for the background, and is used to apply weights to events in any MC sample to model
any other set of anomalous or SM couplings in either \onshell or \offshell production.
This matrix element library also allows reweighting of the signal \POWHEG simulation of the wide resonances
at NLO in QCD in either gluon fusion, \VBF, or triple-gauge-boson production to model
the signal, background, or their interference.

The main background in this analysis, $\qqbar\to\ZZ/\PZ\gamma^*\to 4\ell$, is estimated from simulation with \POWHEG.
A fully differential cross section has been computed at NNLO in QCD~\cite{Grazzini:2015hta}, but it is not yet available
in a partonic level event generator. Therefore the NNLO/NLO QCD correction is applied as a function of \mell.
Additional NLO EW corrections are also applied to this background process in the region
$\mell>2 m_{\PZ}$~\cite{Gieseke:2014gka,Baglio:2013toa}.
The parton distribution functions (PDFs) used in this paper belong to the NNPDF~3.0 PDF sets~\cite{Ball:2011uy}.
All MC samples are interfaced to \PYTHIA~8~\cite{Sjostrand:2014zea} for parton
showering, using version 8.212 for the simulation of the 2016 data period and 8.230 for the simulation of the 2017 data period.
Simulated events include the contribution from additional $\Pp\Pp$ interactions within the same or adjacent bunch crossings (pileup),
and are weighted to reproduce the observed pileup distribution.
The MC samples are further processed through a dedicated simulation of the CMS detector based
on \GEANTfour~\cite{Agostinelli2003250}.

The selection of $4\ell$ events and associated particles closely follows the methods used in the analyses
of the Run~1~\cite{Chatrchyan:2013mxa} and Run~2~\cite{Sirunyan:2017exp} data sets.
The main triggers for the Run~2 analysis select either a pair of electrons or muons, or an electron and a muon.
The minimal transverse momentum of the leading electron (muon) is 23\,(17)\GeV,
while that of the subleading lepton is 12\,(8)\GeV. To maximize the signal acceptance,
triggers requiring three leptons with lower $\pt$ thresholds and no isolation requirement are also used,
as are isolated single-electron and single-muon triggers with thresholds of 27 and 22\GeV
in 2016, or 35 and 27\GeV in 2017, respectively. The overall trigger efficiency for
simulated signal events that pass the full selection chain of this analysis is larger than 99\%.
The trigger efficiency is measured in data using a sample of $4\ell$ events collected
by the single-lepton triggers and is found to be consistent with the expectation from simulation.

Event reconstruction is based on the particle-flow (PF) algorithm~\cite{Sirunyan:2017ulk},
which exploits information from all the CMS subdetectors to identify and reconstruct individual particles in the event.
The PF candidates are classified as charged hadrons, neutral hadrons, photons, electrons, or muons, and they are
then used to build higher-level objects such as jets and lepton isolation quantities.
Electrons (muons) are reconstructed within the geometrical acceptance defined by a requirement
on the pseudorapidity $\abs{\eta} < 2.5\,(2.4)$ for transverse momentum $\PT > 7\,(5)\GeV$ with
an algorithm that combines information from the ECAL (muon system) and the tracker.
A dedicated algorithm is used to collect the final-state radiation (FSR) of leptons~\cite{Sirunyan:2017exp}.

The reconstructed vertex with the largest value of summed physics-object $\pt^2$ is taken to be the
primary $\Pp\Pp$ interaction vertex. The physics objects are the jets and the associated missing transverse
momentum, taken as the negative vector sum of the \pt of those jets.
The jets are clustered using the anti-$\kt$ jet finding algorithm~\cite{Cacciari:2008gp,Cacciari:2011ma}
with a distance parameter of 0.4 and the associated tracks assigned to the vertex as inputs.
Jets must satisfy $\pt>30\GeV$ and $\abs{\eta}<4.7$ and must be separated
from all selected lepton candidates and any selected FSR photons with a requirement on the
distance parameter $\Delta R(\ell/\cPgg,{\mathrm{jet}})>0.4$, where $(\Delta R)^2 = (\Delta \phi)^2 + (\Delta \eta)^2$.
For event categorization, jets are tagged as $\PQb$-jets using the Combined Secondary Vertex
algorithm~\cite{Chatrchyan:2012jua,Sirunyan:2017ezt}, which combines information about impact
parameter significance, the secondary vertex, and jet kinematics.

Each lepton track is required to have the ratio of the impact parameter in three dimensions, 
which is computed with respect to the chosen primary vertex position, and its uncertainty to be less than 4.
To discriminate between leptons from prompt $\PZ$ boson decays and those arising from
hadron decays within jets, an isolation requirement for leptons is imposed in the analysis 
of the 2016 data~\cite{Sirunyan:2017exp}. For electrons, the isolation variable is included as part 
of the multivariate training inputs for electron identification in 2017.

We consider three mutually exclusive channels: $\PH\to4\Pe$, $4\Pgm$, and $2\Pe 2\Pgm$.
At least two leptons are required to have $\pt > 10$\GeV, and at least one is required to have $\pt > 20$\GeV.
All four pairs of oppositely charged leptons that can be built with the four leptons are required
to satisfy $m_{\ell^{+}\ell'^{-}} > 4$\GeV regardless of lepton flavor. The $\PZ$ candidates are required to satisfy 
the condition $12 < \mlplm < 120$\GeV, where the invariant mass of at least one of the $\PZ$ candidates must
be larger than 40\GeV. The region between 105 and 140\GeV in the four-lepton invariant mass \mell
is identified as the \onshell region, and the region above 220\GeV is identified as the \offshell region.

Different sources of leptons such as the decays of heavy flavor jets or light mesons may produce additional 
background to the \Hboson signal in any of these decay channels, or the \onshell and \offshell regions. 
We denote this background collectively as the \ZX background, and employ a data-driven method for its 
estimation and \mell dependence. The lepton misidentification rates are first derived using $\PZ+1\ell$ control 
regions with relaxed selection requirements on the third lepton, and the extracted rates are then applied 
on $\PZ+2\ell$ control regions, where the two additional leptons with relaxed selection requirements 
have the same lepton flavor but may have opposite charge~\cite{Chatrchyan:2013mxa,Sirunyan:2017exp}.

\section{Analysis techniques and categorization of events}
\label{sec:AnalysisStrategyIntro}

The full kinematic information from each event using either the \Hboson decay or associated
particles in its production is extracted using discriminants from matrix element calculations.
These discriminants use a complete set of mass and angular
input observables $\boldsymbol{\Omega}$~\cite{Gao:2010qx,Anderson:2013afp,Gritsan:2016hjl}
to describe kinematics at LO in QCD. The \PT of either the combined \Hboson and two-jet system for the
production discriminant (\eg, $\mathcal{D}^{\VBF/\VH}$), or the \Hboson itself for the decay
discriminants (\eg, $\mathcal{D}^\text{dec}$), or for their combination (\eg, $\mathcal{D}^{\VBF/\VH+{\text{dec}}}$)
is not included in the input observables.
This information is not used in the analysis of the \Hboson width and anomalous couplings, as
the \pt of the overall system is sensitive to QCD, parton shower, and underlying event uncertainties.

The kinematic discriminants used in this study are computed using the same \mela package that is utilized in simulation.
The signal includes both the four-lepton decay kinematics in the processes
$\PH \to \ZZ$ / $\PZ\gamma^*$ / $\gamma^*\gamma^*\to4\ell$, and kinematics of
associated particles in production $\PH$+jet, $\PH$+2jets, \VBF, \WH, \ZH, \ttH, \tqH, or \bbH.
The background includes \glufu or $\qqbar \to \ZZ$ / $\PZ\gamma^*$ / $\gamma^*\gamma^*$ / $\PZ\to 4\ell$ processes,
and \VBF or associated production with a \V boson of the \ZZ system.
Analytical algorithms are available for the cross-checks of the four-lepton kinematics in $\PH$ decay
and \VH associated production within the \mela framework and were adopted in the previous
CMS analyses~\cite{Chatrchyan:2012xdj,Chatrchyan:2013lba,Chatrchyan:2012jja}.

\begin{figure*}[!tbhp]
\centering
\includegraphics[width=0.32\textwidth]{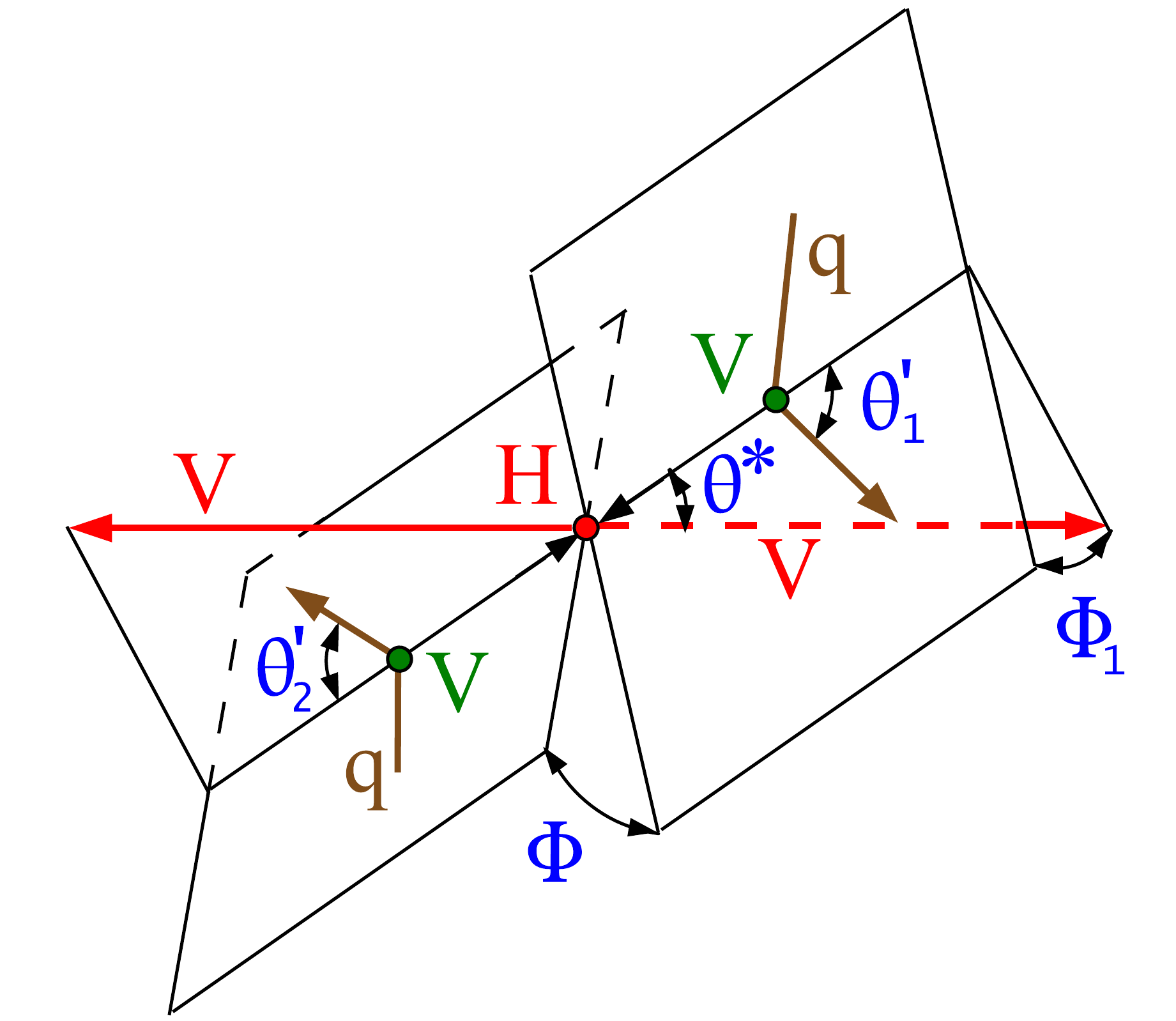}
\includegraphics[width=0.32\textwidth]{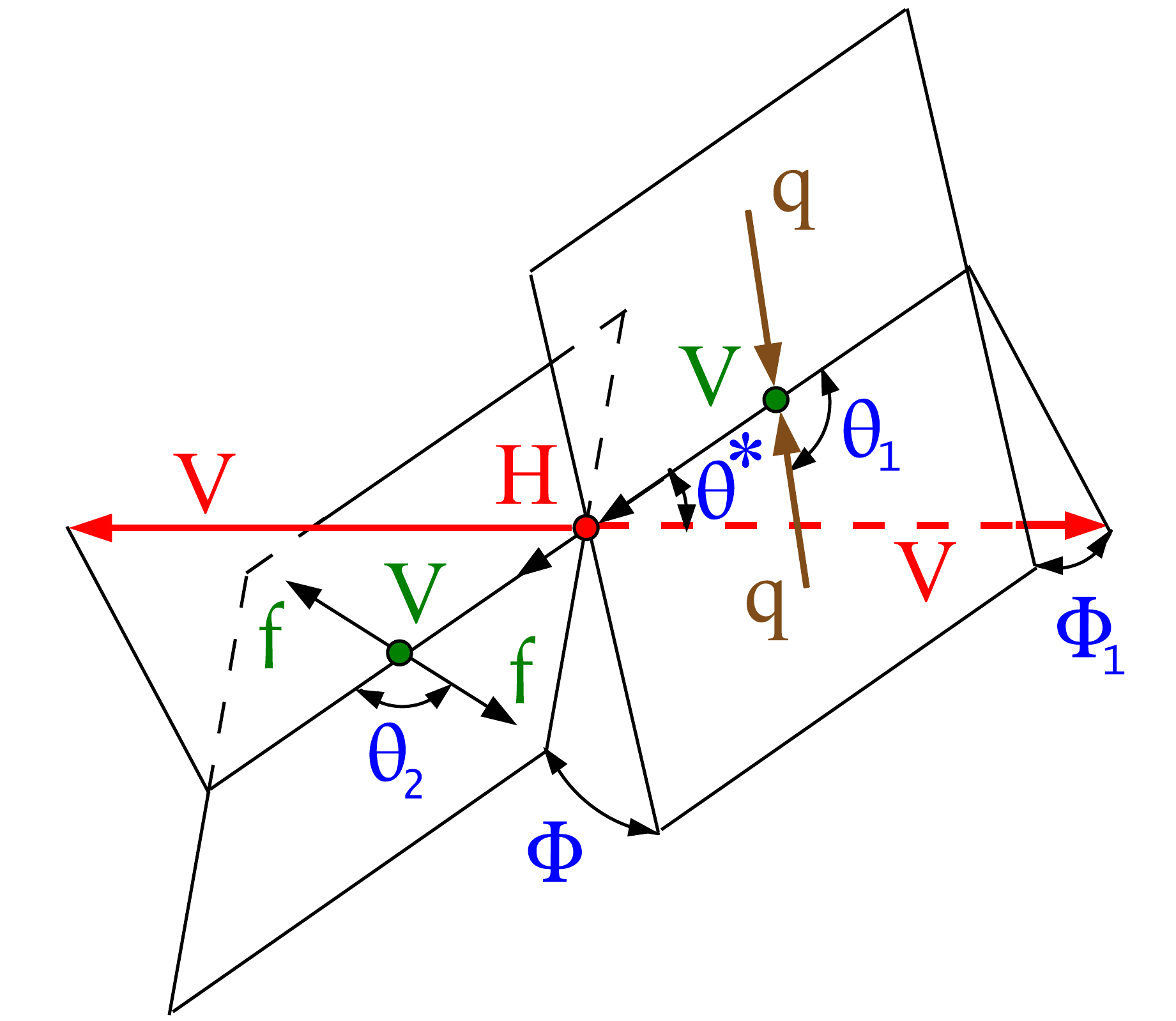}
\includegraphics[width=0.32\textwidth]{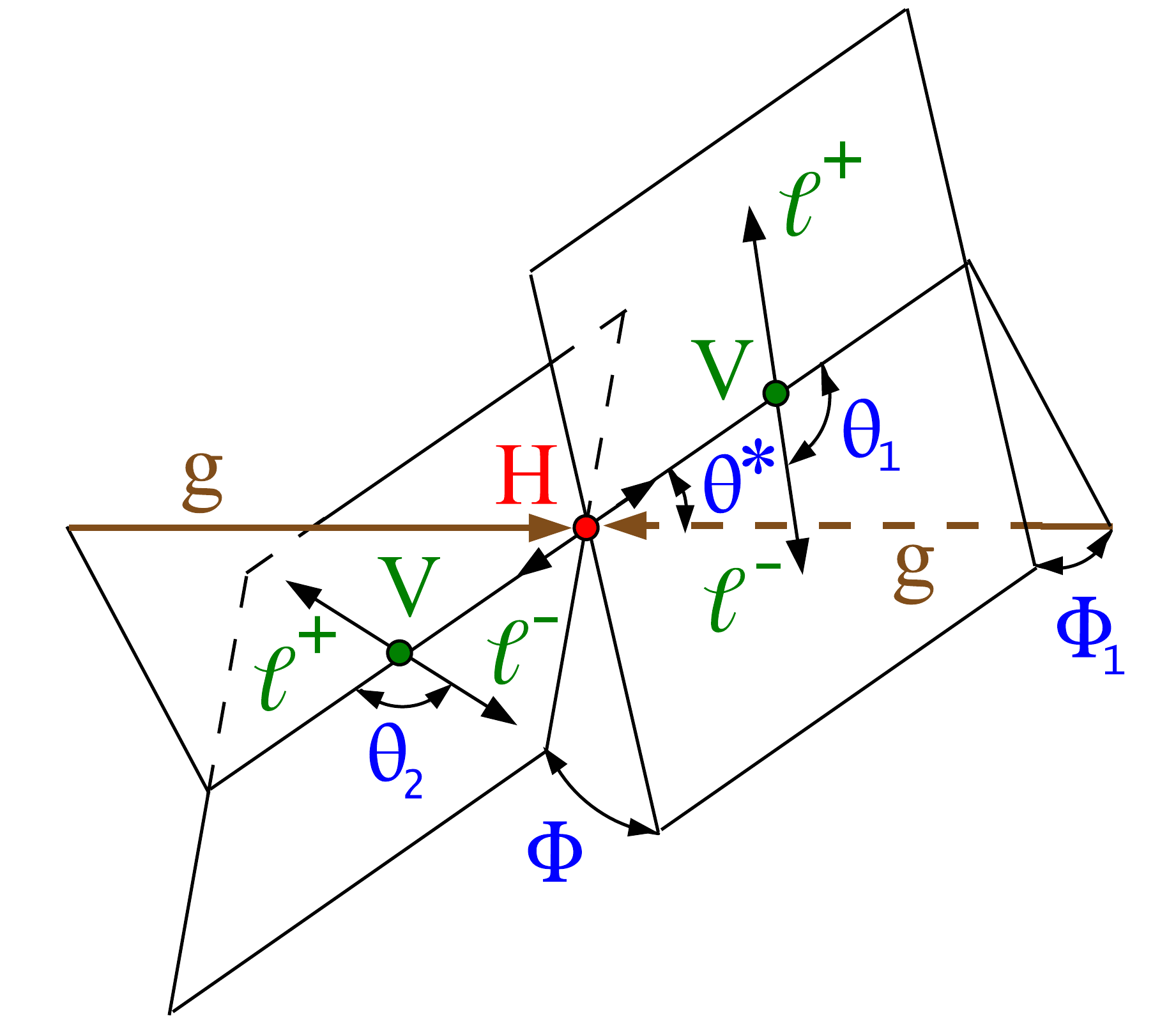}
\caption{
Three topologies of the \Hboson production and decay: vector boson fusion $\qq\to \VV (\qq) \to \PH (\qq) \to \VV (\qq)$ (left);
associated production $\qq\to \V \to \V\PH \to (\ffbar)\ \PH \to (\ffbar)\ \VV$ (middle); and gluon fusion $\Pg\Pg \to \PH \to \VV \to 4\ell$ (right)
representing the topology without associated particles.
The incoming particles are shown in brown, the intermediate vector bosons and their fermion daughters are shown in green,
the \Hboson and its vector boson daughters are shown in red, and angles are shown in blue.
In the first two cases the production and decay $\PH \to \VV$ are followed by the same four-lepton decay shown in the third case.
The angles are defined in either the $\PH$ or $\V$ boson rest frames~\cite{Gao:2010qx,Anderson:2013afp}.
}
\label{fig:kinematics}
\end{figure*}

Kinematic distributions of particles produced in the \Hboson decay or in association with \Hboson production
are sensitive to the quantum numbers and anomalous couplings of the \Hboson.
In the $1\to4$ process of the $\PH \to 4 \cmsSymbolFace{f}$ decay, six observables fully
characterize kinematics of the decay products
$\boldsymbol{\Omega}^{\text{decay}}=\{\theta_1, \theta_2, \Phi, m_1, m_2, m_{4\cmsSymbolFace{f}} \}$,
while two other angles relate orientation of the decay frame with respect to the production axis,
$\boldsymbol{\Omega}^{\text{prod}}=\{\theta^*, \Phi_1 \}$, as described in Ref.~\cite{Gao:2010qx}.
Moreover, two sets of observables,
$\boldsymbol{\Omega}^{\text{assoc,}\,\VBF}=\{\theta_1^{\VBF}, \theta_2^{\VBF}, \Phi^{\VBF}, q_1^{2,\VBF}, q_2^{2,\VBF} \}$
for the \VBF process and
$\boldsymbol{\Omega}^{\text{assoc,}\,\VH}=\{\theta_1^{\VH}, \theta_2^{\VH}, \Phi^{\VH}, q_1^{2,{\VH}}, q_2^{2,\VH} \}$
for the \VH process, can also be defined in a similar way to $\boldsymbol{\Omega}^{\mathrm{decay}}$ for \Hboson
associated production~\cite{Anderson:2013afp}.  As a result, 13 kinematic observables,
illustrated in Fig.~\ref{fig:kinematics}, are defined for the $2\to 6$ associated production process
with subsequent \Hboson decay to a four-fermion final state.

With up to 13 observables, $\boldsymbol{\Omega}$, sensitive to the \Hboson anomalous couplings
in \Eq{eq:formfact-fullampl-spin0}, it is a challenging task to perform an optimal analysis in a multidimensional
space of observables. The \mela approach introduced earlier is designed to reduce the number of observables
to the minimum while retaining all essential information. Two types of discriminants were defined for either
the production or decay process, and we also combine them into a joint discriminant for the full $2\to 6$ process
where relevant.

These types of discriminants are
\begin{equation}
\mathcal{D}_\mathrm{alt}\left(\boldsymbol{\Omega}\right) = \frac{\mathcal{P}_\text{sig}\left(\boldsymbol{\Omega}\right) }
        {\mathcal{P}_\text{sig}\left(\boldsymbol{\Omega}\right) +\mathcal{P}_\mathrm{alt}\left(\boldsymbol{\Omega}\right) }
\label{eq:melaD}
\end{equation}
and
\begin{equation}
\mathcal{D}_\mathrm{int}\left(\boldsymbol{\Omega}\right) =
\frac{\mathcal{P}_\mathrm{int}\left(\boldsymbol{\Omega}\right) }
{2 \ \sqrt{{\mathcal{P}_\text{sig}\left(\boldsymbol{\Omega}\right) \ \mathcal{P}_\mathrm{alt}\left(\boldsymbol{\Omega}\right) }}},
\label{eq:melaDint}
\end{equation}
where the probability of a certain process $\mathcal{P}$ is calculated using the full kinematics characterized
by $\boldsymbol{\Omega}$ for the processes denoted as ``sig'' for a signal model and ``alt'' for an alternative model,
which could be an alternative \Hboson production mechanism (used to categorize events),
background (to isolate signal), or an alternative \Hboson coupling model (to measure coupling parameters).
The ``int'' label represents the interference between the two model contributions.
The probabilities $\mathcal{P}$ are calculated from the matrix elements provided by the \mela package and
are normalized to give the same integrated cross sections in the relevant phase space of each process.
Such normalization leads to a balanced distribution of events in the range between 0 and 1
of the $\mathcal{D}_\mathrm{alt}$ discriminants, and between $-1$ and $1$ of $\mathcal{D}_\mathrm{int}$.
One can apply the Neyman-Pearson lemma to prove that the two discriminants in \Eqs{eq:melaD}{eq:melaDint}{and}
become the minimal and complete set of optimal observables for the purpose of separating the two processes
``sig'' and ``alt'' while including their interference as well~\cite{Anderson:2013afp,Gritsan:2016hjl}.

\begin{figure*}[!tbhp]
\centering
\includegraphics[width=0.45\textwidth]{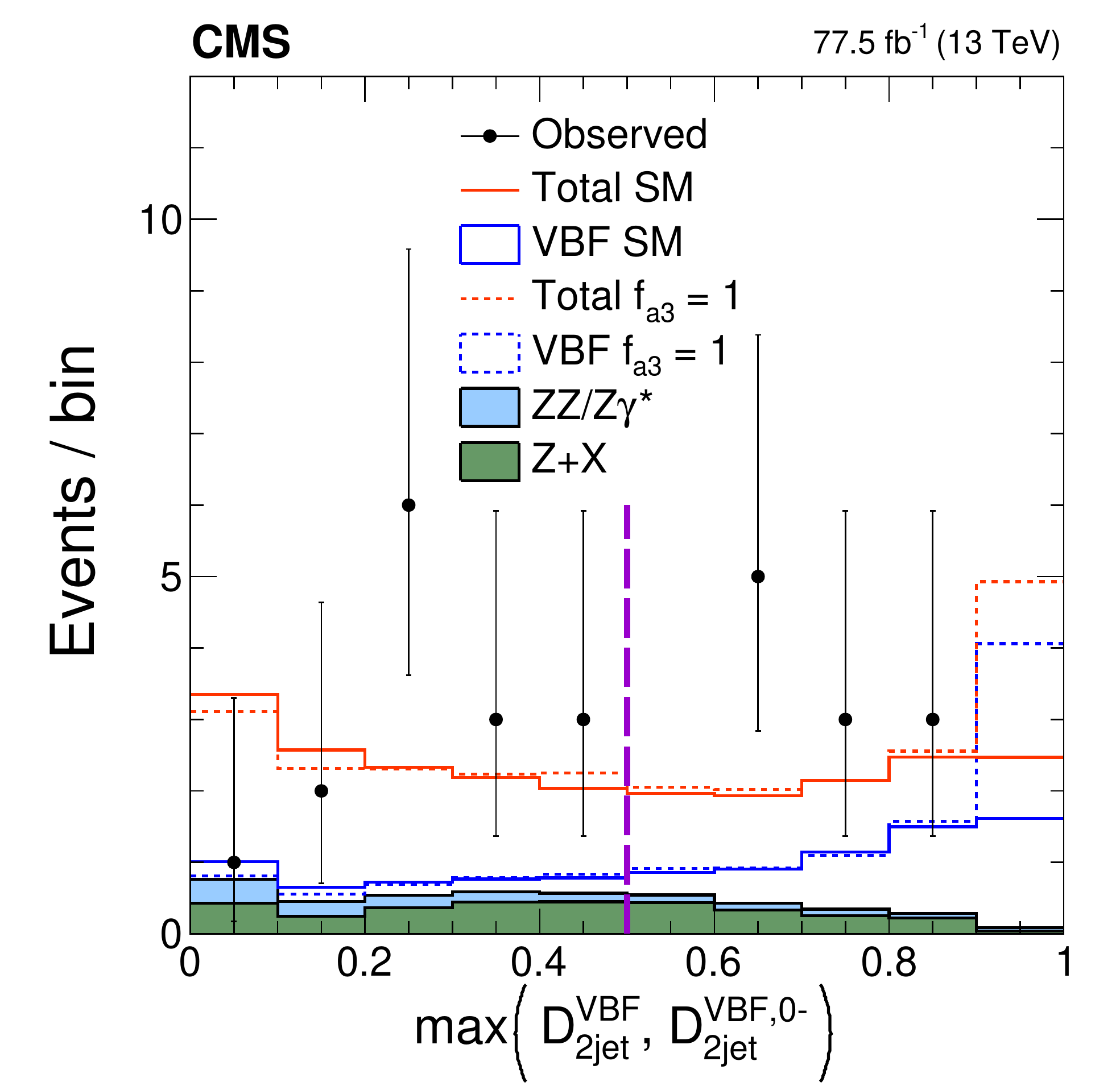}
\includegraphics[width=0.45\textwidth]{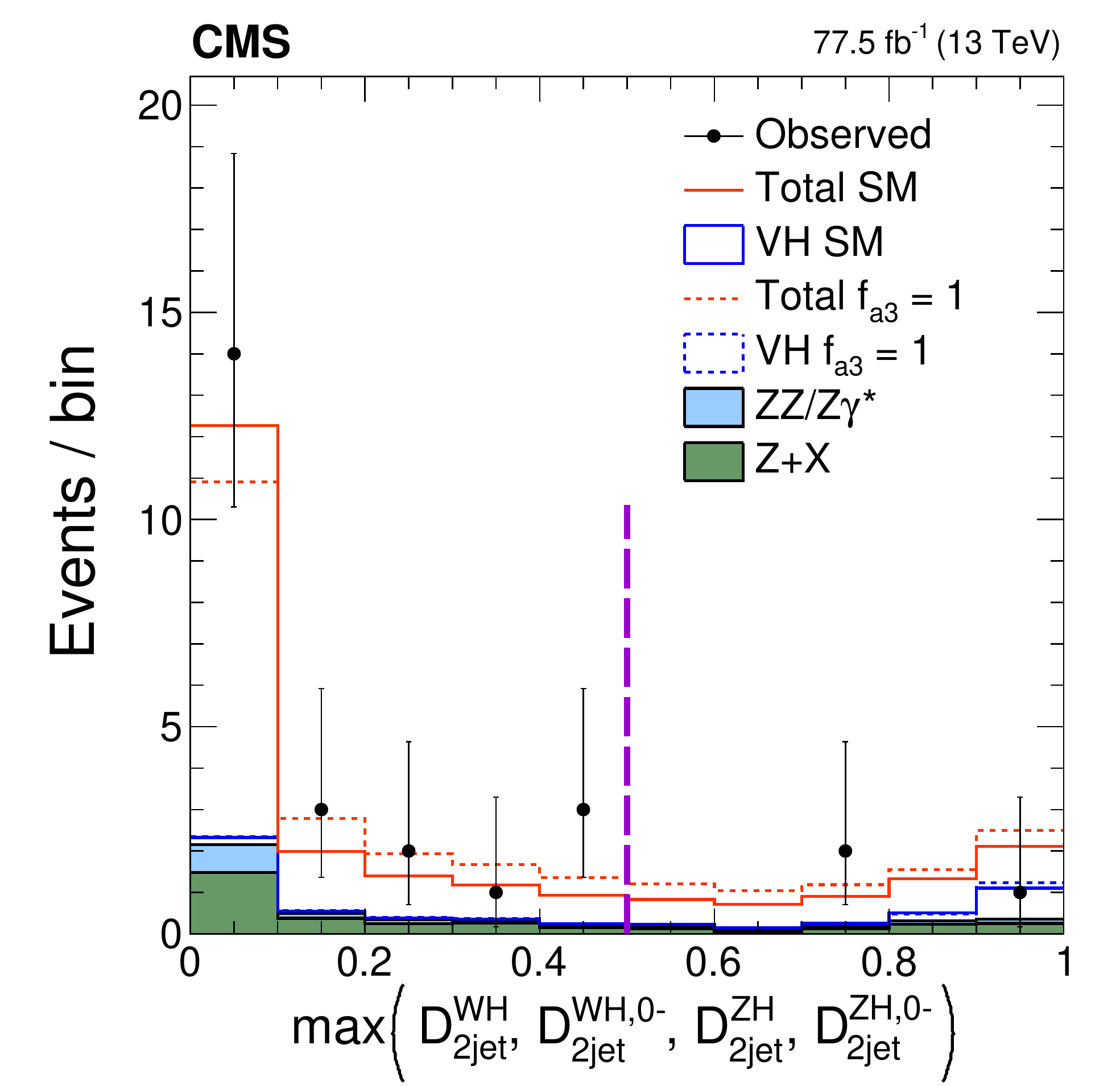}
\caption{
The distributions of events
for $\max \left( \mathcal{D}_\text{2jet}^{\VBF}, \mathcal{D}_\text{2jet}^{\VBF,{\mathrm{0-}}} \right)$ (left)
and
$\max \left( \mathcal{D}_\text{2jet}^{\WH}, \mathcal{D}_\text{2jet}^{\WH,{\mathrm{0-}}},
\mathcal{D}_\text{2jet}^{\ZH}, \mathcal{D}_\text{2jet}^{\ZH,{\mathrm{0-}}} \right)$ (right)
in the \onshell region in the data from 2016 and 2017 from the analysis of the \AC{3} coupling 
for a pseudoscalar contribution. The requirement $\Dbkg>0.5$ is applied in order to enhance 
the signal contribution over the background. The \VBF signal under both the SM and pseudoscalar 
hypotheses is enhanced in the region above $0.5$ for the former variable, and the \WH and \ZH 
signals are similarly enhanced in the region above $0.5$ for the latter variable. 
}
\label{fig:d2jet}
\end{figure*}

The selected events are split into three categories: \VBF-tagged, \VH-tagged, and untagged.
A set of discriminants $\mathcal{D}_\text{2jet}$ is constructed, following \Eq{eq:melaD},
where $\mathcal{P}_\text{sig}$ corresponds to the signal probability for the VBF ($\WH$ or $\ZH$)
production hypothesis in the \VBF-tagged (\VH-tagged) category, and $\mathcal{P}_\mathrm{alt}$
corresponds to that of \Hboson production in association with two jets via gluon fusion.
When more than two jets pass the selection criteria, the two jets with the highest $\PT$ are chosen 
for the matrix element calculations. Thereby, the $\mathcal{D}_\text{2jet}$ discriminants separate the 
target production mode of each category from gluon fusion production,
in all cases using only the kinematics of the \Hboson and two associated jets.
Figure~\ref{fig:d2jet} illustrates these discriminants, designed for the \VBF
or \VH signal enhancement in the \AC{3} coupling analysis for a pseudoscalar contribution. 
A selection based on the $\Dbkg$ observable, which utilizes information from the $4\ell$ decay kinematics 
and invariant mass, and which is discussed in more detail below, is applied in order to enhance the contribution 
of the signal over the background.

The three \onshell and \offshell categories are summarized
in Tables~\ref{table:categoriesonshell} and~\ref{table:categoriesoffshell},
and their sequential selection criteria are as follows:
\begin{itemize}
\item  ${\VBF}$\textit{-tagged} requires exactly four leptons, either two or three jets of which at most one
is $\PQb$-quark flavor-tagged, or more if none are $\PQb$-tagged jets,
and $\DjjVBF>0.5$ using either the SM or BSM signal hypothesis for the VBF production.
\item  ${\VH}$\textit{-tagged} requires exactly four leptons,
either two or three jets, or more if none are $\PQb$-tagged jets,
and $\DjjVH=\max \left( \DjjWH,\DjjZH \right)>0.5$ using either the SM or BSM signal hypothesis for the \VH production.
\item  \textit{Untagged} consists of the remaining events.
\end{itemize}
The requirements on the number of $\PQb$-tagged jets are applied to reduce crossfeed from $\ttH$ production.
Even though \VH cross sections are significantly lower with respect to VBF for $\mell>220$\GeV,
the \VH cross section becomes comparable to the \VBF cross section in the presence of anomalous couplings.
Therefore, the \offshell analysis also benefits from
featuring the \VH-tagged category with hadronic decays of the associated \V.
In either the \onshell or \offshell regions, events are not tagged for the smaller \VH contribution
with leptonic \V decays explicitly, but this contribution is taken into account in the simulation and
parametrization of the \VH process in the three different categories.
The expected and observed numbers of events are listed in Table~\ref{table:category-yields-onshell}
for the \onshell region and Table~\ref{table:category-yields-offshell} for the \offshell region.

\begin{table*}[!bht]
\centering
\topcaption{
Summary of the three production categories in the \onshell \mell region.
The selection requirements on the $\mathcal{D}_\text{2jet}$ discriminants are quoted for each category,
and further requirements can be found in the text.
Two or three observables (abbreviated as obs.) are listed for each analysis and for each category.
All discriminants are calculated with the \jhugen signal matrix elements and \MCFM background matrix elements.
The discriminants \Dbkg in the tagged categories also include probabilities using
associated jets and decay in addition to the \mell probability.
The \VH interference discriminants in the hadronic \VH-tagged categories are defined
as the simple average of the ones corresponding to the \WH and \ZH processes.
}
\renewcommand*{\arraystretch}{1.5}
\begin{scotch}{lccc}
   Category              & \VBF-tagged & \VH-tagged  & Untagged \\
\hline\\[-3ex]
Selection
 & $ \mathcal{D}_\text{2jet}^{\VBF}$ or $ \mathcal{D}_\text{2jet}^{\VBF,{\mathrm{BSM}}} >0.5$
 & $ \mathcal{D}_\text{2jet}^{\WH}$  or $ \mathcal{D}_\text{2jet}^{\WH,{\mathrm{BSM}}}$, or
 & Rest of events \\
 &
 & $ \mathcal{D}_\text{2jet}^{\ZH}$  or $ \mathcal{D}_\text{2jet}^{\ZH,{\mathrm{BSM}}} >0.5$
 &   \\
  SM obs.
 &  \mell, $\mathcal{D}^{{\VBF}+{\text{dec}}}_\text{bkg}$
 &  \mell, $\mathcal{D}^{{\VH}+{\text{dec}}}_\text{bkg}$
 &  \mell, $\mathcal{D}^\text{kin}_\text{bkg}$ \\
   $a_3$ obs.
 &  $\mathcal{D}_\text{bkg}$, $\mathcal{D}_{0-}^{\VBF+{\text{dec}}}$, $\mathcal{D}_{C\!P}^{\VBF}$
 &  $\mathcal{D}_\text{bkg}$, $\mathcal{D}_{0-}^{{\VH}+{\text{dec}}}$, $\mathcal{D}_{C\!P}^{\VH}$
 &  $\mathcal{D}_\text{bkg}$, $\mathcal{D}_{0-}^\text{dec}$, $\mathcal{D}_{C\!P}^\text{dec}$  \\
  $a_2$ obs.
 &  $\mathcal{D}_\text{bkg}$, $\mathcal{D}_{0h+}^{\VBF+{\text{dec}}}$, $\mathcal{D}_\mathrm{int}^{\VBF}$
 &  $\mathcal{D}_\text{bkg}$, $\mathcal{D}_{0h+}^{{\VH}+{\text{dec}}}$, $\mathcal{D}_\mathrm{int}^{\VH}$
 &  $\mathcal{D}_\text{bkg}$, $\mathcal{D}_{0h+}^\text{dec}$, $\mathcal{D}_\mathrm{int}^\text{dec}$  \\
  $\Lambda_1$ obs.
 &  $\mathcal{D}_\text{bkg}$, $\mathcal{D}_{\Lambda1}^{\VBF+{\text{dec}}}$, $\mathcal{D}_{0h+}^{\VBF+{\text{dec}}}$
 &  $\mathcal{D}_\text{bkg}$, $\mathcal{D}_{\Lambda1}^{{\VH}+{\text{dec}}}$, $\mathcal{D}_{0h+}^{{\VH}+{\text{dec}}}$
 &  $\mathcal{D}_\text{bkg}$, $\mathcal{D}_{\Lambda1}^\text{dec}$,  $\mathcal{D}_{0h+}^\text{dec}$  \\
  $\Lambda_1^{\PZ\gamma}$ obs.
 &  $\mathcal{D}_\text{bkg}$,  $\mathcal{D}^{\PZ\gamma,{\VBF+{\text{dec}}}}_{\Lambda1}$,  $\mathcal{D}_{0h+}^{\VBF+{\text{dec}}}$
 &  $\mathcal{D}_\text{bkg}$,  $\mathcal{D}^{\PZ\gamma,{{\VH}+{\text{dec}}}}_{\Lambda1}$,  $\mathcal{D}_{0h+}^{{\VH}+{\text{dec}}}$
 &  $\mathcal{D}_\text{bkg}$, $\mathcal{D}^{\PZ\gamma,{\text{dec}}}_{\Lambda1}$,  $\mathcal{D}_{0h+}^\text{dec}$  \\
\end{scotch}
\label{table:categoriesonshell}
\end{table*}

\begin{table*}[!bthp]
\centering
\topcaption{
Summary of the three production categories in the \offshell \mell region, listed in a similar manner, as in Table~\ref{table:categoriesonshell}.
All discriminants are calculated with the \jhugen or \MCFM/\jhugen signal, and \MCFM background matrix elements.
The \VH interference discriminant in the SM-like analysis hadronic \VH-tagged category is defined as the simple average
of the ones corresponding to the \WH and \ZH processes.
}
\renewcommand*{\arraystretch}{1.5}
\begin{scotch}{lccc}
   Category              & \VBF-tagged & \VH-tagged  & Untagged \\
\hline\\[-3ex]
  Selection
 & $ \mathcal{D}_\text{2jet}^{\VBF}$ or $ \mathcal{D}_\text{2jet}^{\VBF,{\mathrm{BSM}}} >0.5$
 & $ \mathcal{D}_\text{2jet}^{\WH}$  or $ \mathcal{D}_\text{2jet}^{\WH,{\mathrm{BSM}}}$, or
 & Rest of events \\
 &
 & $ \mathcal{D}_\text{2jet}^{\ZH}$  or $ \mathcal{D}_\text{2jet}^{\ZH,{\mathrm{BSM}}} >0.5$
 &   \\
  SM obs.
  &  \mell, $\mathcal{D}^{{\VBF}+{\text{dec}}}_\text{bkg}$, $\mathcal{D}_\mathrm{bsi}^{{\VBF}+{\text{dec}}}$
 &  \mell, $\mathcal{D}^{{\VH}+{\text{dec}}}_\text{bkg}$, $\mathcal{D}_\mathrm{bsi}^{{\VH}+{\text{dec}}}$
 &  \mell, $\mathcal{D}^\text{kin}_\text{bkg}$, $\mathcal{D}_\mathrm{bsi}^{{\glufu},{\text{dec}}}$  \\
   $a_3$ obs.
 &  \mell, $\mathcal{D}^{{\VBF}+{\text{dec}}}_\text{bkg}$, $\mathcal{D}_{0-}^{\VBF+{\text{dec}}}$
 &  \mell, $\mathcal{D}^{{\VH}+{\text{dec}}}_\text{bkg}$, $\mathcal{D}_{0-}^{{\VH}+{\text{dec}}}$
 &  \mell, $\mathcal{D}^\text{kin}_\text{bkg}$, $\mathcal{D}_{0-}^\text{dec}$  \\
  $a_2$ obs.
 &  \mell, $\mathcal{D}^{{\VBF}+{\text{dec}}}_\text{bkg}$, $\mathcal{D}_{0h+}^{\VBF+{\text{dec}}}$
 &  \mell, $\mathcal{D}^{{\VH}+{\text{dec}}}_\text{bkg}$, $\mathcal{D}_{0h+}^{{\VH}+{\text{dec}}}$
 &  \mell, $\mathcal{D}^\text{kin}_\text{bkg}$, $\mathcal{D}_{0h+}^\text{dec}$  \\
  $\Lambda_1$ obs.
 &  \mell, $\mathcal{D}^{{\VBF}+{\text{dec}}}_\text{bkg}$, $\mathcal{D}_{\Lambda1}^{\VBF+{\text{dec}}}$
 &  \mell, $\mathcal{D}^{{\VH}+{\text{dec}}}_\text{bkg}$, $\mathcal{D}_{\Lambda1}^{{\VH}+{\text{dec}}}$
 &  \mell, $\mathcal{D}^\text{kin}_\text{bkg}$, $\mathcal{D}_{\Lambda1}^\text{dec}$  \\
\end{scotch}
\label{table:categoriesoffshell}
\end{table*}

\begin{table*}[!tbhp]
\centering
\topcaption{
The numbers of events expected in the SM (or $\fAC{3}=1$ in parentheses) for the different
signal and background contributions and the total numbers of observed events are listed across
the three \AC{3} analysis categories in the \onshell region for the combined 2016 and 2017 data set.
}
\begin{scotch}{lccc}
 & \VBF-tagged & \VH-tagged & Untagged \\
\hline
\VBF signal & 4.7 (3.4) & 0.3 (0.2) & 5.7 (0.8)\\
\WH signal & 0.3 (0.6) & 0.7 (1.9) & 2.1 (5.3)\\
\ZH signal & 0.2 (0.4) & 0.5 (1.0) & 1.5 (2.5)\\
\VV background & 0.2 & 0.1 & 0.5\\[\cmsTabSkip]
\glufu signal & 5.5 (5.8) & 3.2 (3.3) & 98.9 (98.4)\\
\glufu background & 0.8 & 0.3 & 12.7\\
\ttH signal & 0.2 (0.2) & 0.1 (0.1) & 1.1 (1.2)\\
\bbH signal & 0.1 (0.1) & 0.1 (0.1) & 1.1 (1.1)\\[\cmsTabSkip]
$\qqbar\to4\ell$ background& 1.6 & 1.5 & 120.3\\
\ZX background & 5.2 & 3.0 & 46.3\\[\cmsTabSkip]
Total expected & 18.8 (18.2) & 9.7 (11.4) & 290.3 (289.1)\\
Total observed & 19 & 9 & 332\\
\end{scotch}
\label{table:category-yields-onshell}
\end{table*}

\begin{table*}[!bthp]
\centering
\topcaption{
The numbers of events expected in the SM-like analysis (or $\fAC{3}=0$ in the \AC{3} analysis categorization, divided with a vertical bar)
for the different signal and background contributions and the total observed numbers of events are listed across the three
SM~$|$~\AC{3} analysis categories in the \offshell region for the combined 2016 and 2017 data set.
The signal, background, and interference contributions are shown separately for the gluon fusion (\glufu)
and EW processes (\VV) under the $\GH=\GHSM$ assumption.
}
\begin{scotch}{lccc}
 & \VBF-tagged & \VH-tagged & Untagged \\
\hline
\VV signal & 1.0~$|$~1.2 & 0.3~$|$~0.3 & 3.3~$|$~3.1 \\
\VV background & 7.3~$|$~9.9 & 2.5~$|$~2.8 & 16.2~$|$~13.3 \\
\VV interference & $-$1.8~$|$~$-$2.1 & 0.06~$|$~0.03 & $-$2.4~$|$~$-$2.2 \\[\cmsTabSkip]
\glufu signal & 1.0~$|$~1.6 & 0.8~$|$~1.0 & 20.3~$|$~19.5 \\
\glufu background & 10.4~$|$~16.4 & ~~8.7~$|$~10.4 & 245.9~$|$~238.1 \\
\glufu interference & $-$1.6~$|$~$-$2.6 & $-$1.4~$|$~$-$1.6 & $-$34.4~$|$~$-$33.0 \\[\cmsTabSkip]
$\qqbar\to4\ell$ background & 15.8~$|$~33.5 & 27.8~$|$~31.2 & 992.0~$|$~970.8 \\
\ZX background & 2.4~$|$~6.4 & 2.8~$|$~3.3 & 45.4~$|$~40.8 \\[\cmsTabSkip]
Total expected & 34.4~$|$~64.8 & 41.6~$|$~47.5 & 1286.3~$|$~1251.0 \\
Total observed & 36~$|$~92 & 46~$|$~51 & 1325~$|$~1264 \\
\end{scotch}
\label{table:category-yields-offshell}
\end{table*}

In each category of events, typically three observables $\vec{x}$ are defined following \Eqs{eq:melaD}{eq:melaDint}{and},
as summarized in Tables~\ref{table:categoriesonshell} and~\ref{table:categoriesoffshell}.
In the \onshell region, except for the SM-like analysis, these are $\vec{x} = \{ \Dbkg, \mathcal{D}_{ai}, \Dint \}$.
The first observable, \Dbkg, is calculated differently in the three tagged categories. In the untagged category,
$\mathcal{P}_\text{bkg}$ is calculated for the dominant $\qqbar\to4\ell$ background process.  The signal and background
probabilities include both the matrix element probability based on the four-lepton kinematics and
the \mell probability parametrization extracted from simulation of detector effects.
The signal \mell parametrization assumes $\mH=125$\GeV.
In the \VBF-tagged and \VH-tagged categories, $\mathcal{P}_\text{bkg}$ and $\mathcal{P}_\text{sig}$ include
four-lepton kinematics and the \mell probability parametrization, but they also include kinematics of the
two associated jets. The $\mathcal{P}_\text{bkg}$ probability density represents the EW and QCD background processes $4\ell+2$\,jets,
while $\mathcal{P}_\text{sig}$ represents EW processes \VBF and \VH. It was found that jet kinematics in the
\Dbkg calculation improves separation of the targeted signal production both against background
and against the \Hboson gluon fusion production.
However, in the \offshell region and in the SM-like \onshell analysis, the four-lepton invariant mass \mell
is one of the most important observables, because the mass parametrization becomes an important feature of the analysis.
Therefore, the \mell parametrization is not used in the \Dbkg calculation in these cases,
and this is reflected with the superscript denoting which information is used, either with decay only information
in \Dbkgkin or with both decay and production in \DbkgVBFdec and \DbkgVHdec.

The other observable, \Dai, separates the SM hypothesis $\fai=0$ as $\mathcal{P}_\text{sig}$
from the alternative hypothesis $\fai=1$ as $\mathcal{P}_\mathrm{alt}$, following \Eq{eq:melaD}.
In the untagged category, the probabilities are calculated using only the decay information,
and the \Dai observable is called $\mathcal{D}_\mathrm{0-}$ in the \AC{3},
$\mathcal{D}_\mathrm{0h+}$ in the \AC{2}, $\mathcal{D}_{\Lambda1}$ in the \LC{1}, and
$\mathcal{D}_{\Lambda1}^{\PZ\gamma}$ in the \LZGs analyses~\cite{Khachatryan:2014kca}.
In the \VBF-tagged and \VH-tagged categories, both the production and decay probabilities are used,
with the matrix elements calculated as the product of the decay component and the component
from either \VBF production or $(\WH+\ZH)$ associated production, respectively~\cite{Sirunyan:2017tqd}.
The resultant set of \Dai discriminants are called in a similar manner to their counterparts in the untagged
category but indicating the production assumption in their upper index.

The last observable, \Dint defined in \Eq{eq:melaDint}, separates the interference
of the two amplitudes corresponding to the SM-like \Hboson coupling and the alternative \Hboson coupling model,
or the SM-like \Hboson coupling and background as an alternative model
in the case of \Dbsi for the signal-background interference in the \offshell region.
In the case of the \AC{3} analysis, this observable is called \Dcp because if $CP$ is violated
it would exhibit a distinctive forward-backward asymmetry.
In the untagged category, decay information is used in the calculation of \Dint.
In the \VBF-tagged and \VH-tagged categories, production information with the two associated jets is used.
The \Dbsi discriminant extends the idea of the \Dgg discriminant
introduced in Ref.~\cite{Khachatryan:2014iha} for the \Hboson width measurement, but allows independent
treatment of the interference component. It is used only in the SM-like analysis.

\begin{figure*}[htbp!]
\centering
\includegraphics[width=0.3\textwidth]{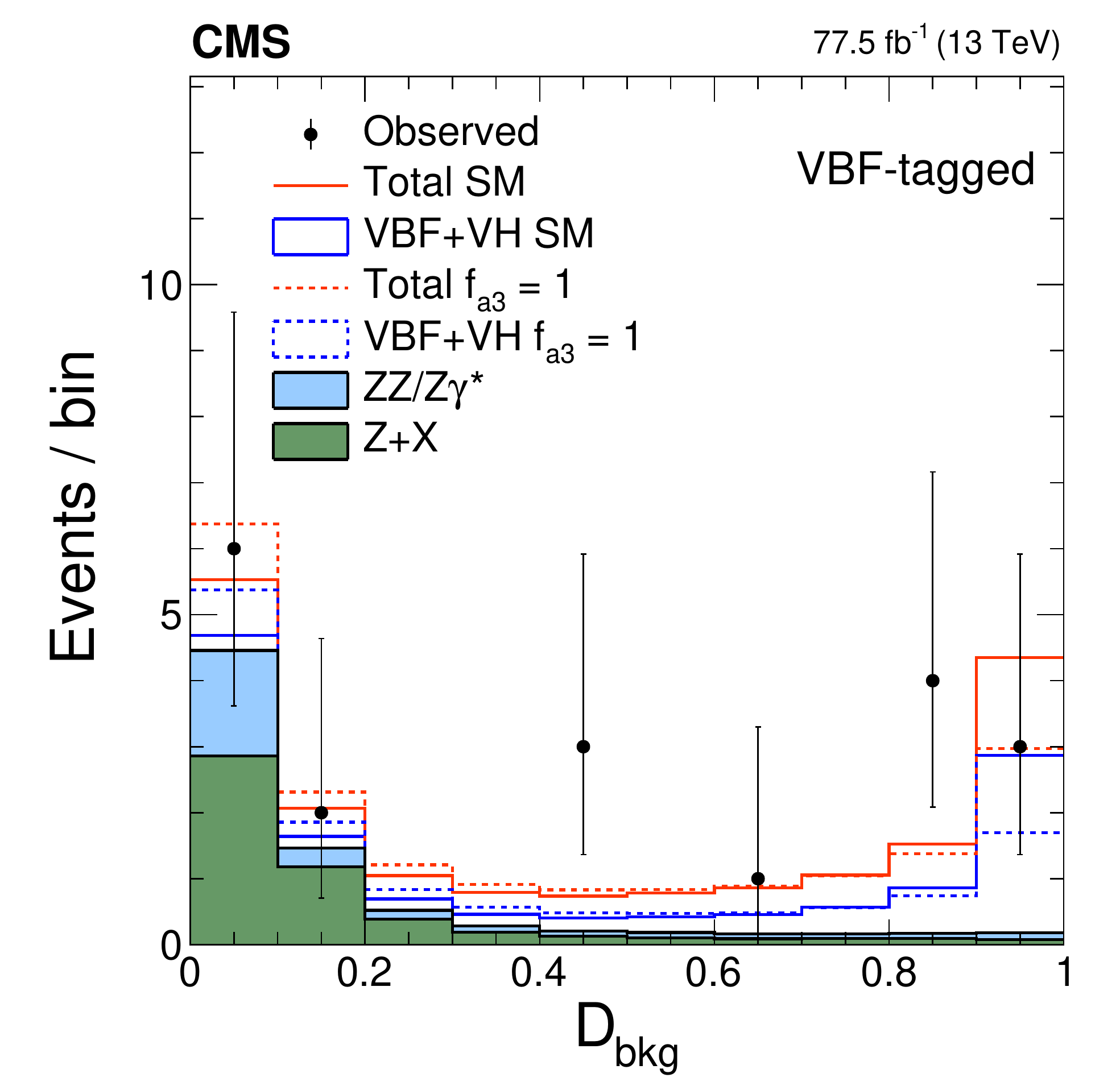}
\includegraphics[width=0.3\textwidth]{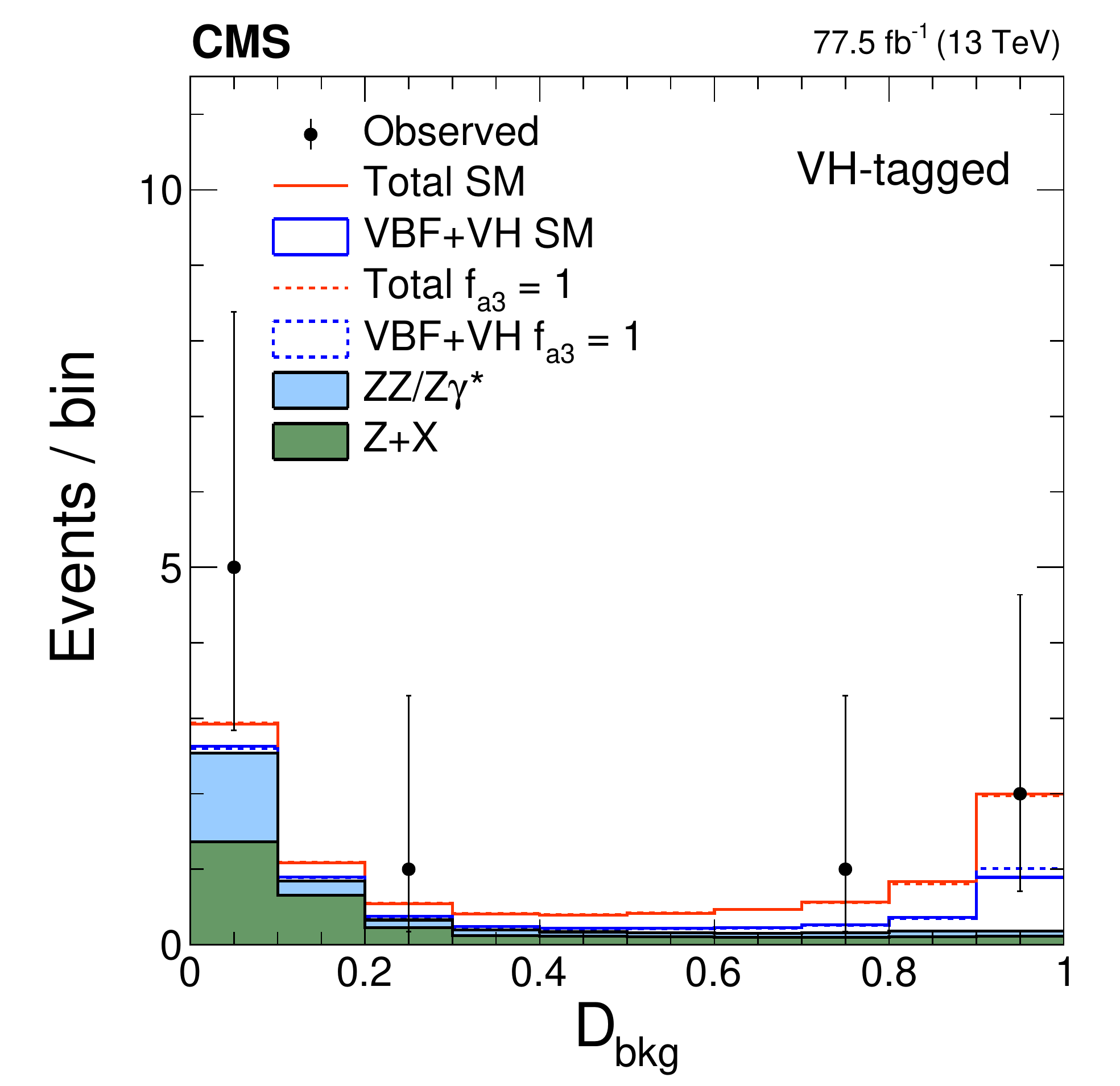}
\includegraphics[width=0.3\textwidth]{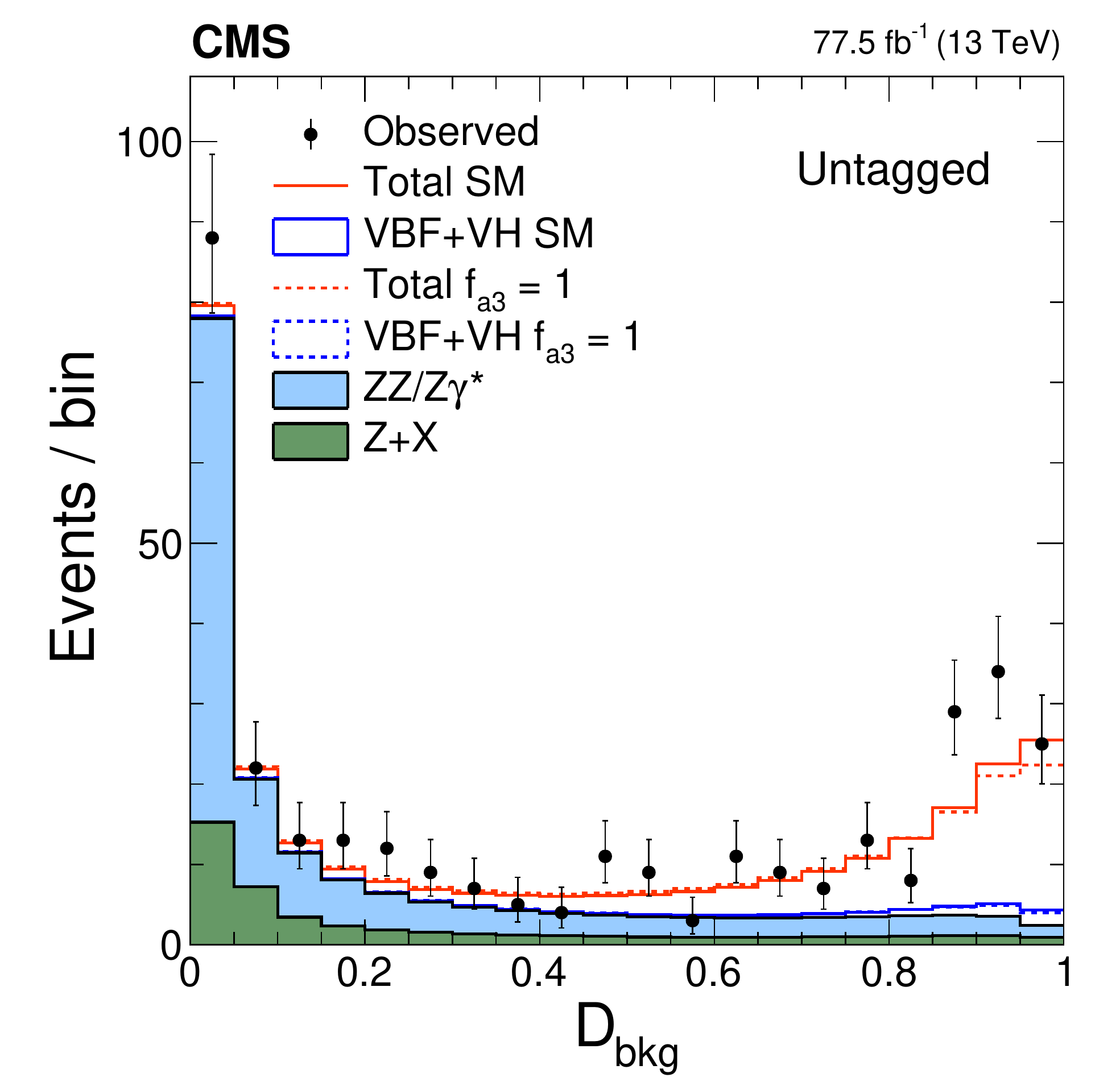} \\
\includegraphics[width=0.3\textwidth]{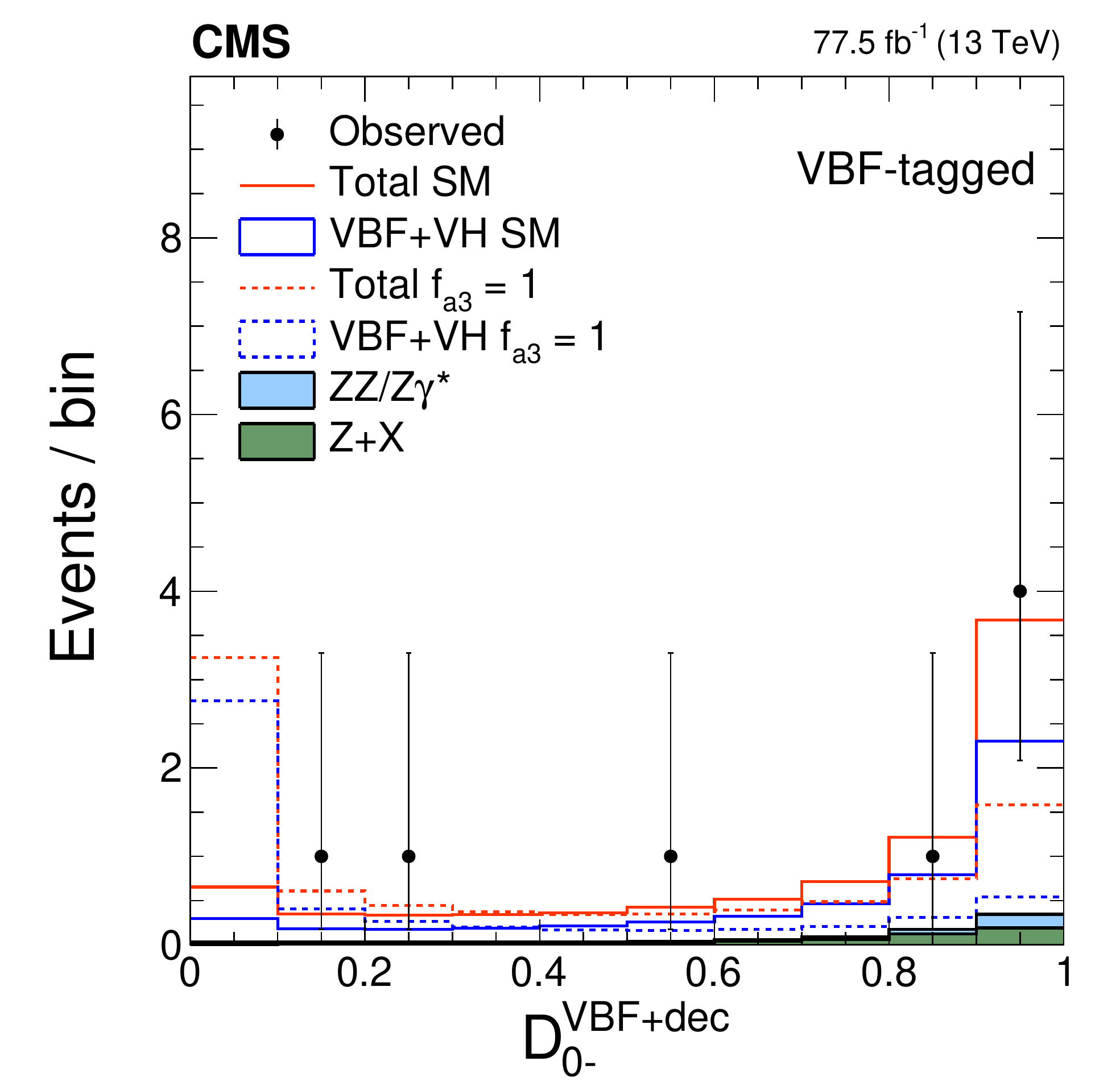}
\includegraphics[width=0.3\textwidth]{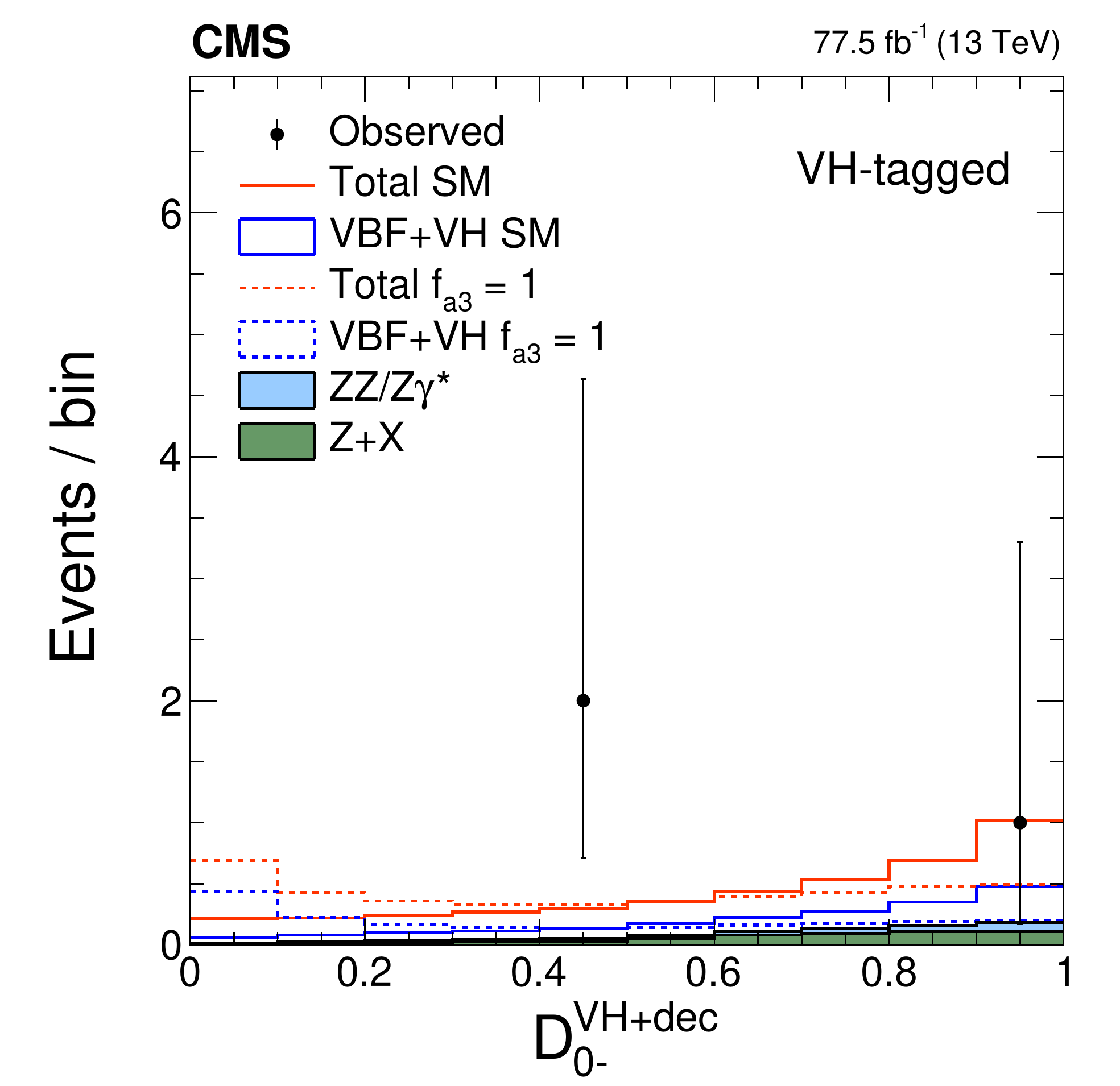}
\includegraphics[width=0.3\textwidth]{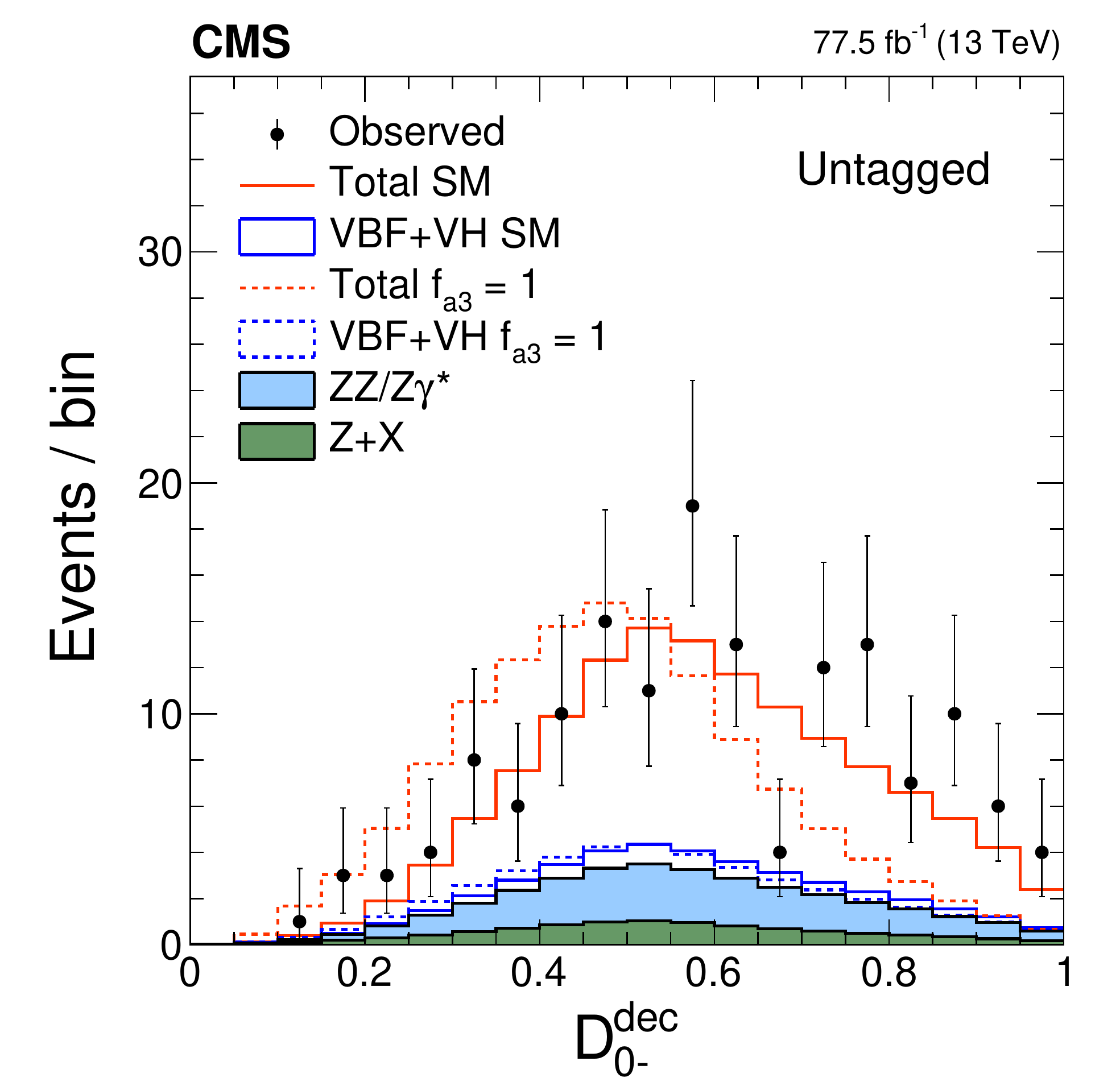} \\
\includegraphics[width=0.3\textwidth]{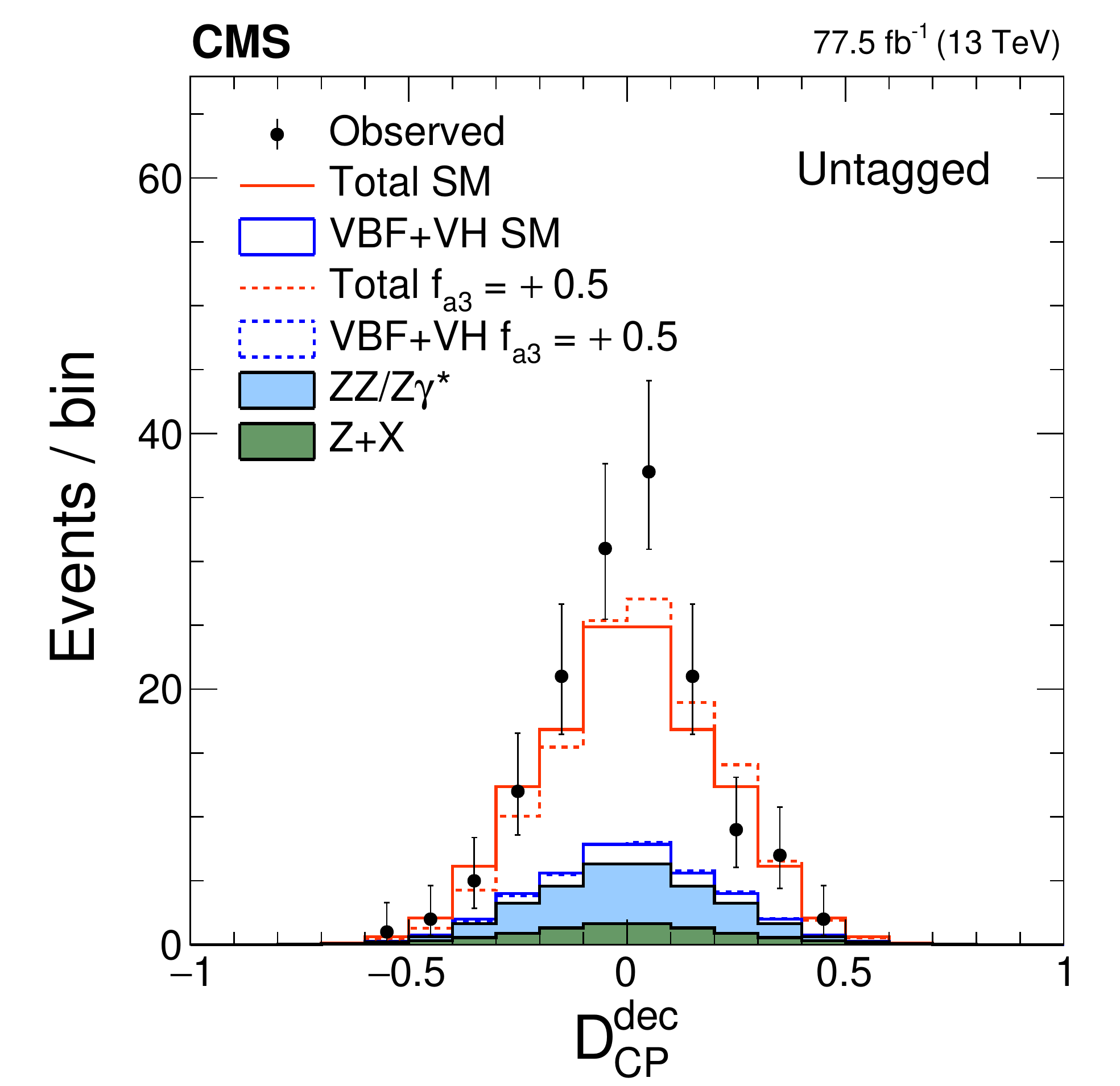}
\includegraphics[width=0.3\textwidth]{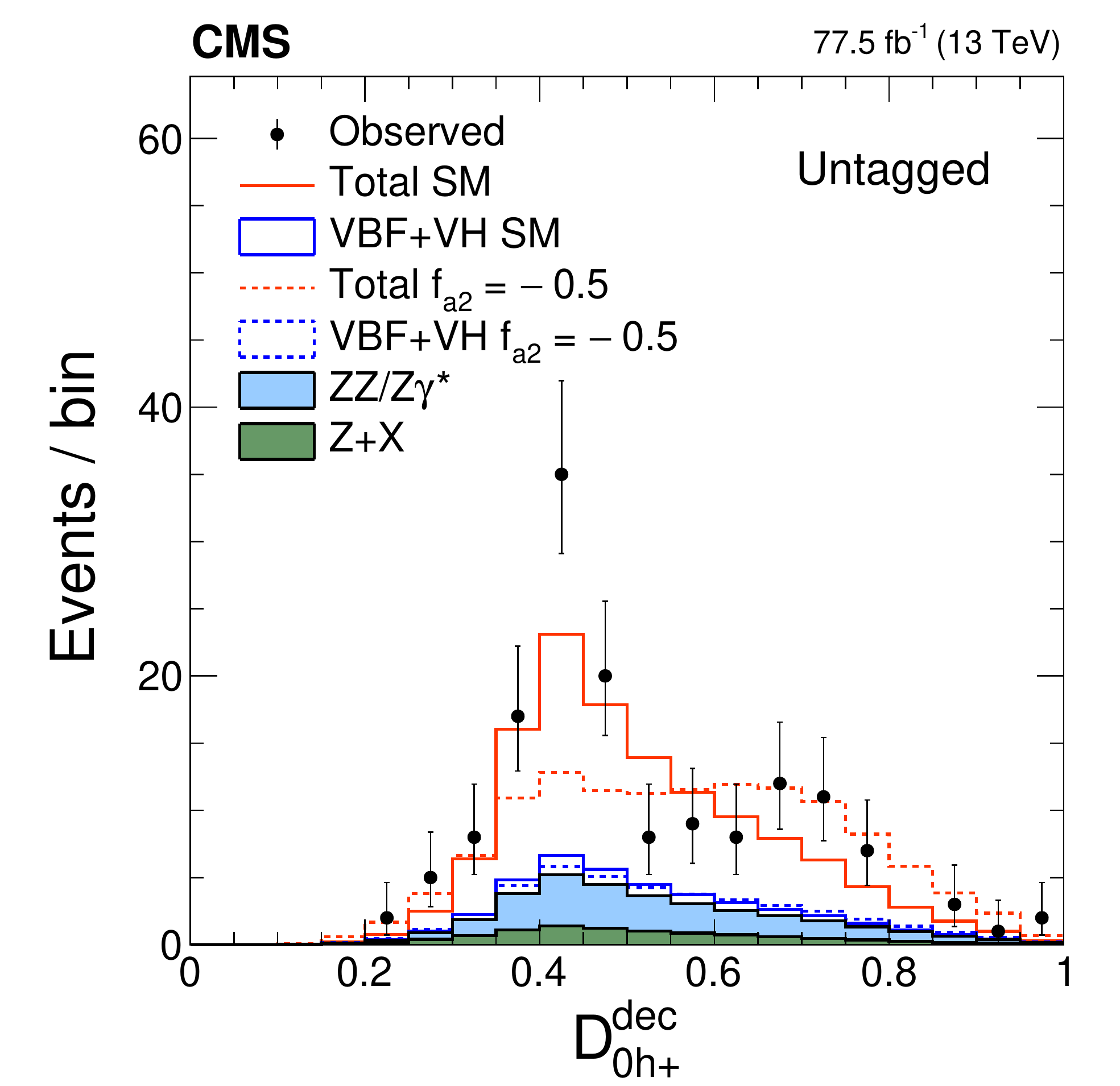}
\includegraphics[width=0.3\textwidth]{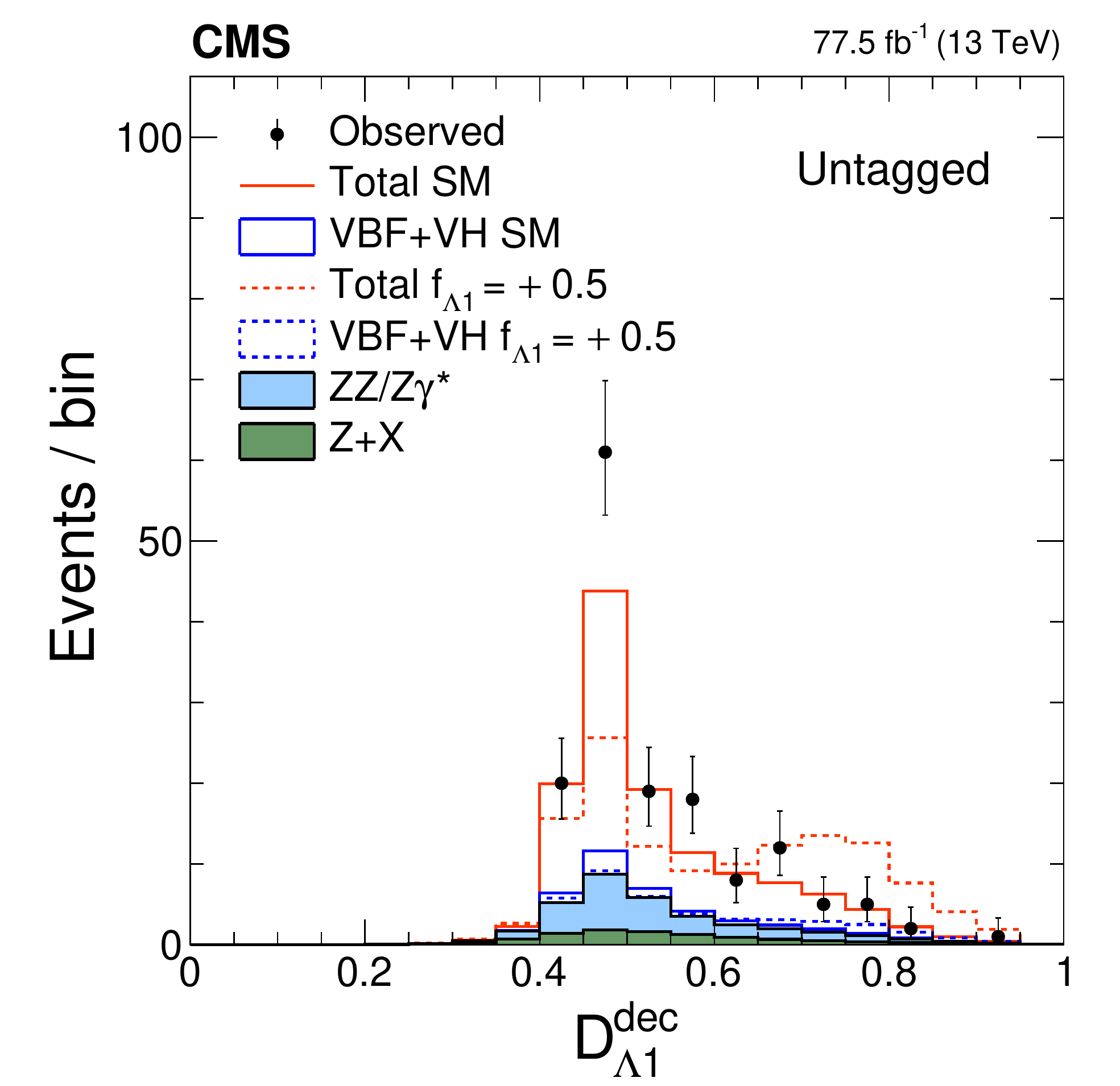}
\caption{
The distributions of events in the \onshell region in the data from 2016 and 2017.
The top row shows \Dbkg in the \VBF-tagged (left), \VH-tagged (middle), and untagged (right) categories
of the analysis of the \AC{3} coupling for a pseudoscalar contribution.
The rest of the distributions are shown with the requirement $\Dbkg>0.5$ in order to enhance signal
over background contributions.
The middle row shows $\mathcal{D}_{0-}$ in the corresponding three categories.
The bottom row shows $\mathcal{D}_{C\!P}^\text{dec}$ of the \AC{3},  $\mathcal{D}_{0h+}^\text{dec}$ of the \AC{2}, and
$\mathcal{D}_{\Lambda1}^\text{dec}$ of the \LC{1} analyses in the untagged categories.
}
\label{fig:stackPlotsOnshell}
\end{figure*}

\begin{figure*}[htbp!]
\centering
\includegraphics[width=0.3\textwidth]{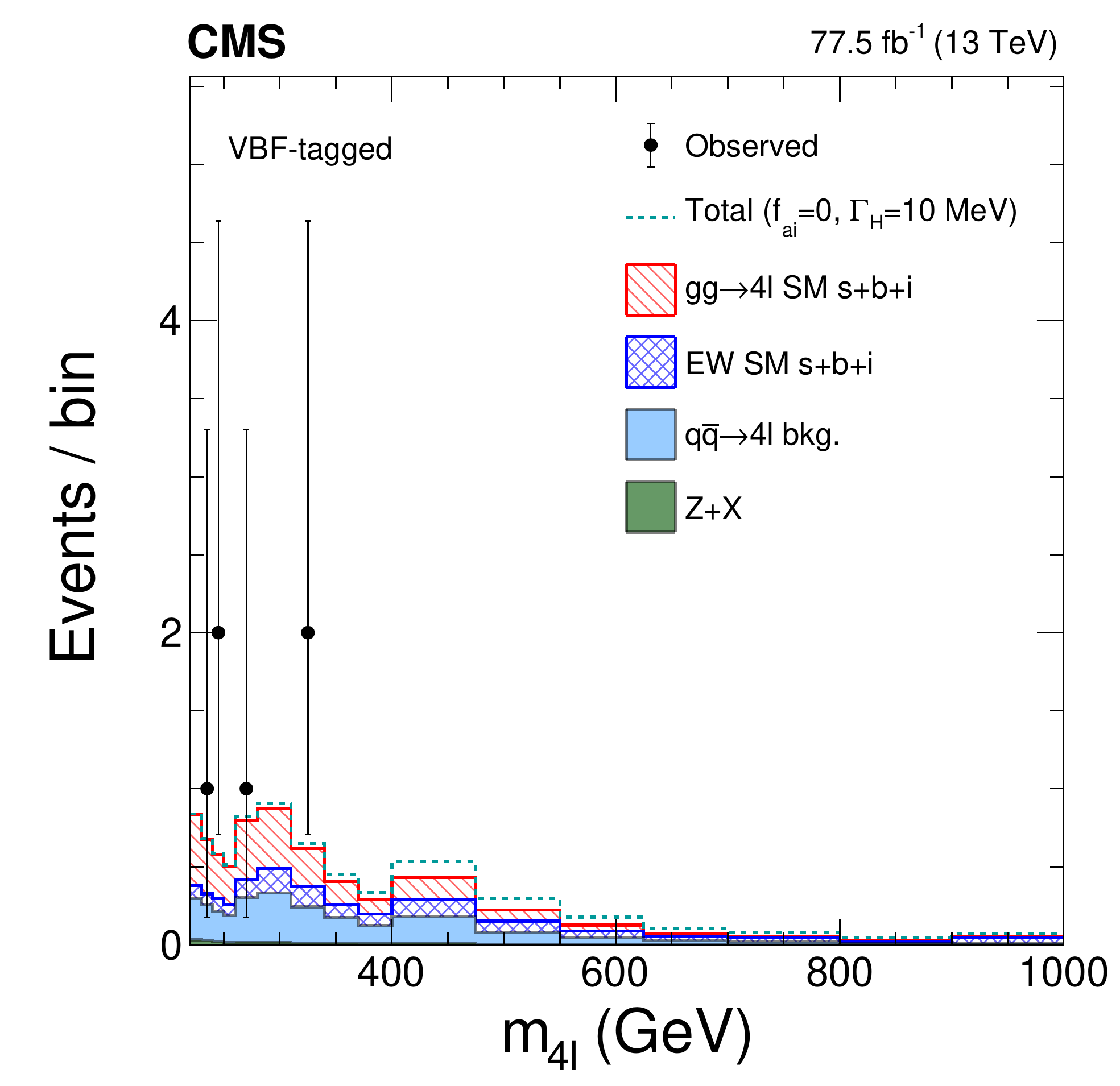}
\includegraphics[width=0.3\textwidth]{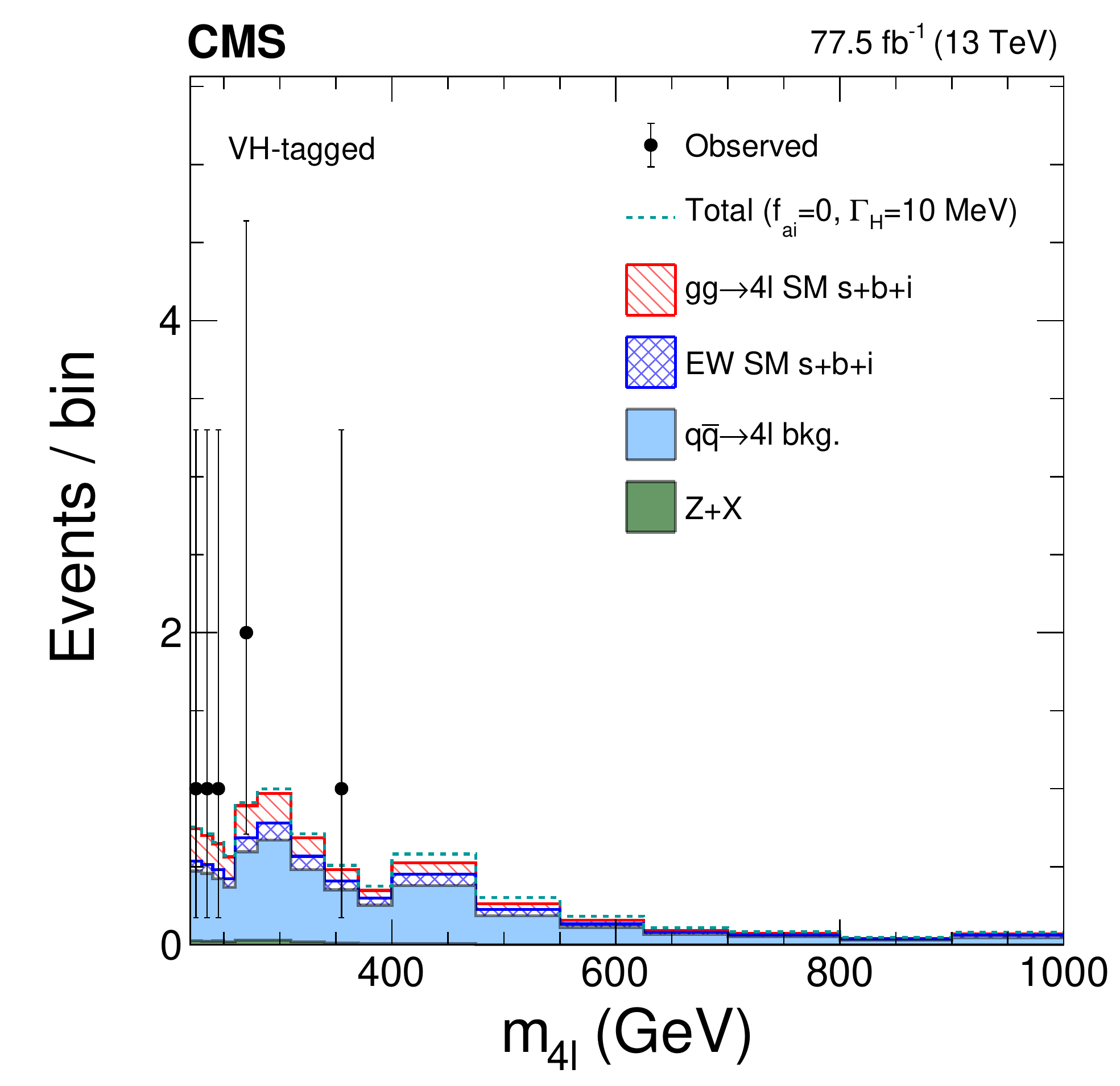}
\includegraphics[width=0.3\textwidth]{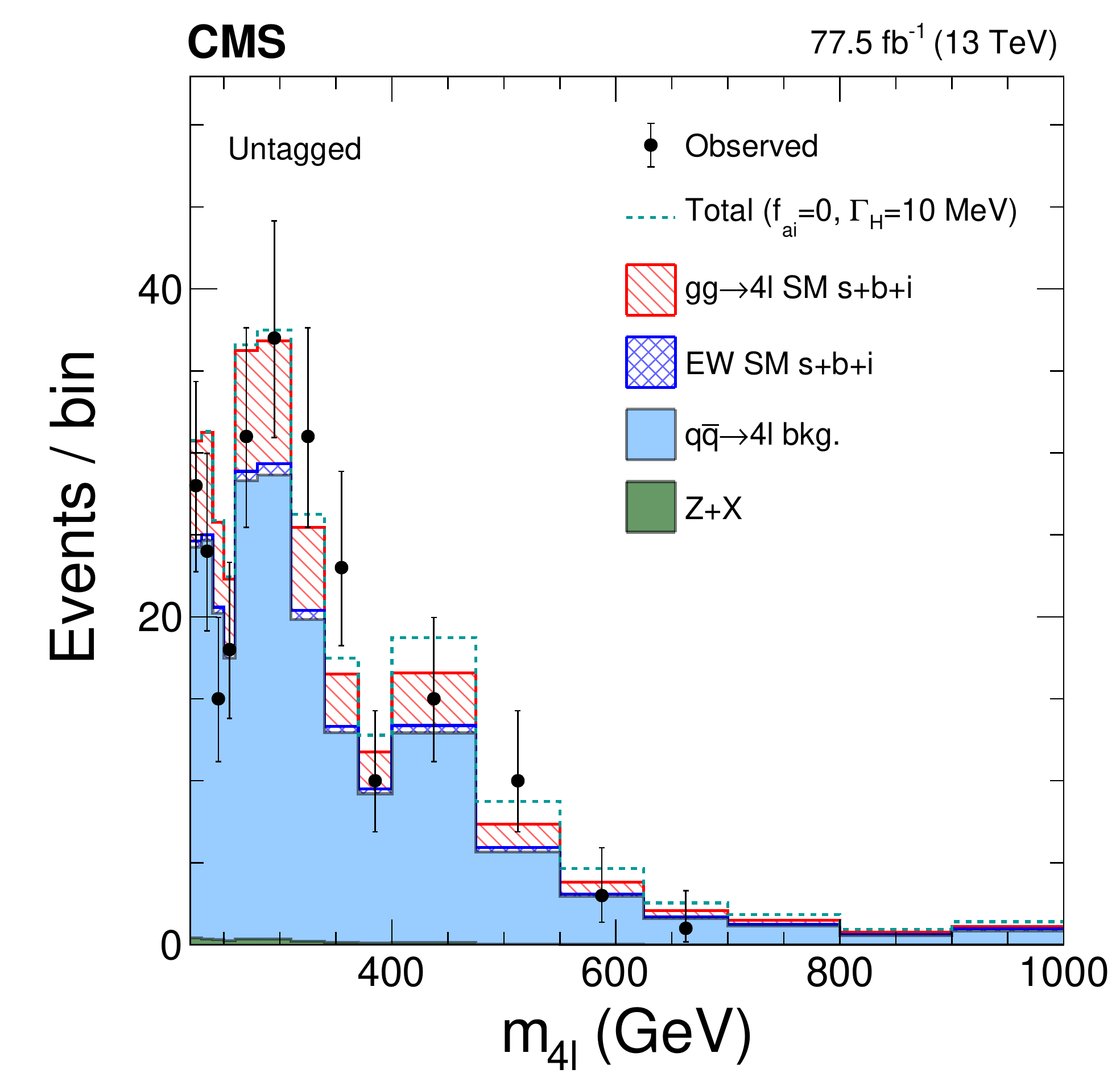} \\
\includegraphics[width=0.3\textwidth]{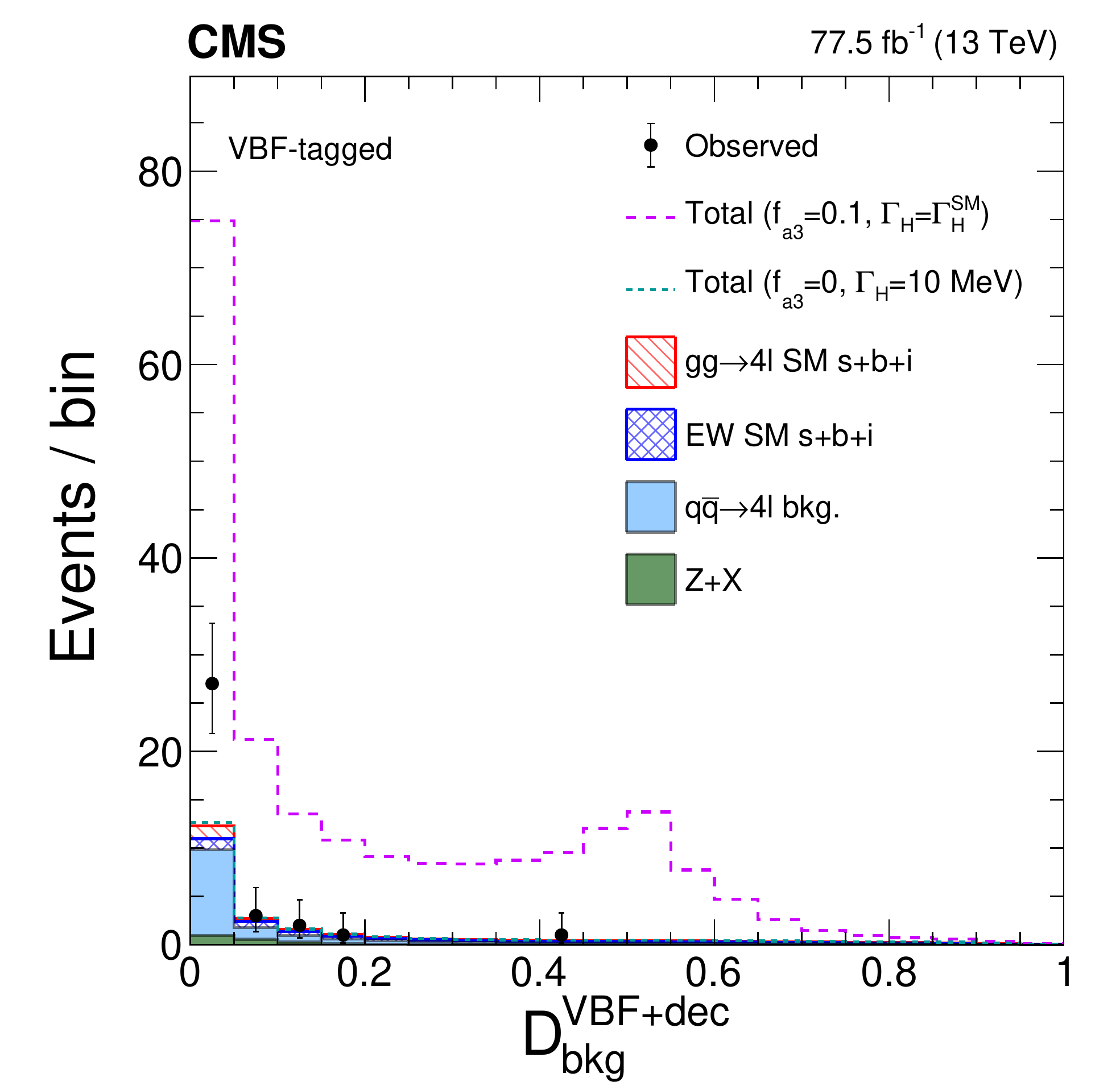}
\includegraphics[width=0.3\textwidth]{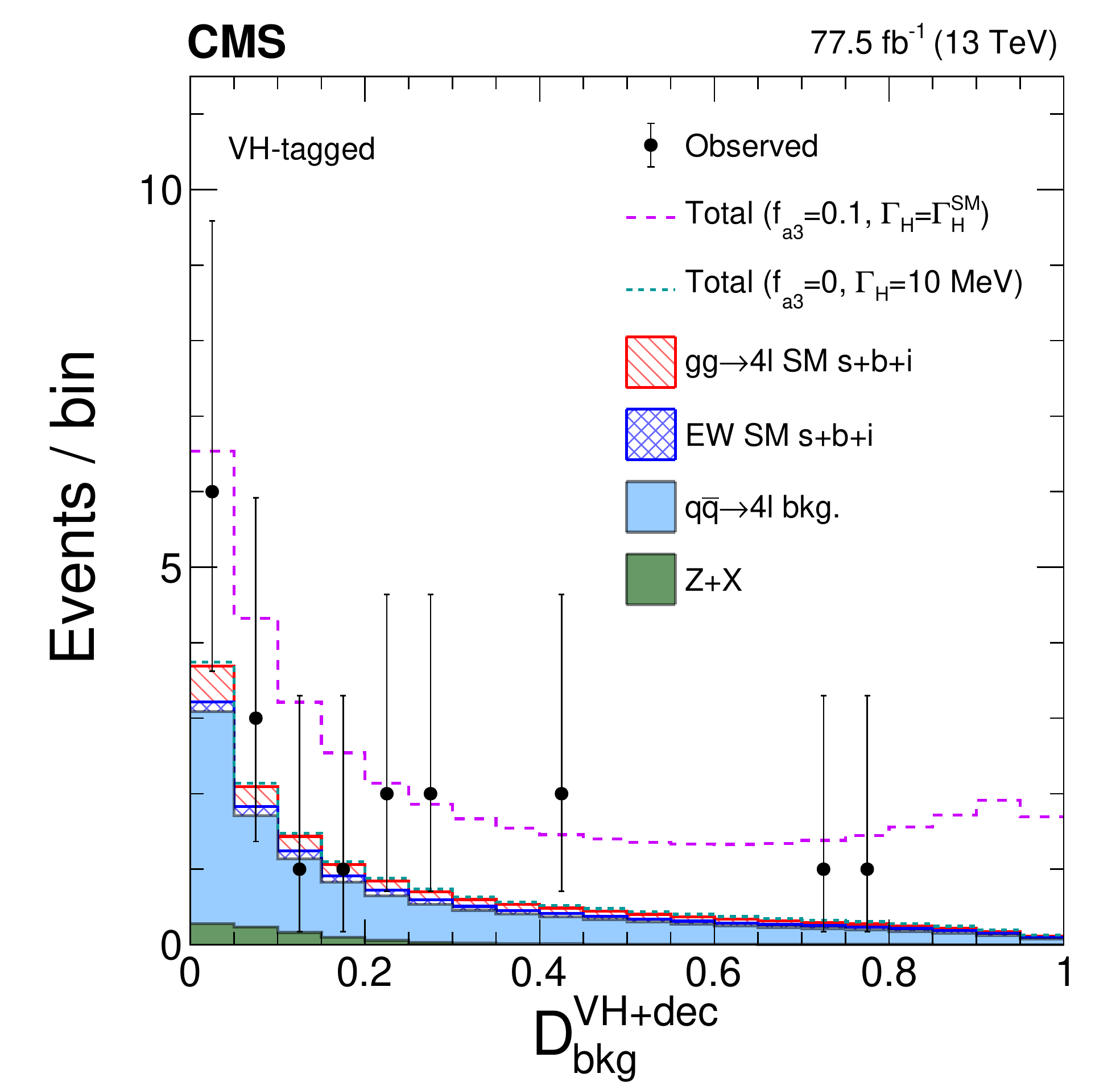}
\includegraphics[width=0.3\textwidth]{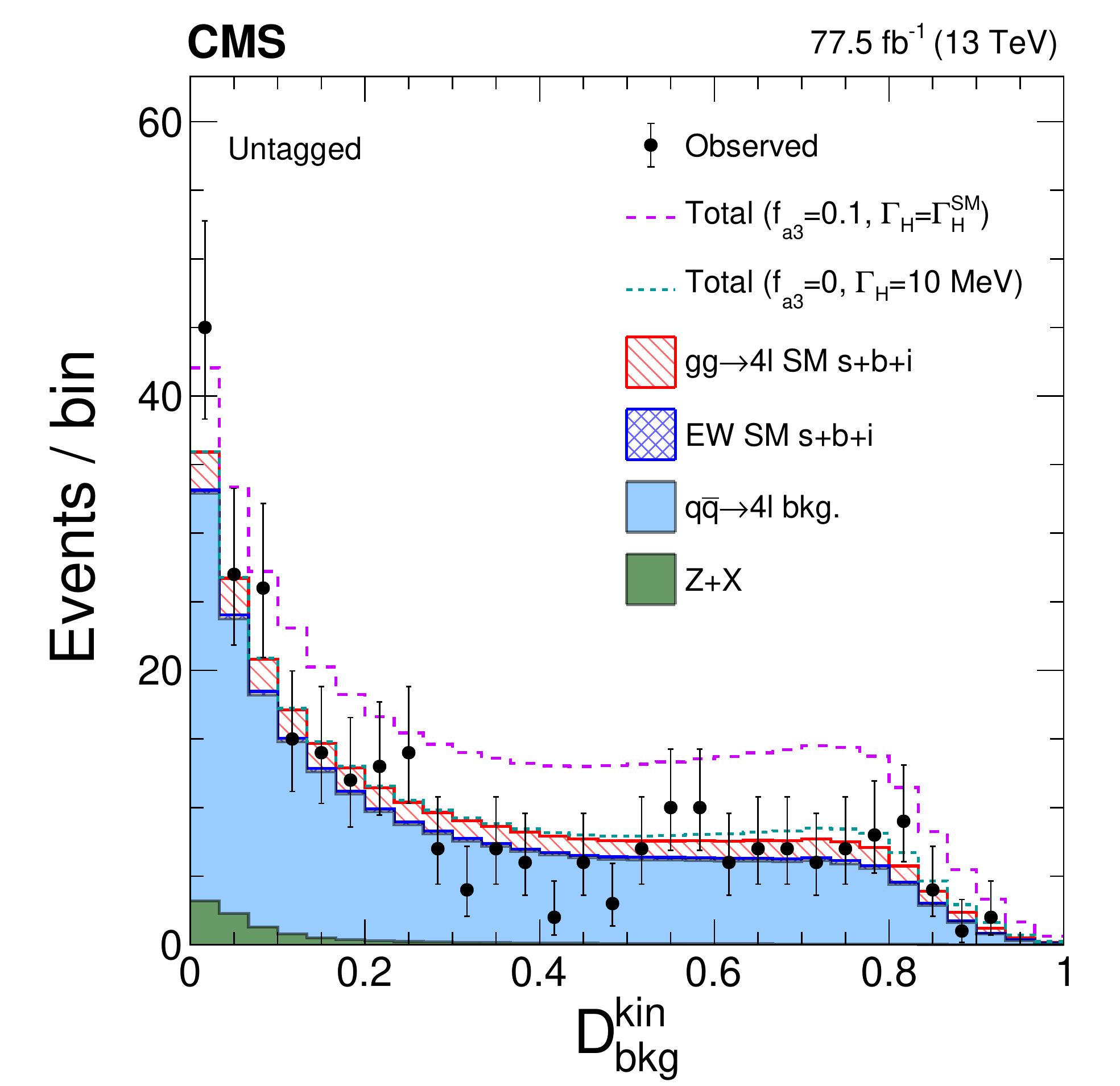} \\
\includegraphics[width=0.3\textwidth]{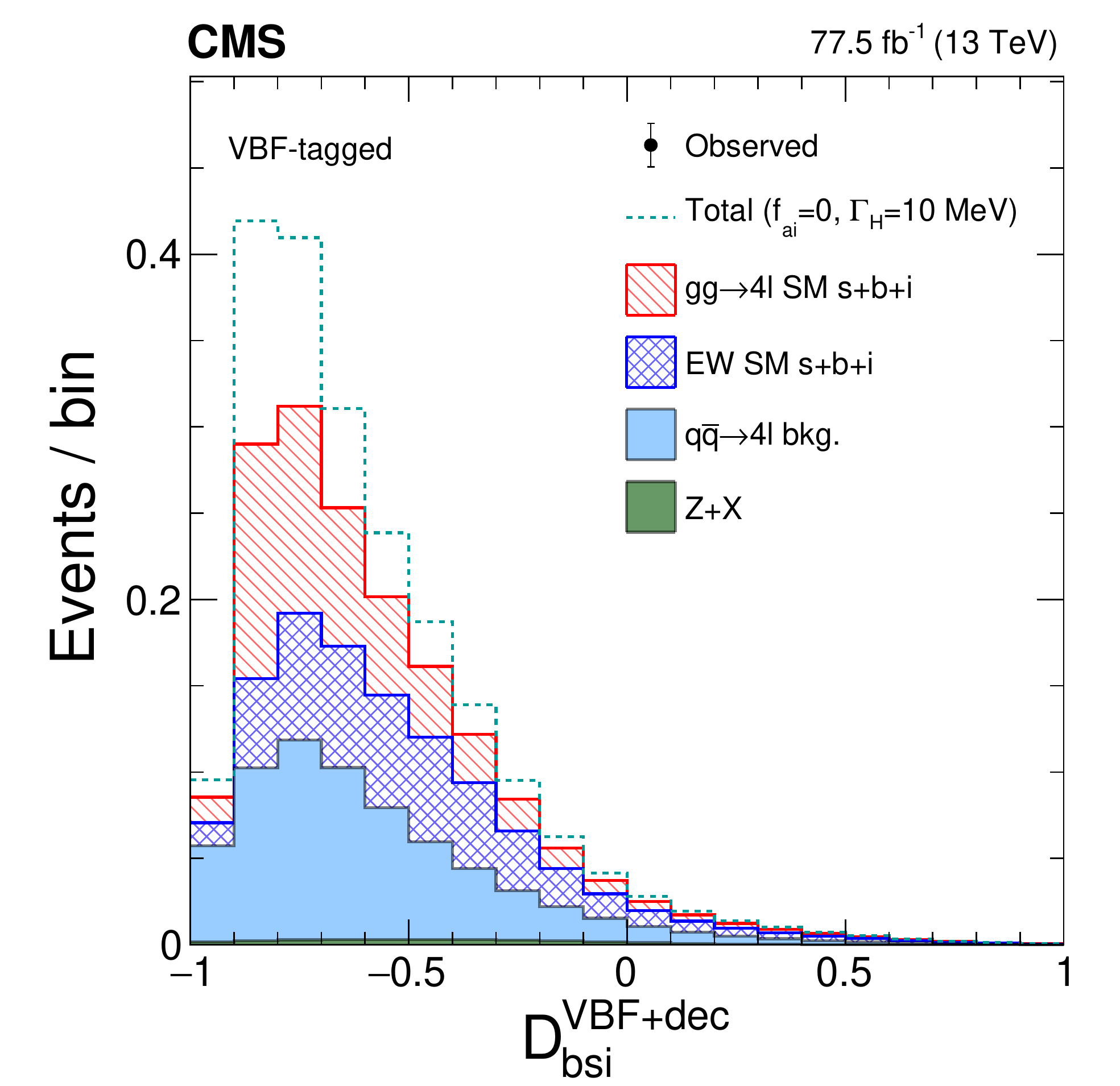}
\includegraphics[width=0.3\textwidth]{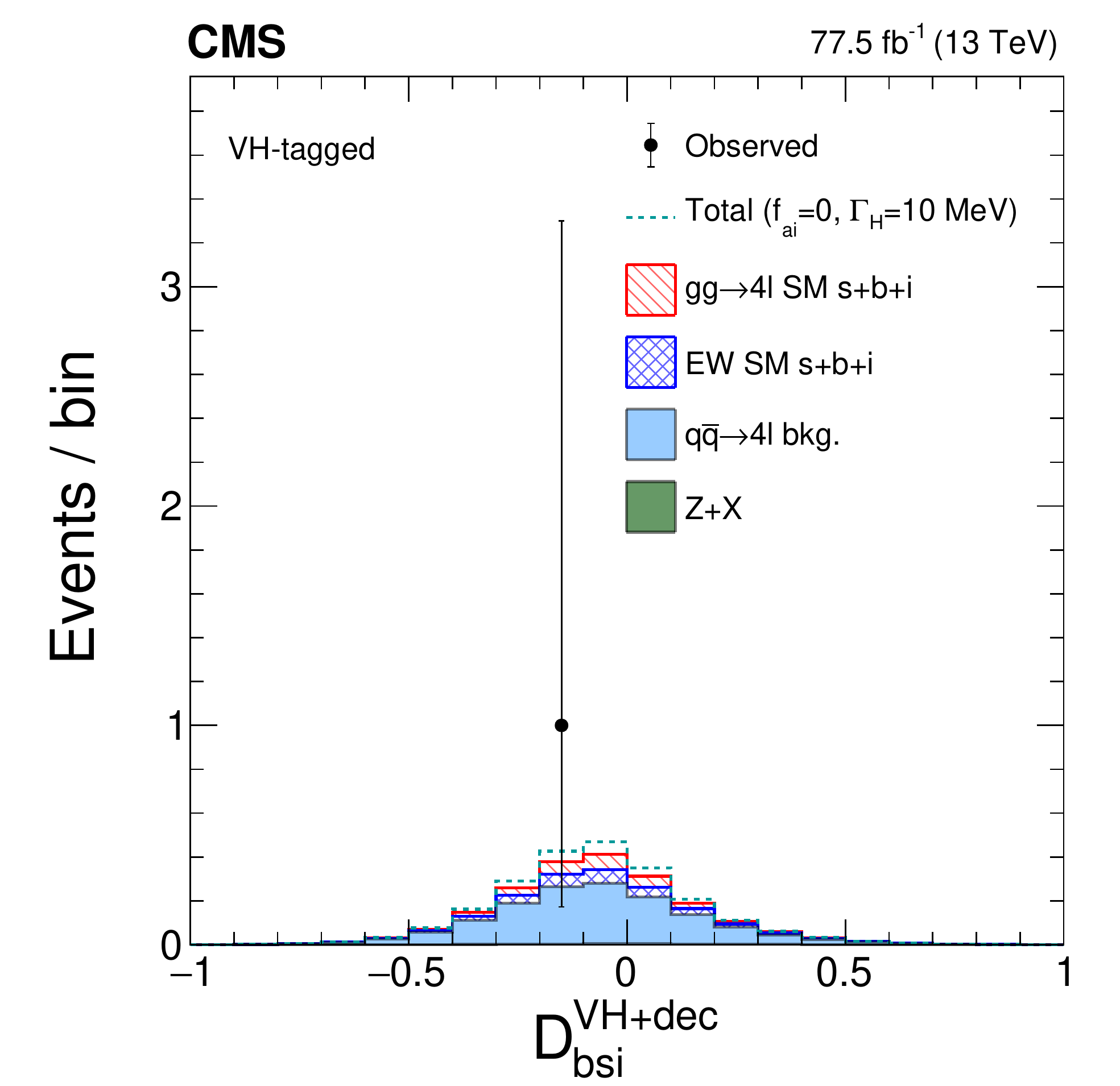}
\includegraphics[width=0.3\textwidth]{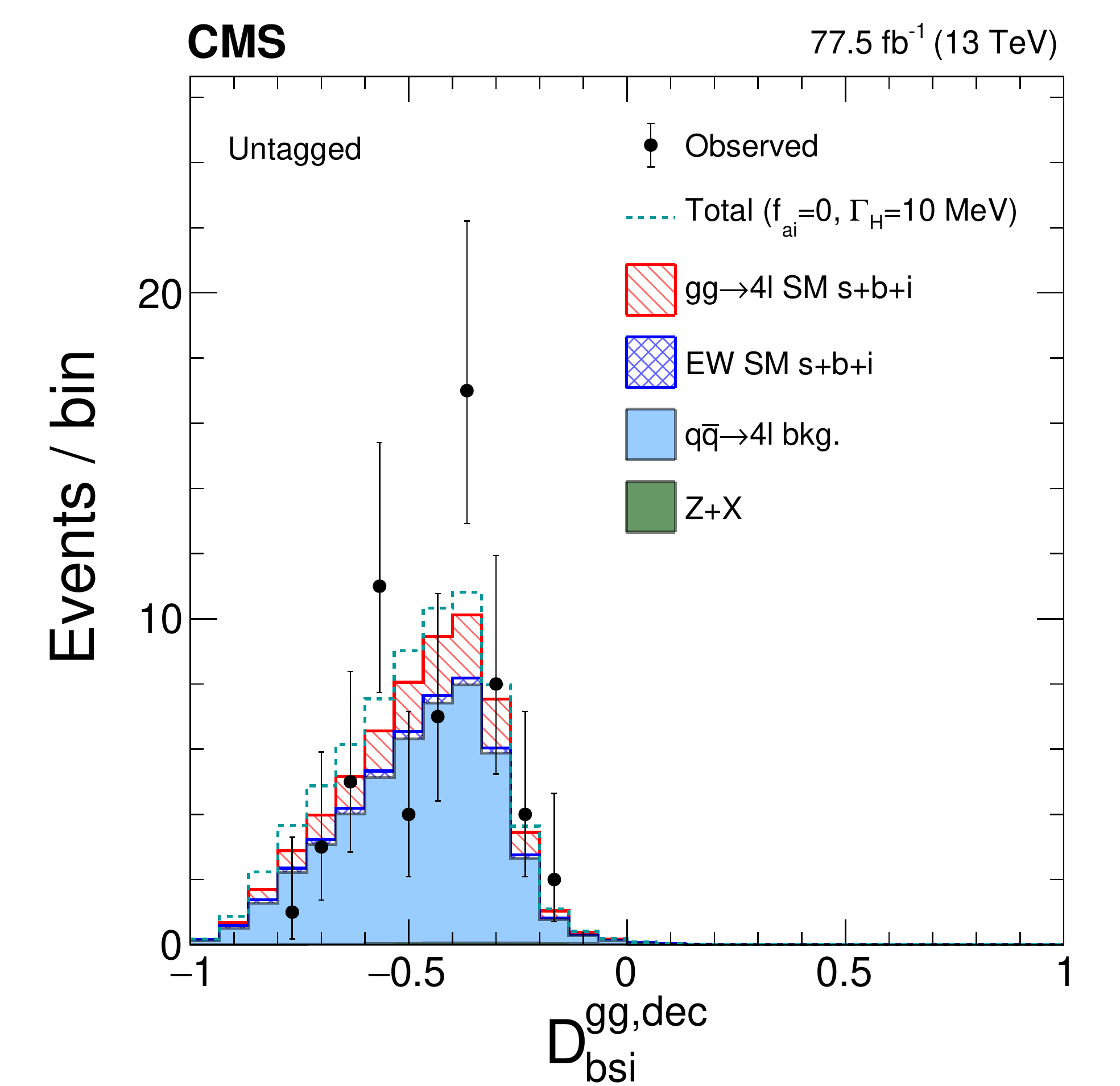}
\caption{
The distributions of events in the \offshell region in the data from 2016 and 2017.
The top row shows $\mell$ in the \VBF-tagged (left), \VH-tagged (middle), and untagged (right) categories
in the dedicated SM-like width analysis where a requirement on
$\mathcal{D}^{{\VBF}+{\text{dec}}}_\text{bkg}$, $\mathcal{D}^{{\VH}+{\text{dec}}}_\text{bkg}$, or $\Dbkgkin>0.6$
is applied in order to enhance signal over background contributions.
The middle row shows $\mathcal{D}^{{\VBF}+{\text{dec}}}_\text{bkg}$ (left), $\mathcal{D}^{{\VH}+{\text{dec}}}_\text{bkg}$ (middle),
$\mathcal{D}^\text{kin}_\text{bkg}$ (right) of the \AC{3} analysis in the corresponding three categories.
The requirement $\mell>340$\GeV is applied in order to enhance signal over background contributions.
The bottom row shows $\mathcal{D}_\mathrm{bsi}$ in the corresponding three categories in the dedicated
SM-like width analysis with both of the \mell and \Dbkgkin requirements enhancing the signal contribution.
The acronym $\mathrm{s}+\mathrm{b}+\mathrm{i}$ designates the sum of the signal (s), background (b),
and their interference contributions (i).
}
\label{fig:stackPlotsOffshell}
\end{figure*}

The distributions of events for several of the observables $\vec{x}$ from Tables~\ref{table:categoriesonshell} and~\ref{table:categoriesoffshell}
are illustrated in Fig.~\ref{fig:stackPlotsOnshell} for the \onshell and in Fig.~\ref{fig:stackPlotsOffshell} for the \offshell regions.
In Figs.~\ref{fig:stackPlotsOnshell} and~\ref{fig:stackPlotsOffshell}, cross sections of all background processes are fixed
to the SM expectations, except for the \ZX background estimated from the data control regions discussed above.
Cross sections of all signal processes, including BSM, are normalized to the SM expectations in the \onshell region.

\section{The fit implementation}
\label{sec:offshell}

We perform an unbinned extended maximum likelihood fit~\cite{Barlow:1990vc}
to the events split into several categories (enumerated with an index $k$ below) according to the
three lepton flavor combinations ($4\Pe$, $4\Pgm$, and $2\Pe 2\Pgm$),
three production categories (\VBF-tagged, \VH-tagged, and untagged),
five data periods (2011, 2012, 2015, 2016, and 2017),
and two mass ranges (\onshell and \offshell).
Therefore, there could be up to 90 categories of events.
However, not all categories are used in each independent measurement
because of the simpler categorization approach applied to the earlier data.
Here we focus on discussion of the 2016 and 2017 data analyses, while treatment of the earlier
data can be found in Refs.~\cite{Khachatryan:2014kca, Khachatryan:2015mma, Sirunyan:2017tqd}.

An independent fit is performed for each of the four anomalous \HVV coupling parameters \fcospai using the \onshell region only.
These fits avoid any assumptions on how the behavior of each process considered in the analysis changes from the \onshell region 
to the \offshell region. Four independent joint fits to the \onshell and \offshell regions are performed in order to determine the width 
of the \Hboson under the SM-like assumption or in the presence of the three anomalous couplings \AC{3}, \AC{2}, and \LC{1}.
These fits are also used to constrain the three corresponding anomalous coupling parameters \fcospai.
When a certain anomalous coupling is tested, all other anomalous couplings are assumed to be zero,
and only real couplings in \Eq{eq:formfact-fullampl-spin0} are tested, that is with $\AC{1} \ge 0$ and $\cospai=\pm1$.

The \onshell analysis with the study of the \AC{3}, \AC{2}, \LC{1}, and \LZGs couplings
has been presented in Ref.~\cite{Sirunyan:2017tqd} using a partial data set. This part of the analysis
remains essentially unchanged, except for a small change in the definition of the interference discriminant
in \Eq{eq:melaDint} and the inclusion of information from the kinematics of the two associated jets in
the \Dbkg calculation discussed in Sec.~\ref{sec:AnalysisStrategyIntro}.
The SM-like \onshell analysis is similar to the one presented in Ref.~\cite{Sirunyan:2017exp} in methodology,
but it uses the observables $\vec{x}$ and categorization $k$ described in Table~\ref{table:categoriesonshell} and
Sec.~\ref{sec:AnalysisStrategyIntro}.
The \onshell probability density is normalized to the total event yield in each process $j$ and category $k$ according to
\ifthenelse{\boolean{cms@external}}{
\begin{multline}
\mathcal{P}_{jk}(\vec{x};\vec{\xi}_{jk},\vec\zeta) \\
=
\mu_j \mathcal{P}_{jk}^\text{sig} \left( \vec{x};\vec{\xi}_{jk},\fai,\pai \right)
+ \mathcal{P}_{jk}^\text{bkg} \left( \vec{x};\vec{\xi}_{jk} \right),
\label{eq:ponshell}
\end{multline}
}{
\begin{equation}
\mathcal{P}_{jk}(\vec{x};\vec{\xi}_{jk},\vec\zeta) =
\mu_j \mathcal{P}_{jk}^\text{sig} \left( \vec{x};\vec{\xi}_{jk},\fai,\pai \right)
+ \mathcal{P}_{jk}^\text{bkg} \left( \vec{x};\vec{\xi}_{jk} \right),
\label{eq:ponshell}
\end{equation}
}
where $\vec\zeta=(\muF, \muV, \GH, \fcospai)$ are the unconstrained parameters of interest,
$\vec{\xi}_{jk}$ are the constrained nuisance parameters for a particular parametrization,
and $\vec{x}$ are the observables listed in Table~\ref{table:categoriesonshell}, specific to each \ai.
The \onshell signal strength $\mu_j$ in \Eq{eq:ponshell} is defined in references to \Eq{eq:resonant}
as either \muF or \muV according to the process
type $j$ (\glufu, \VBF, \WH, \ZH, \ttH, \bbH,  $\qqbar\to4\ell$, and \ZX).
Each process includes both signal (sig) and background (bkg) components, but may contain only signal
(\ttH and \bbH) or only background ($\qqbar\to4\ell$ and $\ZX$) contributions in the particular cases.
The interference between the signal and background components, when both are present,
is negligible in the \onshell region because of the very small width $\GH$ compared to the mass range of interest.
This also leads to the \onshell parametrization in \Eq{eq:ponshell} being independent from the width $\GH$.

The \offshell probability density follows \Eqs{eq:resonant}{eq:ponshell}{and} closely
but with the additional contribution of interference (int) between the signal and background amplitudes as
\ifthenelse{\boolean{cms@external}}{
\begin{multline}
\-\mathcal{P}_{jk}(\vec{x};\vec{\xi}_{jk},\vec\zeta) =
\frac{\mu_j \GH}{\Gref}~\mathcal{P}_{jk}^\text{sig} \left( \vec{x};\vec{\xi}_{jk},f_{ai},\phi_{ai} \right)\\
+ \sqrt{\frac{\mu_j \GH}{\Gref}}~\mathcal{P}_{jk}^\mathrm{int} \left( \vec{x};\vec{\xi}_{jk},f_{ai},\phi_{ai} \right)
+ \mathcal{P}_{jk}^\text{bkg} \left( \vec{x};\vec{\xi}_{jk} \right),
\label{eq:poffshell}
\end{multline}
}{
\begin{equation}
\-\mathcal{P}_{jk}(\vec{x};\vec{\xi}_{jk},\vec\zeta) =
\frac{\mu_j \GH}{\Gref}~\mathcal{P}_{jk}^\text{sig} \left( \vec{x};\vec{\xi}_{jk},f_{ai},\phi_{ai} \right)
+ \sqrt{\frac{\mu_j \GH}{\Gref}}~\mathcal{P}_{jk}^\mathrm{int} \left( \vec{x};\vec{\xi}_{jk},f_{ai},\phi_{ai} \right)
+ \mathcal{P}_{jk}^\text{bkg} \left( \vec{x};\vec{\xi}_{jk} \right),
\label{eq:poffshell}
\end{equation}
}
where the notation remains the same as for \Eq{eq:ponshell}.
The $\vec{x}$ observables are listed in Table~\ref{table:categoriesoffshell} and are specific to each coupling analysis.
They include \mell and two other discriminants.
The process type $j$ does not include \ttH and \bbH because of their negligible
contribution in the \offshell region, while the \VBF, \WH, and \ZH processes are combined into one EW process.
The parametrization in \Eq{eq:poffshell} depends on the width $\GH$ explicitly and
the reference value is taken to be $\Gref=4.07$\MeV, which
determines the relative strength of $\mathcal{P}_{jk}^\text{sig}$ and $\mathcal{P}_{jk}^\mathrm{int}$
with respect to $\mathcal{P}_{jk}^\text{bkg}$ in the parametrization.

The EW \Hboson production (\VBF and \VH) or production via gluon fusion have different dependence on anomalous
\HVV couplings, equally in the \onshell or \offshell regions. There are two \HVV vertices in the former production
mechanism with the subsequent $\PH\to\V\V\to 4\ell$ decay while there is only one \HVV decay vertex in the latter case.
In addition, there is interference with the background in the \offshell region.
This leads to the following general expressions for the signal (sig) or interference (int) contributions
appearing in \Eqs{eq:ponshell}{eq:poffshell}{and}:
\ifthenelse{\boolean{cms@external}}{
\begin{multline}
\mathcal{P}_{jk}^{\mathrm{sig/int}}\left(\vec{x};\vec{\xi}_{jk},f_{ai},\phi_{ai} \right) \\
=\sum_{m=0}^{M}
\mathcal{P}^{\mathrm{sig/int}}_{jk,m}\left(\vec{x};\vec{\xi}_{jk}\right)
f_{ai}^\frac{m}{2} (1-f_{ai})^\frac{M-m}{2} \cos^m(\phi_{ai}),
\label{eq:poffshellACsimplified}
\end{multline}
}{
\begin{equation}
\mathcal{P}_{jk}^{\mathrm{sig/int}}\left(\vec{x};\vec{\xi}_{jk},f_{ai},\phi_{ai} \right) =
\sum_{m=0}^{M}
\mathcal{P}^{\mathrm{sig/int}}_{jk,m}\left(\vec{x};\vec{\xi}_{jk}\right)
f_{ai}^\frac{m}{2} (1-f_{ai})^\frac{M-m}{2} \cos^m(\phi_{ai}),
\label{eq:poffshellACsimplified}
\end{equation}
}
where the sum over the index $m$ runs up to
$M=4$ in the case of the EW signal process;
$M=2$ in the case of the gluon fusion, \ttH, and \bbH signal processes,
or the interference between the signal and background in the EW process; and
$M=1$ in the case of the interference between the signal and background in the gluon fusion process.
In this expression, the index $m$ corresponds to the exponent of \ai in the squared scattering amplitude 
from \Eq{eq:formfact-fullampl-spin0}, which may contain contributions from production and decay, 
and the factor $\cospai=\pm1$ affects only the sign of the terms that scale with an odd power of \ai.

The $\mathcal{P}_{jk,m}^{\mathrm{sig/int}}$ and $\mathcal{P}_{jk}^\text{bkg}$ probability densities
are normalized to the expected number of events, and are binned histograms (templates)
of the observables $\vec{x}$ listed in Tables~\ref{table:categoriesonshell} and~\ref{table:categoriesoffshell},
except for the signal \mell parametrization in the \onshell region as discussed below.
These templates are obtained by reweighting the existing signal or background samples
for different couplings and then finding their linear combination.
Since \mell is treated directly as an observable in the \onshell SM-like fit, the signal \mell shape
for each process $j$ and category $k$ is parametrized using a double-sided crystal-ball function~\cite{Oreglia:1980cs},
and the full signal probability density is parametrized as the product of the parametric \mell shape and
a template of other discriminants conditional in \mell.
In all cases, the \Hboson mass $\mH=125$\GeV is assumed.

The final constraints on \fcospai and \GH are placed using the profile likelihood method using the RooFit toolkit~\cite{Verkerke:2003ir}
within the \ROOT~\cite{Brun:1997pa} framework. The extended likelihood function is constructed using the probability densities in
\Eqs{eq:ponshell}{eq:poffshell}{and} with each event characterized by the discrete category $k$ and typically three continuous
observables $\vec{x}$. The likelihood $\mathcal{L}$ is maximized with respect to the nuisance parameters $\vec{\xi}_{jk}$ describing
the systematic uncertainties discussed below and the yield parameters \muF and \muV. The allowed 68\% and 95\%\CL
intervals are defined using the profile likelihood function, $-2\Delta\ln\mathcal{L} = 1.00$ and $3.84$, for which exact coverage is expected
in the asymptotic limit~\cite{Wilks:1938dza}.

Several systematic uncertainties are featured in the vectors of constrained parameters $\vec{\xi}_{jk}$.
The template shapes describing probability distributions in \Eqss{eq:ponshell}{eq:poffshell}{eq:poffshellACsimplified}{and} 
are varied separately within either theoretical or experimental uncertainties.
In the following, a range of uncertainties affecting the template distributions is given for the $\mell$ values
from around 100\GeV (typical for the \onshell range) to around 1\TeV (in the \offshell range), respectively.
The factorization (or renormalization) scale uncertainties
are evaluated by multiplying the central scale by $2$ or $1/2$, and the uncertainties range
from $\pm$0.7\% ($^{+1.2}_{-1.4}\%$)  to $^{-1.0}_{+0.6}\%$ ($^{+5}_{-4}\%$)  in the \glufu process,
from $^{+0.6}_{-0.1}\%$ ($^{+5}_{-4}\%$) to $\pm$5\% ($^{+30}_{-25}\%$) in the \VBF process,
from $^{+3}_{-5}\%$ ($^{+5}_{-4}\%$) to $\pm$6\% ($^{+30}_{-25}\%$) in the processes with an associated EW boson,
and from $^{+3.5}_{-5.5}\%$ to $\pm$1\% ($\pm$3\%) in the $\qqbar\to4\ell$ background.
PDF parametrization uncertainties are evaluated by taking the envelope of the 100
alternative NNPDF variations. Variations due to PDF parametrization uncertainties
[or due to uncertainties in $\alpS(m_{\PZ})=0.1180\pm0.0015$]
range from $^{+1.2}_{-1.4}\%$ ($^{+2.0}_{-2.5}\%$) to $^{+5}_{-4}\%$ ($^{+2.4}_{-1.0}\%$) in the \glufu process,
from $^{+5}_{-4}\%$ to about $^{+30}_{-25}\%$ in the EW processes,
and are approximately $\pm$3\% (from $^{+1.0}_{-1.8}\%$ to $\pm$0.5\%) for the $\qqbar\to4\ell$ background.
The signal processes, and the backgrounds that interfere with the signal, feature the uncertainties as a function
of the multiplicity and kinematics of associated jets due to the hadronization scale used in \PYTHIA and the
underlying event variations, obtained with the variations of the \PYTHIA tune.
In the \VBF-tagged categories, the correlated template variations for
the hadronization scale (underlying event) range
from $\pm$11\% ($\pm$45\%) to $\mp$8\% ($\mp$40\%) in the \glufu process,
from $\pm$8\% ($\pm$24\%) to $\mp$6\% ($\mp$8\%) in the \VBF process, and
from $\pm$13\% ($\pm$20\%) to $\mp$10\% ($\mp$32\%) in the processes with an associated EW boson.
In the \VH-tagged categories, these correlated template variations instead range
from $\pm$15\% ($\pm$50\%) to $\mp$9\% ($\mp$45\%) in the \glufu process,
from $\pm$8\% ($\pm$25\%) to $\mp$7\% ($\mp$30\%) in the \VBF process, and
from $\pm$4\% ($\pm$19\%) to $\mp$4\% ($\mp$13\%) in the processes with an associated EW boson.
Template shapes in the \glufu processes are also varied to account for a second jet in the hard process, and these correlated 
variations range from $\pm$18\% ($\pm$32\%) to $\mp$15\% ($\mp$14\%) in the \VBF-tagged (\VH-tagged) category.
The $\qqbar\to 4\ell$ background further features an uncertainty in the NLO EW corrections applied to the
simulation~\cite{Gieseke:2014gka,Baglio:2013toa}, which are significant at higher \mell values, reaching up to $20\%$ at 1\TeV.

Experimental uncertainties involve jet energy calibration (JEC) uncertainties, which are only relevant when production 
categories are considered, and lepton efficiency and momentum uncertainties, which are similar for the different
processes and categories. Systematic uncertainties in the JEC account for variations in the \VBF-tagged (\VH-tagged) category, 
and range from $\pm$13\% ($\pm$4\%) to $\pm$8\% ($\pm$1\%) in the \glufu process,
from $\pm$5\% ($^{-10}_{+2}\%$) to about $\pm$11\% ($\pm$6\%) in the \VBF process,
from $\pm$9\% ($\pm$4\%) to $\pm$12\% ($\pm$1\%) in processes with an associated EW boson,
and from $\pm$17\% ($\pm$8\%) to $\pm$15\% ($^{+2.0}_{-0.5}\%$) for the $\qqbar\to4\ell$ background.
The cross-section uncertainties due to electron (muon) efficiency range
from $^{+6}_{-7}\%$ ($^{+3.0}_{-4.5}\%$) to $^{+3.5}_{-4.5}\%$ ($^{+0.8}_{-2.0}\%$) to $^{+7}_{-8}\%$ ($^{+0.8}_{-2.0}\%$)
in the $2\cPe2\cPm$ channel, and roughly double for the $4\cPe$ ($4\cPm$) channel,
from $\mell\sim100$\GeV to $230$\GeV to around 1\TeV.

In the estimation of the \ZX background, the flavor composition of hadronic jets misidentified as leptons may
be different in the $\PZ+1\ell$ and $\PZ+2\ell$ control regions, and together with the statistical uncertainty
in the $\PZ+2\ell$ region, this uncertainty accounts for about $\pm$30\% variation in the background
estimate from the 2017 data set. The uncertainty on the modeling of this misidentification as a function of \pt and $\eta$,
combined with the $\PZ+1\ell$ control region statistical uncertainty, leads to
a $^{+20}_{-12}\%$ to $^{+30}_{-27}\%$ variation in the $4\cPe$ channel,
$\pm(10-20)\%$ variation in the \mell shape in the $2\cPe2\cPm$ channel,
and $\pm$4\% to $^{+14}_{-17}\%$ variation in the $4\cPm$ channel.
Uncertainties in the \ZX background in the 2016 data set are only slightly larger.
The normalization of the background processes derived from the MC simulation is affected by the uncertainties in the integrated luminosity
of 2.5\%~\cite{CMS-PAS-LUM-17-001} and 2.3\%~\cite{CMS-PAS-LUM-17-004} in the 2016 and 2017 data sets, respectively.
The integrated luminosity is measured using data from the CMS silicon pixel detector, drift tubes, and the forward hadron calorimeters,
or from the fast beam conditions monitor and pixel luminosity telescope.
All systematic uncertainties are treated as correlated between different time periods except for the luminosity
and jet-related uncertainties which originate from statistically independent sources.

\section{Results}
\label{sec:results}

\begin{figure*}[htbp!]
\centering
\includegraphics[width=0.45\textwidth]{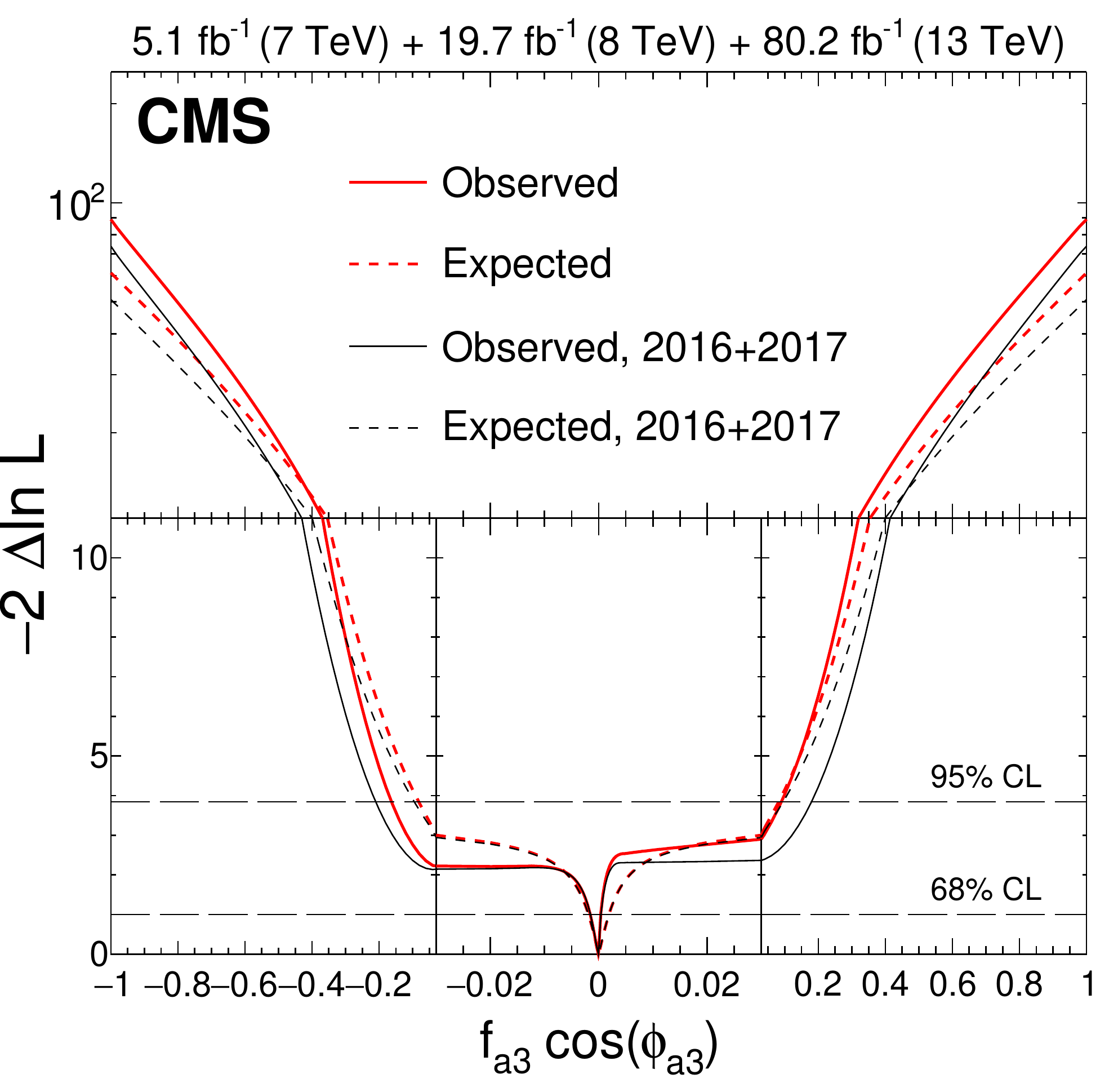}
\includegraphics[width=0.45\textwidth]{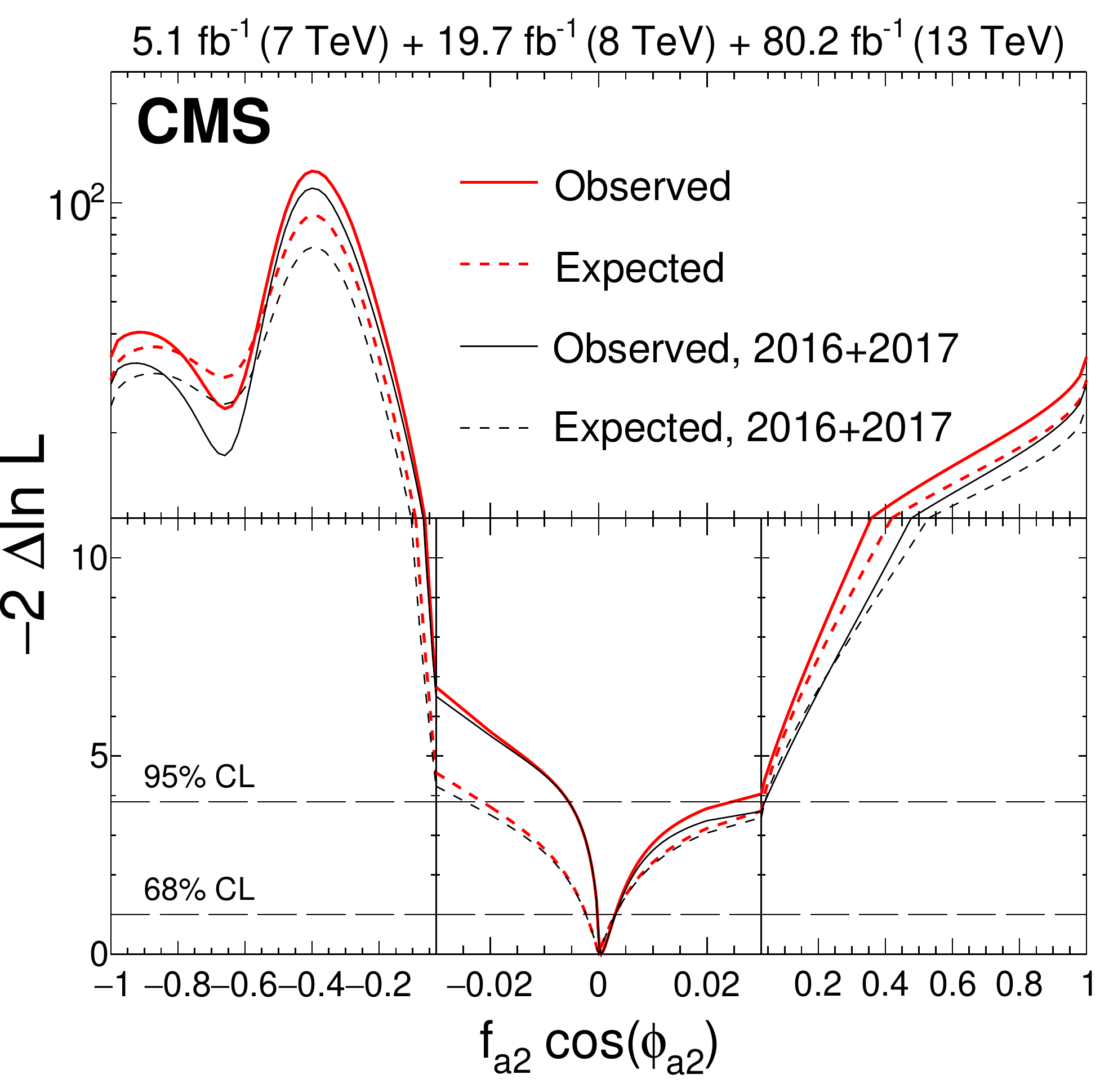} \\
\includegraphics[width=0.45\textwidth]{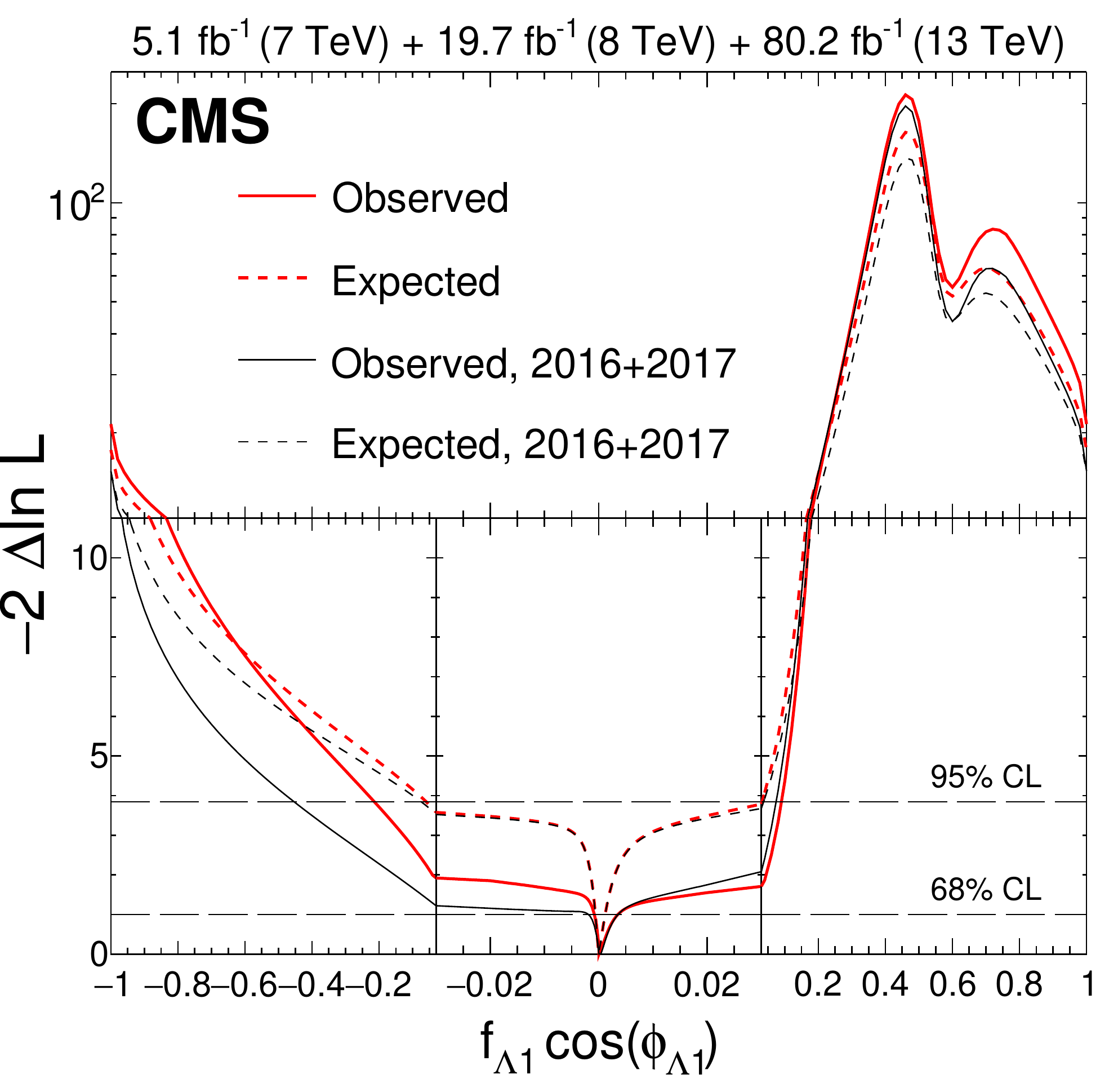}
\includegraphics[width=0.45\textwidth]{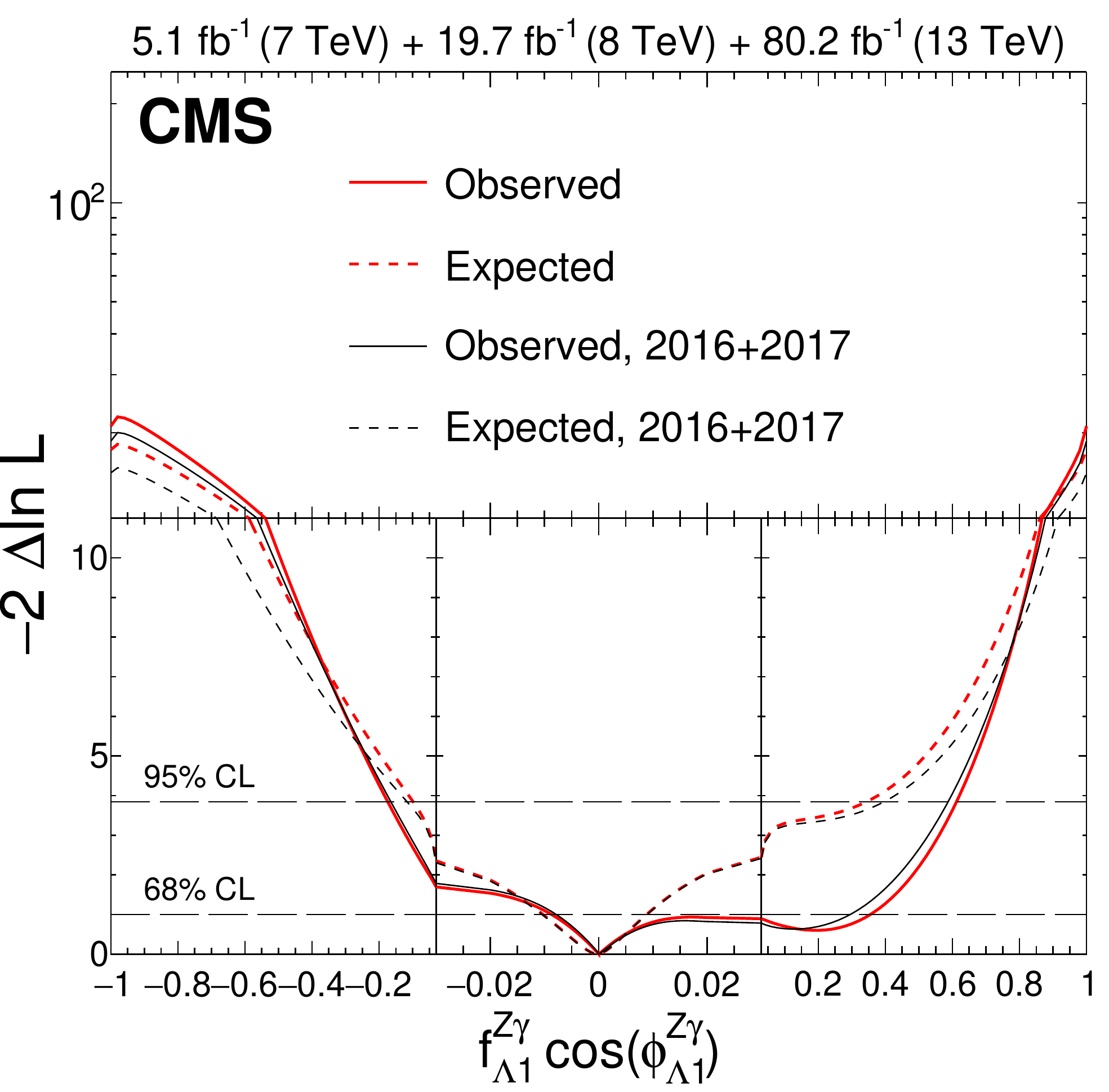}
\caption{
Observed (solid) and expected (dashed) likelihood scans of
\fcospAC{3} (top left),  \fcospAC{2} (top right), \fcospLC{1}  (bottom left), and \fcospLZGs (bottom right)
using \onshell events only.
Results of analysis of the data from 2016 and 2017 only (black) and the combined Run~1 and Run~2 analysis (red)
are shown. The dashed horizontal lines show the 68 and 95\%~\CL regions.
}
\label{fig:resultsfan}
\end{figure*}

Four \fcospai parameters sensitive to anomalous \HVV interactions, as defined in \Eqs{eq:formfact-fullampl-spin0}{eq:fa_definitions}{and},
are tested in the \onshell data sample using the probability densities defined in \Eq{eq:ponshell}.
Since only the real couplings are tested, $\cospai=\pm1$.
Figure~\ref{fig:resultsfan} shows the results of the likelihood scans of these parameters
for the 2016 and 2017 periods of the 13\TeV run and for the full combined data set from collisions at 7, 8, and 13\TeV.
The analysis of the 2016 and 2017 data uses the approach presented here with the observables sensitive
to anomalous couplings in both production and decay.
Because of the smaller numbers of events, the data from the 2015 period of the 13\TeV run and from the 2011 and 2012
periods of the 7 and 8\TeV runs are analyzed using only the decay information as in
Refs.~\cite{Khachatryan:2014kca, Sirunyan:2017tqd}, which is equivalent to having all events
in the untagged category of this analysis.
The results from \onshell events in the combined data set are listed in Table~\ref{tab:summary_fai}.
These results supersede our previous measurements of these parameters in Refs.~\cite{Khachatryan:2014kca, Sirunyan:2017tqd}.

\begin{table*}[!tbh]
\centering
\topcaption{
Summary of allowed 68\%~\CL (central values with uncertainties) and 95\%~\CL (in square brackets)
intervals for the anomalous coupling parameters \fcospai obtained from the analysis of
the combination of Run~1 (only \onshell) and Run~2 (\onshell and \offshell) data sets.
Three constraint scenarios are shown: using only \onshell events, using both \onshell and \offshell events
with the $\GH$ left unconstrained, or with the constraint $\GH=\GHSM$.
}
\renewcommand{\arraystretch}{1.25}
\begin{scotch}{ccll}
 Parameter     & Scenario & \multicolumn{1}{c}{Observed} & \multicolumn{1}{c}{Expected}  \\
\hline
  \fcospAC{3} & \onshell & $-0.0001^{+0.0004}_{-0.0015}$ $[-0.163,  0.090] $ & $0.0000^{+0.0019}_{-0.0019}$ $[-0.082, 0.082]$ \\
                   & any \GH & ~~~$0.0000_{-0.0010}^{+0.0003}$ $[-0.0165, 0.0087]$ & $0.0000_{-0.0015}^{+0.0015}$ $[-0.038, 0.038]$  \\
                   & $\GH=\GHSM$ & ~~~$0.0000_{-0.0009}^{+0.0003}$ $[-0.0067, 0.0050]$ & $0.0000_{-0.0014}^{+0.0014}$ $[-0.0098, 0.0098]$  \\
 \fcospAC{2} & \onshell & ~~~$0.0004^{+0.0026}_{-0.0006}$ $[-0.0055, 0.0234]$ & $0.0000^{+0.0030}_{-0.0023}$ $[-0.021, 0.035]$ \\
                  & any \GH & ~~~$0.0004_{-0.0006}^{+0.0026}$ $[-0.0035, 0.0147]$ & $0.0000_{-0.0017}^{+0.0019}$ $[-0.015, 0.021]$  \\
                  & $\GH=\GHSM$ & ~~~$0.0005_{-0.0006}^{+0.0025}$ $[-0.0029, 0.0129]$ & $0.0000_{-0.0016}^{+0.0012}$ $[-0.010, 0.012]$  \\
 \fcospLC{1} & \onshell & ~~~$0.0002^{+0.0030}_{-0.0009}$ $[-0.209,  0.089] $ & $0.0000^{+0.0012}_{-0.0006}$ $[-0.059, 0.032]$ \\
                  & any \GH & ~~~$0.0001_{-0.0006}^{+0.0015}$ $[-0.090, 0.059]$ & $0.0000_{-0.0007}^{+0.0013}$ $[-0.017, 0.019]$   \\
                  & $\GH=\GHSM$ & $~~~0.0001_{-0.0005}^{+0.0015}$ $[-0.016, 0.068]$ & $0.0000_{-0.0006}^{+0.0013}$ $[-0.015, 0.018]$   \\
 \fcospLZGs  & \onshell & ~~~$0.0000^{+0.3554} _{-0.0087} $ $[-0.17,  0.61] $ & $0.0000^{+0.0091} _{-0.0100} $ $[-0.098, 0.343]$   \\
\end{scotch}
\label{tab:summary_fai}
\end{table*}

\begin{table*}[!tbh]
\centering
\topcaption{
Summary of the allowed 95\%~\CL intervals for the anomalous \HVV couplings
using results in Table~\ref{tab:summary_fai}.
The coupling ratios are assumed to be real and include the factor $\cospLC{1}$ or $\cospLZGs=\pm1$.
}
\renewcommand{\arraystretch}{1.25}
\begin{scotch}{ccll}
 Parameter     & Scenario & \multicolumn{1}{c}{Observed} & \multicolumn{1}{c}{Expected}  \\
\hline
 $\AC{3}/\AC{1}$ & \onshell    & $[-1.13, 0.80]$ & $[-0.76, 0.76]$ \\
                   & any \GH          & $[-0.33, 0.24]$ & $[-0.50, 0.50]$ \\
                   & $\GH=\GHSM$ & $[-0.21, 0.18]$ & $[-0.25, 0.25]$ \\
 $\AC{2}/\AC{1}$ & \onshell   & $[-0.12, 0.26]$    & $[-0.24, 0.31]$ \\
                  & any \GH          &  $[-0.098, 0.202]$ & $[-0.21, 0.25]$ \\
                  & $\GH=\GHSM$ &  $[-0.089, 0.189]$ & $[-0.17, 0.18]$ \\
$(\LC{1}\sqrt{\abs{\AC{1}}}) \cospLC{1}$  ~$(\GeV)$
                  & \onshell           &  $[-\infty, -130]\cup[160, \infty]$  &  $[-\infty, -180]\cup[210, \infty]$ \\
                  & any \GH          &  $[-\infty, -160]\cup~[180, \infty]$  &  $[-\infty, -250]\cup[240, \infty]$ \\
                  & $\GH=\GHSM$ &  $[-\infty, -250]\cup[170, \infty]$  & $[-\infty, -260]\cup[250, \infty]$ \\
$(\LZGs\sqrt{\abs{\AC{1}}}) \cospLZGs$ ~$(\GeV)$
                  & \onshell          &  $[-\infty, -170]\cup[100, \infty]$      &$[-\infty, -200]\cup[130, \infty]$ \\
\end{scotch}
\label{tab:summary_aia1}
\end{table*}

The observed and expected 68\%~\CL constraints are significantly tighter than in the Run~1
analysis~\cite{Khachatryan:2014kca} as it is evident from the narrow minima at $\fai=0$
in Fig.~\ref{fig:resultsfan}. This effect comes from utilizing production information
because the cross section in \VBF and \VH production increases quickly with \fai.
Moreover, the minima of the $-2\ln\mathcal{L}$ distributions appear rather sharp because of the
higher order polynomial of the \fai parameters appearing in \Eq{eq:poffshellACsimplified}
in the case of \VBF and \VH production. At the same time, the constraints above $\fai\sim0.02$
are dominated by the decay information from \Hell.
The best fit $(\muF, \muV$) values in the four analyses under the assumption that $\fai=0$ are as follows:
$(1.21^{+0.21}_{-0.17},0.84^{+0.71}_{-0.59})$ at $\fAC{3}=0$,
$(1.19^{+0.21}_{-0.17},0.91^{+0.69}_{-0.55})$ at $\fAC{2}=0$,
$(1.26^{+0.20}_{-0.18},0.53^{+0.64}_{-0.50})$ at $\fLC{1}=0$, and
$(1.24^{+0.19}_{-0.17},0.55^{+0.64}_{-0.51})$ at $\fLZGs=0$.
The values obtained for the different analyses vary because of the different categorization and observables in each \ai analysis.

\begin{figure*}[htbp!]
\centering
\includegraphics[width=0.42\textwidth]{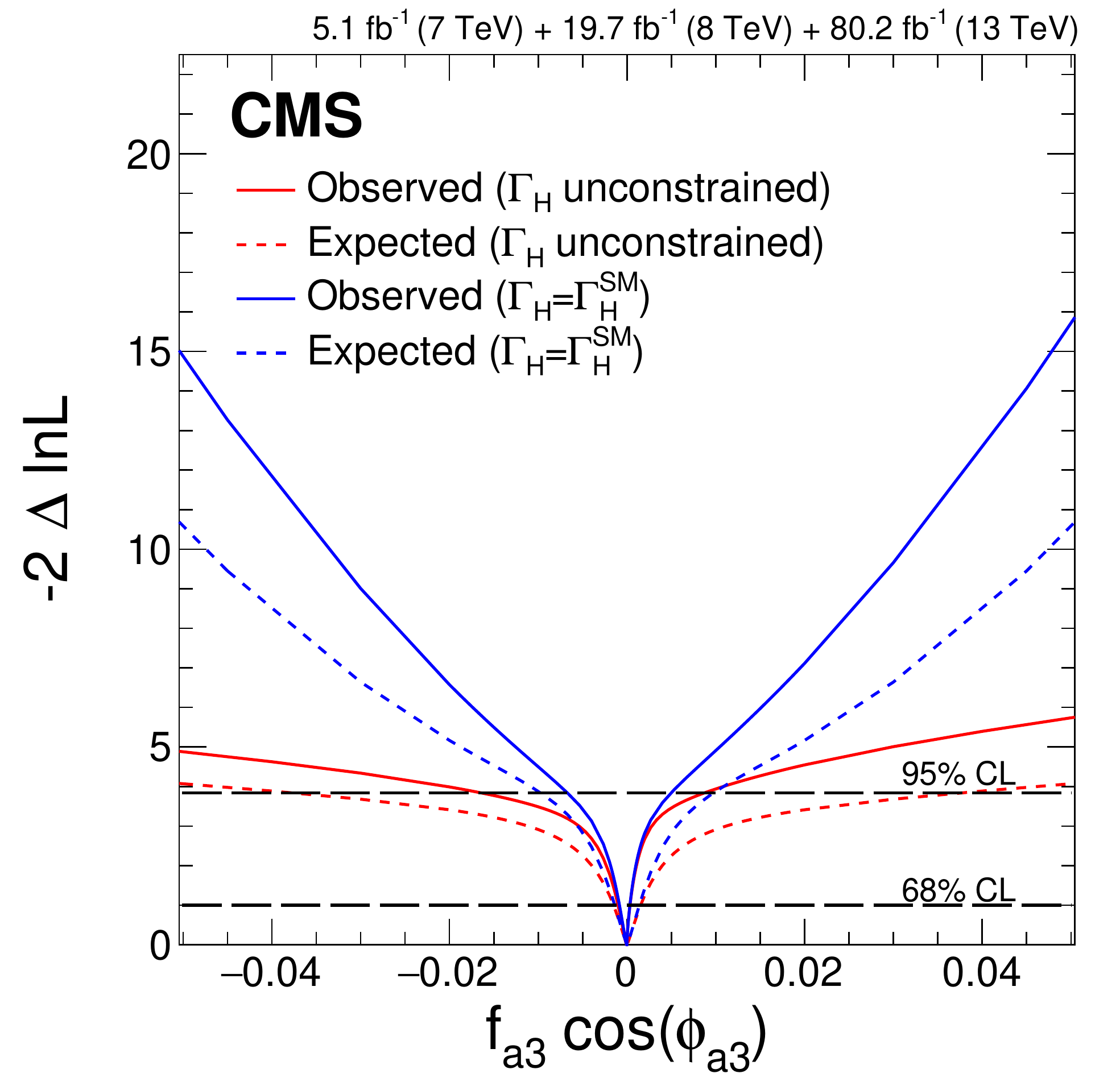}
\includegraphics[width=0.52\textwidth]{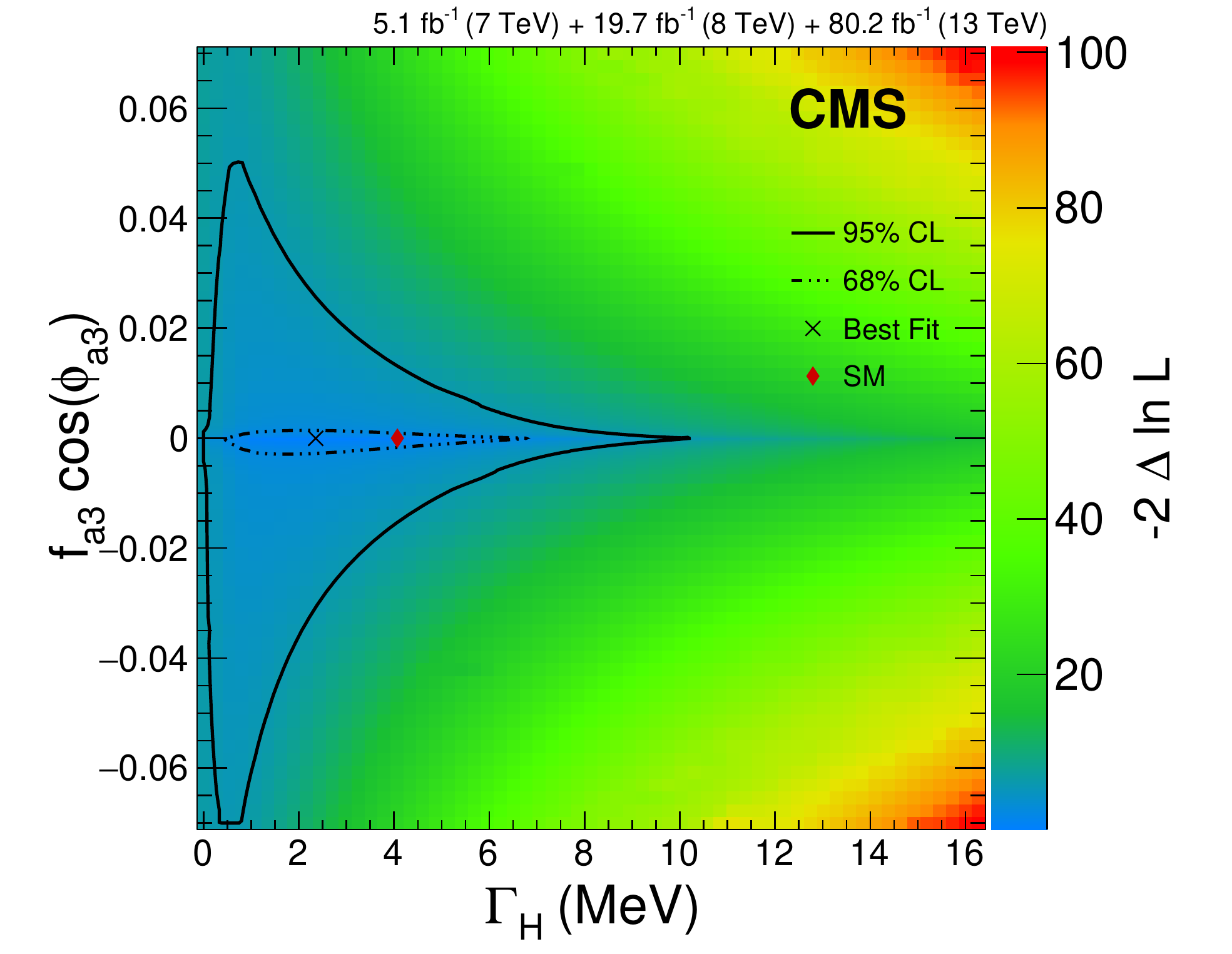} \\
\includegraphics[width=0.42\textwidth]{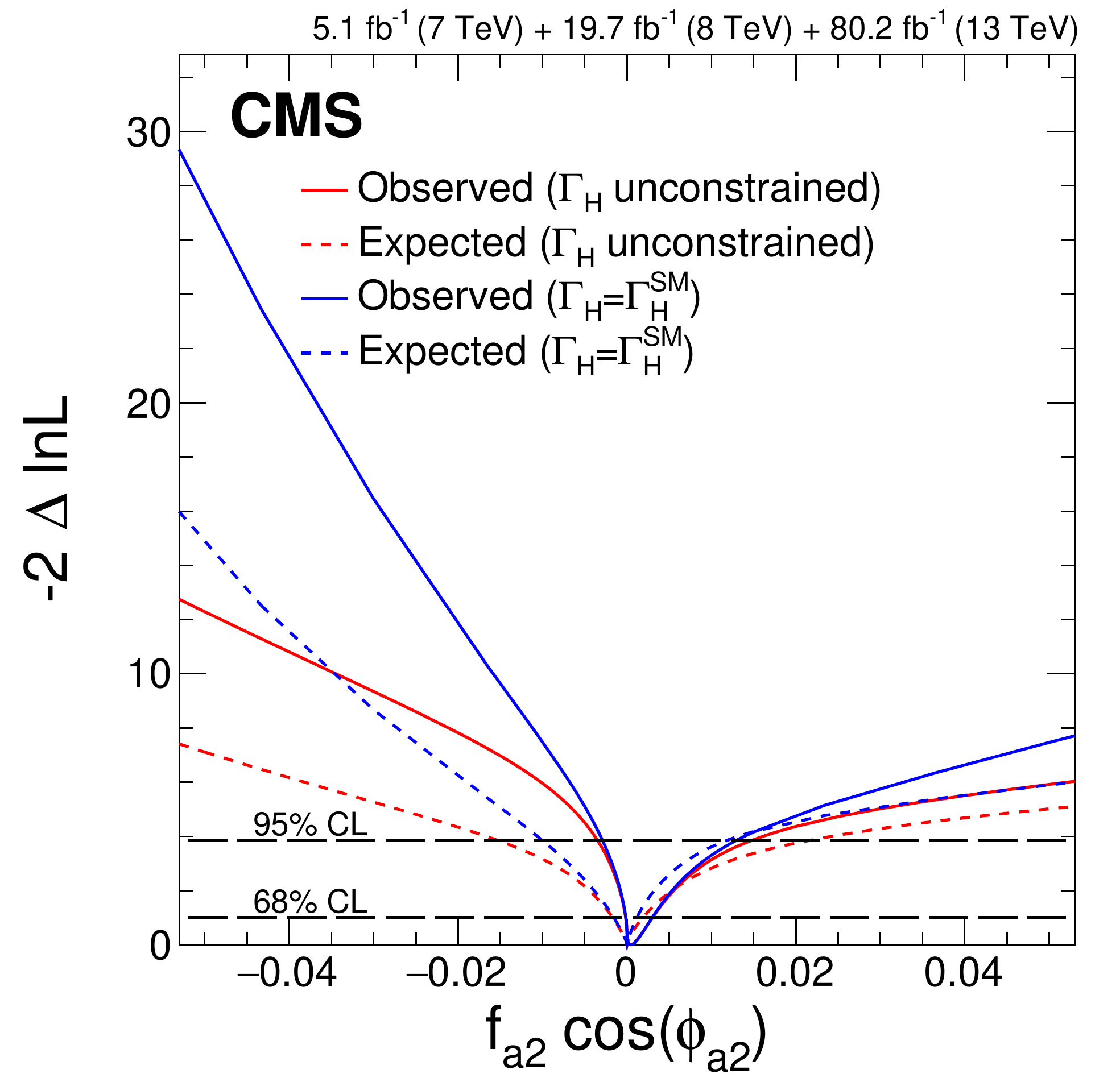}
\includegraphics[width=0.52\textwidth]{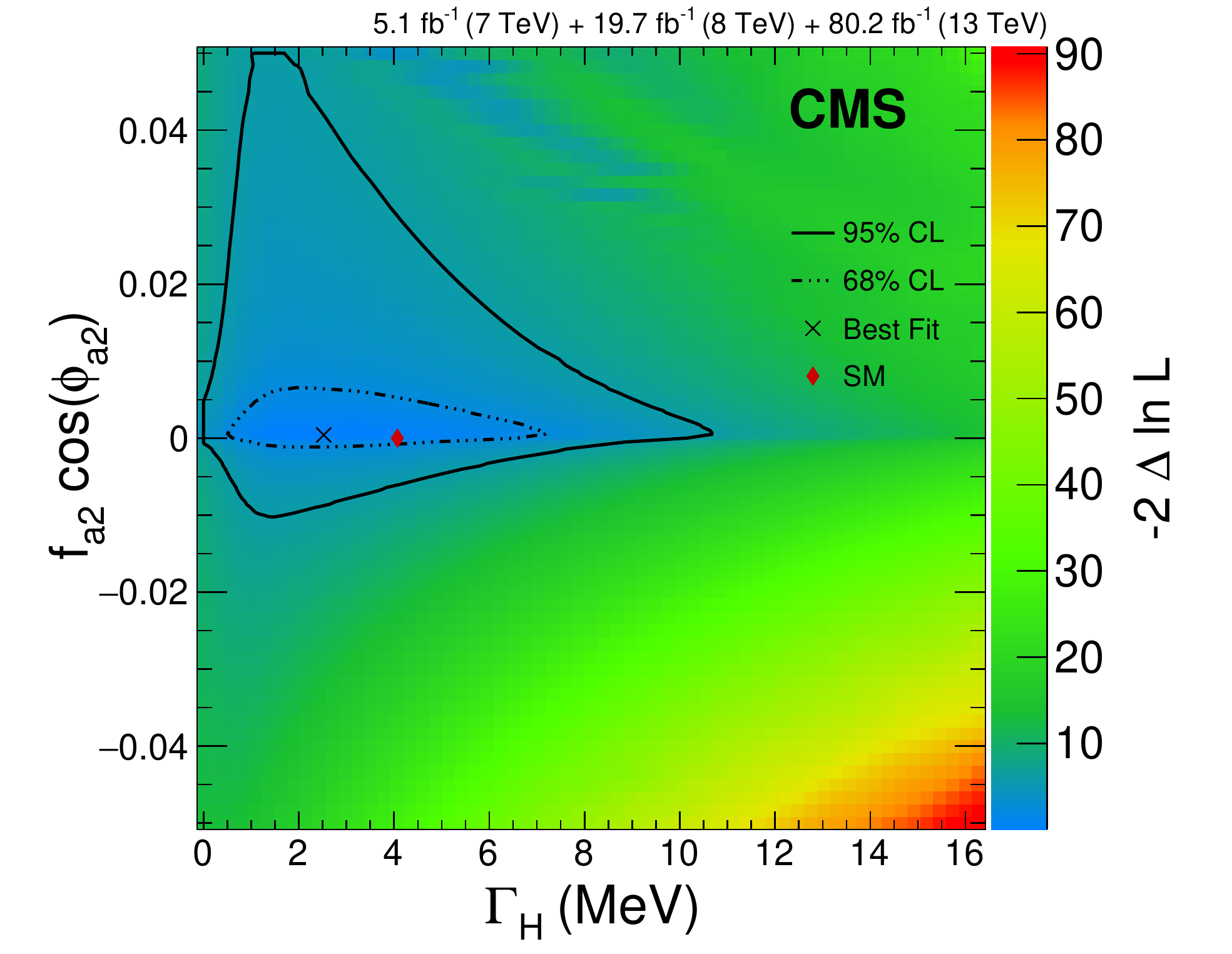} \\
\includegraphics[width=0.42\textwidth]{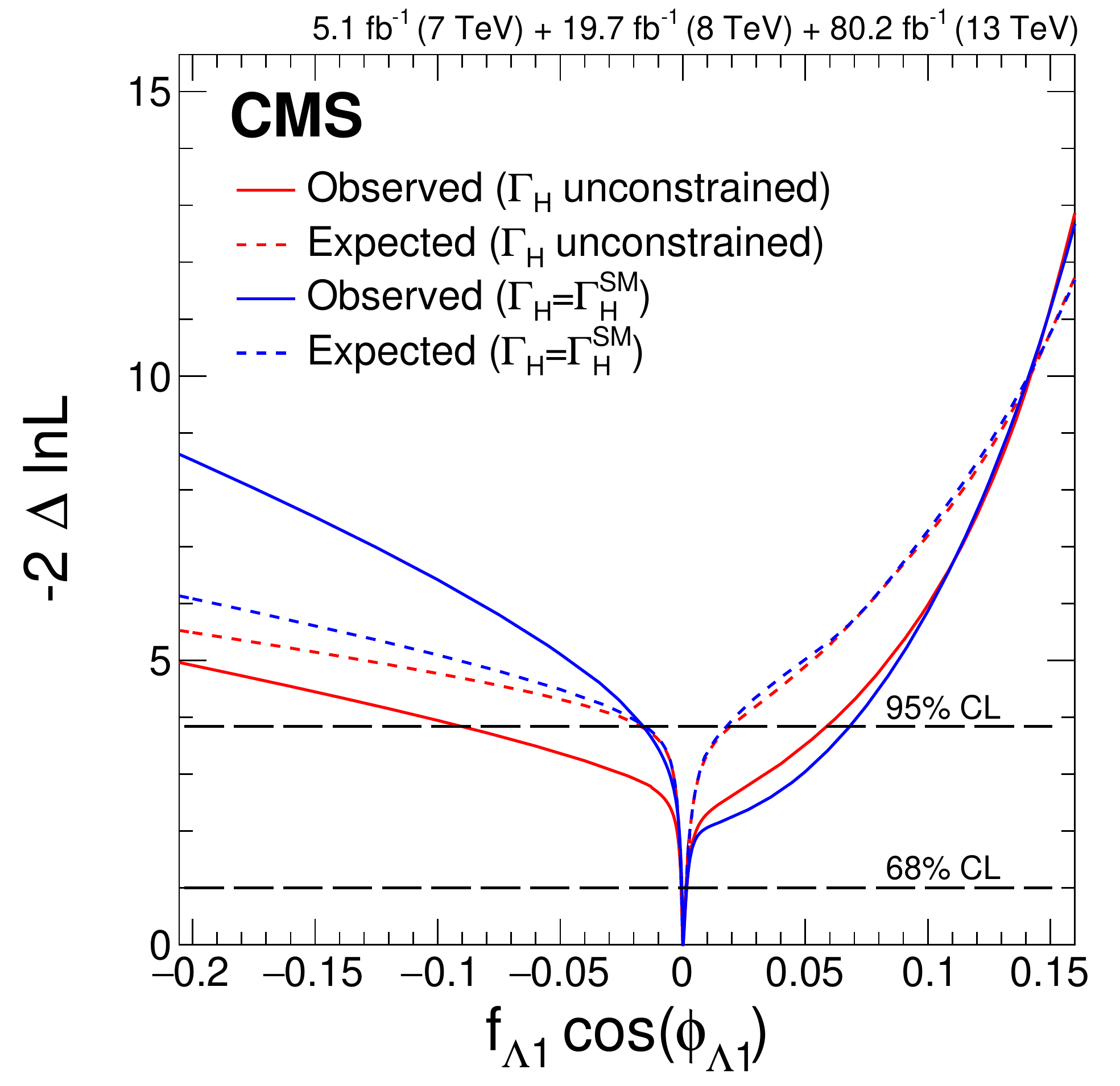}
\includegraphics[width=0.52\textwidth]{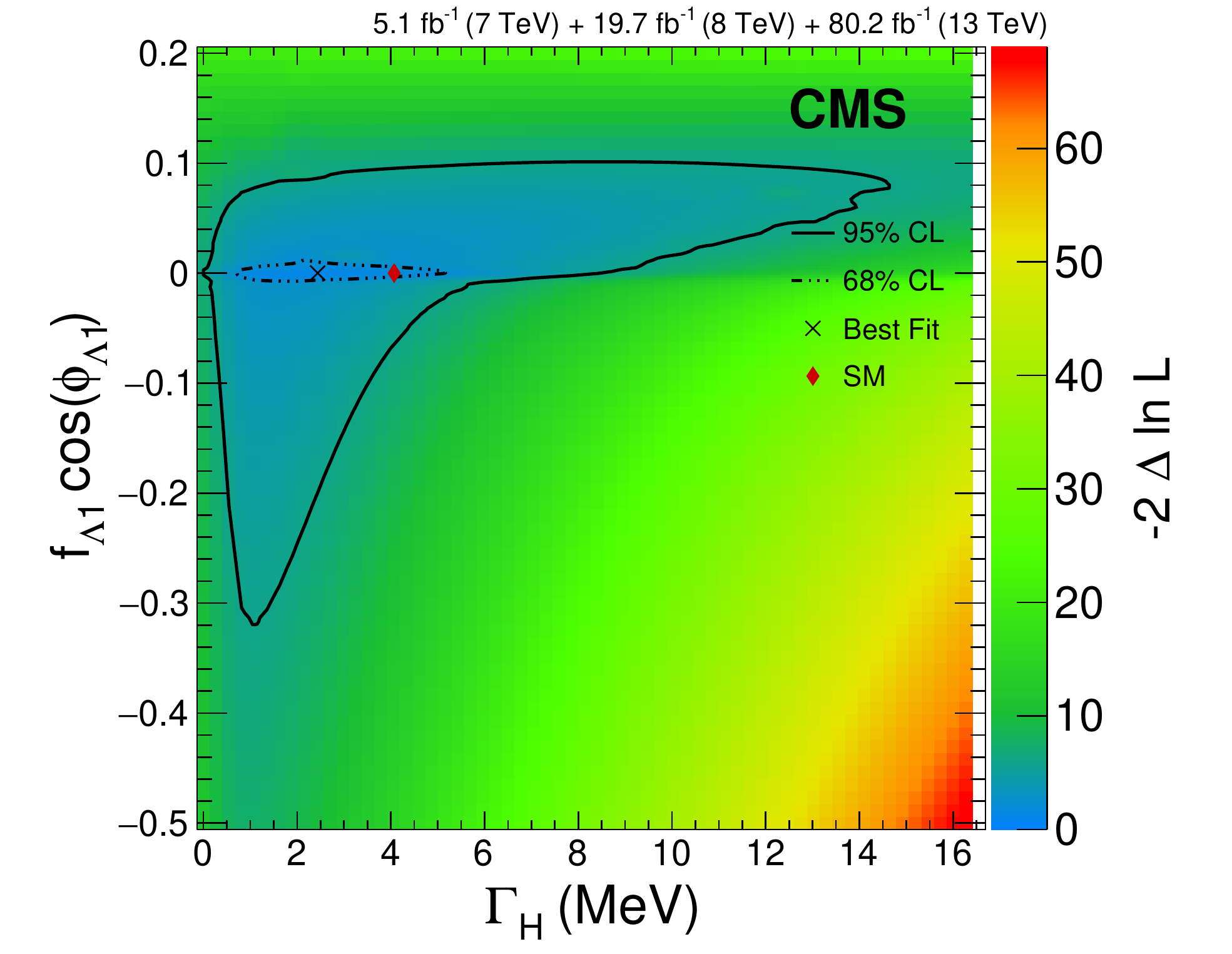}
\caption{
Constraints on \fcospAC{3} (top),  \fcospAC{2} (middle), and \fcospLC{1} (bottom) from the combined
Run~1 and Run~2 data set using both \onshell and \offshell events.
Left plots: likelihood scans of the parameters of interest with unconstrained \GH (red) or assuming $\GH=\GHSM$ (blue).
The dashed horizontal lines show the 68 and 95\%~\CL regions.
Right plots: observed two-parameter (\GH, \fcospai) likelihood scans.
The two-parameter 68 and 95\%~\CL regions are indicated with the dashed and solid curves, respectively.
}
\label{fig:resultsoffshelloneparfixed}
\end{figure*}

The combination of \onshell and \offshell regions allows the setting of tighter constraints on \fcospai
using the probability densities defined in \Eqs{eq:ponshell}{eq:poffshell}{and}.
As discussed above, the \onshell region is analyzed using the 2015, 2016, and 2017 data, and the earlier Run~1 data.
The \offshell region is analyzed using only 2016 and 2017 data because no such analysis of the three
anomalous couplings has been performed with the Run~1 or 2015 data in this region.
The one-parameter likelihood scans of \fcospai combining all such available \onshell and \offshell events is shown
for two cases in Fig.~\ref{fig:resultsoffshelloneparfixed}, either with \GH unconstrained in the fit or with
the constraint $\GH=\GHSM$.
The corresponding 68 and 95\%~\CL constraints are summarized in Table~\ref{tab:summary_fai}.
The full two-parameter likelihood scans of \fcospai and \GH are likewise shown in Fig.~\ref{fig:resultsoffshelloneparfixed}.
Using the transformation in \Eq{eq:fa_a_correspondence}, the \fcospai results can be interpreted for the coupling
parameters used in \Eq{eq:formfact-fullampl-spin0}, as shown in Table~\ref{tab:summary_aia1}.

\begin{figure*}[ht!]
\centering
\includegraphics[width=0.45\textwidth]{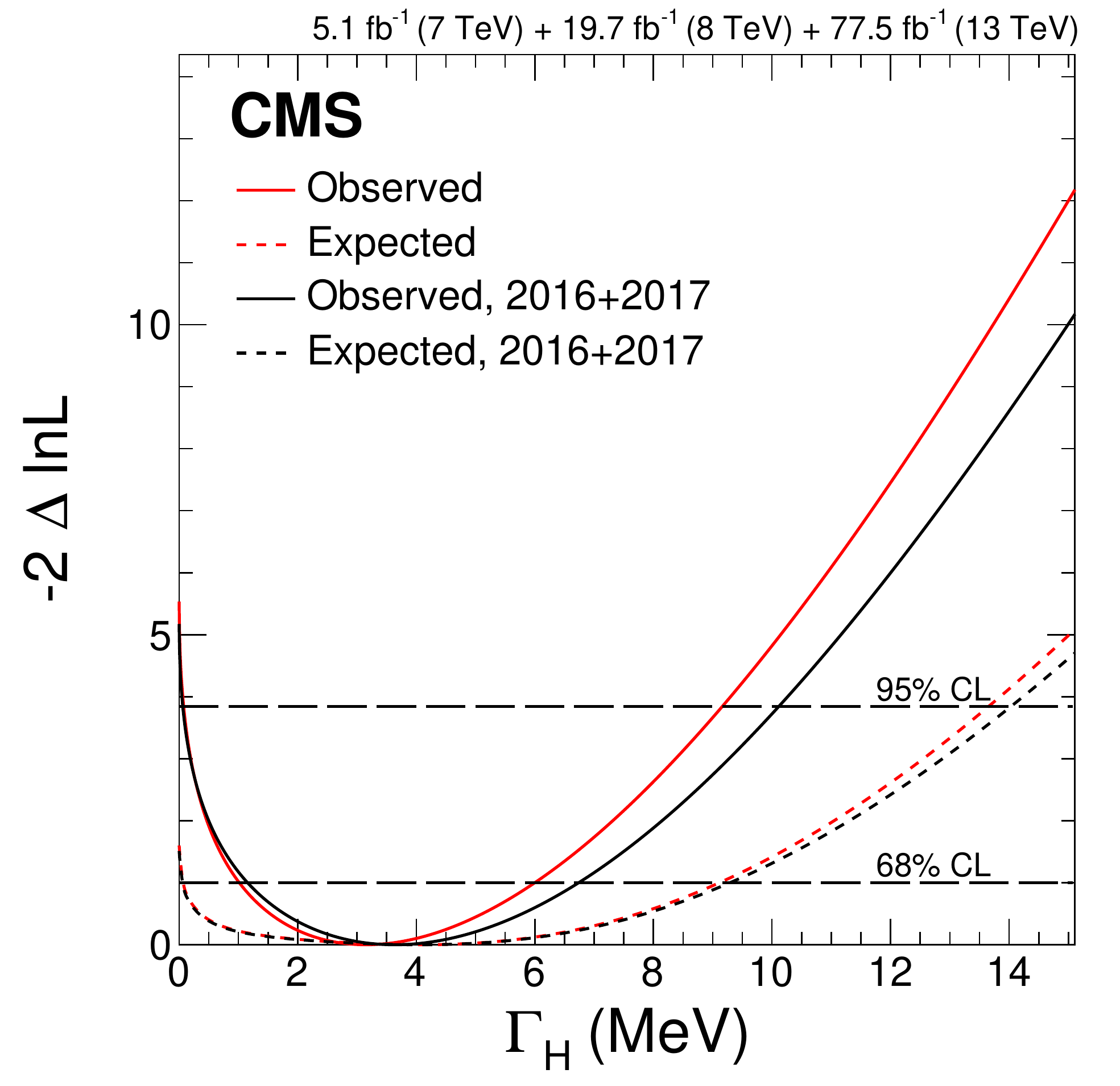}
\includegraphics[width=0.45\textwidth]{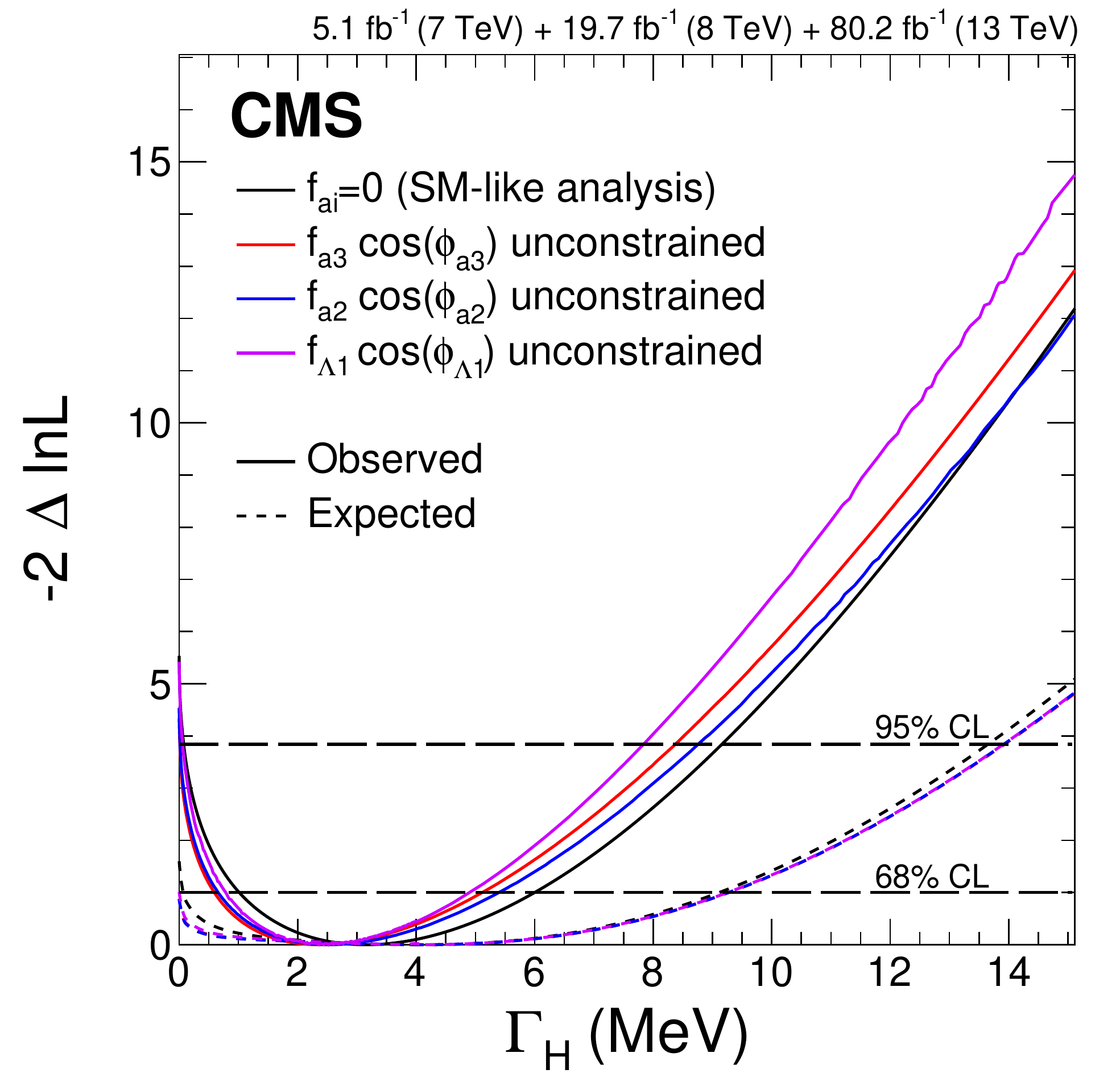}
\caption{
Observed (solid) and expected (dashed) likelihood scans of \GH.
Left plot: results of the SM-like couplings analysis are shown using the data only from 2016 and 2017 (black)
or from the combination of Run~1 and Run~2 (red), which do not include 2015 data.
Right plot: results of the combined Run~1 and Run~2 data analyses, with 2015 data included in the \onshell case,
for the SM-like couplings or with three unconstrained anomalous coupling parameters, \fcospAC{3} (red),
\fcospAC{2} (blue), and \fcospLC{1} (violet). The dashed horizontal lines show the 68\% and 95\%~\CL regions.
}
\label{fig:resultsoffshellwidth}
\end{figure*}

Limits on \GH are set by combining events from the \onshell and \offshell regions.
The left-hand panel of Fig.~\ref{fig:resultsoffshellwidth} shows the results of the likelihood scans of \GH
for the 2016 and 2017 period of the 13\TeV run and for the combined data set from collisions
at 7, 8 and 13\TeV under the assumption of SM-like couplings. The small contribution from
the 2015 data set is not considered in this case, but the Run~1 analysis includes both the
\onshell and \offshell regions in the analysis of the
$\PH\to\ZZ\to4\ell$ decay~\cite{Khachatryan:2014iha,Khachatryan:2015mma}.
The combined results are listed in Table~\ref{table:widthoffshell}.
The best fit $(\muF, \muV$) values in these results are
$(1.20^{+0.19}_{-0.16},0.62^{+0.57}_{-0.43})$ when $\GH=\GHSM$, and
$(1.21^{+0.19}_{-0.17},0.65^{+0.61}_{-0.45})$ when \GH is unconstrained.
The width constraints are also placed with the \fcospAC{3}, \fcospAC{2}, or \fcospLC{1}
parameters unconstrained, and are shown in Fig.~\ref{fig:resultsoffshellwidth} right panel
and summarized in Table~\ref{table:widthoffshell_fai}.
These results are obtained with the same fit configurations as for the study of anomalous couplings
in the combination of the \onshell and \offshell regions.

\begin{table}[!htb]
\centering
\topcaption{
Summary of the total width \GH measurement, showing the allowed 68\%~\CL (central values with uncertainties)
and 95\%~\CL (in square brackets). The limits are reported for the SM-like couplings using the Run~1 and Run~2 combination.
}
\renewcommand{\arraystretch}{1.25}
\begin{scotch}{lcc}
 Parameter                & {Observed}          &  {Expected}   \\
\hline
  \GH (\MeVns{})  &  $3.2_{-2.2}^{+2.8}$ $[0.08, 9.16]$  & $4.1_{-4.0}^{+5.0}$ $[0.0, 13.7]$\\
\end{scotch}
\label{table:widthoffshell}
\end{table}

\begin{table*}[!htb]
\centering
\topcaption{
Summary of the total width \GH measurements, showing allowed 68\%~\CL (central values with uncertainties)
and 95\%~\CL (in square brackets). The \GH limits are reported for the anomalous coupling parameter
of interest unconstrained using the Run~1 and Run~2 combination.
}
\renewcommand{\arraystretch}{1.25}
\begin{scotch}{lccc}
 Parameter               &    Unconstrained parameter           & {Observed}          &  {Expected}   \\
\hline
  \GH (\MeVns{})  & \fcospAC{3}    &  $2.4_{-1.8}^{+2.7}$ $[0.02, 8.38]$  & $4.1_{-4.1}^{+5.2}$ $[0.0, 13.9]$\\
  \GH (\MeVns{})  & \fcospAC{2}    &  $2.5_{-1.8}^{+2.9}$ $[0.02, 8.76]$  & $4.1_{-4.1}^{+5.2}$ $[0.0, 13.9]$\\
  \GH (\MeVns{})  & \fcospLC{1}   &  $2.4_{-1.6}^{+2.5}$ $[0.06, 7.84]$  & $4.1_{-4.1}^{+5.2}$ $[0.0, 13.9]$\\
\end{scotch}
\label{table:widthoffshell_fai}
\end{table*}

\begin{table}[!htb]
\centering
\topcaption
{
Summary of allowed 68\%~\CL (central values with uncertainties) and 95\%~\CL (in square brackets)
intervals for $\mu^\text{\offshell}$, $\muF^\text{\offshell}$, and $\muV^\text{\offshell}$ obtained from
the analysis of the combination of Run~1 and Run~2 \offshell data sets.
}
\renewcommand{\arraystretch}{1.25}
\begin{scotch}{cll}
 Parameter & \multicolumn{1}{c}{Observed} & \multicolumn{1}{c}{Expected}  \\
\hline
 $\mu^\text{\offshell}$ & $0.78_{-0.53}^{+0.72}$ $[0.02, 2.28]$ & $1.00_{-0.99}^{+1.20}$ $[0.0, 3.2]$  \\
$\muF^\text{\offshell}$ & $0.86_{-0.68}^{+0.92}$ $[0.0, 2.7]$ & $1.0_{-1.0}^{+1.3}$ $[0.0, 3.5]$  \\
$\muV^\text{\offshell}$ & $0.67_{-0.61}^{+1.26}$ $[0.0, 3.6]$ & $1.0_{-1.0}^{+3.8}$ $[0.0, 8.4]$  \\
\end{scotch}
\label{table:MuOffshell-Run2}
\end{table}

The systematic uncertainties mostly cancel in the ratios of cross sections in the measurement of fractional
parameters \fcospai, and are therefore negligible. The width constraints are also dominated by the statistical
uncertainties, but because of the non-trivial dependence of systematic uncertainties on \mell, their dominant
contributions may be worth examination. The two leading theoretical and two leading experimental
uncertainties affecting the width constraints (observed and expected at 68\%~\CL) are
the uncertainty on the NLO EW corrections for the $\qqbar\to4\ell$ background ($\pm0.5$ and $\pm1.9$\MeV),
the variation of renormalization scale in gluon fusion ($\pm0.2$ and $\pm0.4$\MeV),
the muon efficiency uncertainty ($\pm0.1$ and $\pm0.4$\MeV),
and the electron efficiency uncertainty ($\pm0.1$ and $\pm0.3$\MeV).

The width constraints could also be reinterpreted as an \offshell signal strength with a change of parameters.
For this interpretation, we perform an SM-like analysis of only the \offshell events, where the signal strength is
modified by the parameter $\mu^\text{\offshell}$ common to all production mechanisms in \Eqs{eq:resonant}{eq:poffshell}{and},
with $\GH=\Gref=\GHSM$ and the SM expectation corresponding to $\mu^\text{\offshell}=1$. In addition,
we also perform a fit of the \offshell events with two unconstrained parameters $\mu^\text{\offshell}_{\F}$ and
$\mu^\text{\offshell}_{\V}$, which express the signal strengths in the gluon fusion and EW processes, respectively.
These constraints are summarized in Table~\ref{table:MuOffshell-Run2}.

\section{Summary}
\label{sec:Summary}

Studies of \onshell and \offshell \Hboson production in the four-lepton final state are presented, using data
from the CMS experiment at the LHC that correspond to an integrated luminosity of 80.2\fbinv at
a center-of-mass energy of 13\TeV. Joint constraints are set on the  \Hboson total width and parameters that
express its anomalous couplings to two electroweak vector bosons. These results are combined with those
obtained from the data collected at center-of-mass energies of 7 and 8\TeV, corresponding to integrated
luminosities of 5.1 and 19.7\fbinv, respectively. Kinematic information from the decay particles
and the associated jets are combined using matrix element techniques to identify the production mechanism
and increase sensitivity to the \Hboson couplings in both production and decay. The constraints on anomalous 
\HVV couplings are found to be consistent with the standard model expectation in both \onshell and \offshell regions,
as presented in Tables~\ref{tab:summary_fai} and~\ref{tab:summary_aia1}.
Under the assumption of a coupling structure similar to that in the standard model, the \Hboson width is constrained
to be $3.2^{+2.8}_{-2.2}$\MeV while the expected constraint based on simulation is $4.1^{+5.0}_{-4.0}$\MeV,
as shown in Table~\ref{table:widthoffshell}.
The constraints on the width remain similar with the inclusion of the tested anomalous \HVV interactions and
are summarized in Table~\ref{table:widthoffshell_fai}. The width results are also interpreted in terms
of the \Hboson signal strength in the \offshell region in Table~\ref{table:MuOffshell-Run2}. The observed \offshell signal strength,
or equivalently a nonzero value of the width, is more than 2 standard deviations away from a background-only hypothesis,
which provides a new direction to measure \Hboson properties when more data are available.

\begin{acknowledgments}
We thank Markus Schulze for optimizing the \textsc{JHUGen} Monte Carlo simulation program and matrix element library for this analysis.

We congratulate our colleagues in the CERN accelerator departments for the excellent performance of the LHC and thank the technical and administrative staffs at CERN and at other CMS institutes for their contributions to the success of the CMS effort. In addition, we gratefully acknowledge the computing centers and personnel of the Worldwide LHC Computing Grid for delivering so effectively the computing infrastructure essential to our analyses. We also acknowledge the Maryland Advanced Research Computing Center (MARCC) for providing computing resources essential for this analysis. Finally, we acknowledge the enduring support for the construction and operation of the LHC and the CMS detector provided by the following funding agencies: BMBWF and FWF (Austria); FNRS and FWO (Belgium); CNPq, CAPES, FAPERJ, FAPERGS, and FAPESP (Brazil); MES (Bulgaria); CERN; CAS, MoST, and NSFC (China); COLCIENCIAS (Colombia); MSES and CSF (Croatia); RPF (Cyprus); SENESCYT (Ecuador); MoER, ERC IUT, and ERDF (Estonia); Academy of Finland, MEC, and HIP (Finland); CEA and CNRS/IN2P3 (France); BMBF, DFG, and HGF (Germany); GSRT (Greece); NKFIA (Hungary); DAE and DST (India); IPM (Iran); SFI (Ireland); INFN (Italy); MSIP and NRF (Republic of Korea); MES (Latvia); LAS (Lithuania); MOE and UM (Malaysia); BUAP, CINVESTAV, CONACYT, LNS, SEP, and UASLP-FAI (Mexico); MOS (Montenegro); MBIE (New Zealand); PAEC (Pakistan); MSHE and NSC (Poland); FCT (Portugal); JINR (Dubna); MON, RosAtom, RAS, RFBR, and NRC KI (Russia); MESTD (Serbia); SEIDI, CPAN, PCTI, and FEDER (Spain); MOSTR (Sri Lanka); Swiss Funding Agencies (Switzerland); MST (Taipei); ThEPCenter, IPST, STAR, and NSTDA (Thailand); TUBITAK and TAEK (Turkey); NASU and SFFR (Ukraine); STFC (United Kingdom); DOE and NSF (USA).

\hyphenation{Rachada-pisek} Individuals have received support from the Marie-Curie program and the European Research Council and Horizon 2020 Grant, contract No. 675440 (European Union); the Leventis Foundation; the A.P.\ Sloan Foundation; the Alexander von Humboldt Foundation; the Belgian Federal Science Policy Office; the Fonds pour la Formation \`a la Recherche dans l'Industrie et dans l'Agriculture (FRIA-Belgium); the Agentschap voor Innovatie door Wetenschap en Technologie (IWT-Belgium); the F.R.S.-FNRS and FWO (Belgium) under the ``Excellence of Science -- EOS" -- be.h project n.\ 30820817; the Ministry of Education, Youth and Sports (MEYS) of the Czech Republic; the Lend\"ulet (``Momentum") Program and the J\'anos Bolyai Research Scholarship of the Hungarian Academy of Sciences, the New National Excellence Program \'UNKP, the NKFIA research grants 123842, 123959, 124845, 124850, and 125105 (Hungary); the Council of Science and Industrial Research, India; the HOMING PLUS program of the Foundation for Polish Science, cofinanced from European Union, Regional Development Fund, the Mobility Plus program of the Ministry of Science and Higher Education, the National Science Center (Poland), contracts Harmonia 2014/14/M/ST2/00428, Opus 2014/13/B/ST2/02543, 2014/15/B/ST2/03998, and 2015/19/B/ST2/02861, Sonata-bis 2012/07/E/ST2/01406; the National Priorities Research Program by Qatar National Research Fund; the Programa Estatal de Fomento de la Investigaci{\'o}n Cient{\'i}fica y T{\'e}cnica de Excelencia Mar\'{\i}a de Maeztu, grant MDM-2015-0509 and the Programa Severo Ochoa del Principado de Asturias; the Thalis and Aristeia programs cofinanced by EU-ESF and the Greek NSRF; the Rachadapisek Sompot Fund for Postdoctoral Fellowship, Chulalongkorn University and the Chulalongkorn Academic into Its 2nd Century Project Advancement Project (Thailand); the Welch Foundation, contract C-1845; and the Weston Havens Foundation (USA).
\end{acknowledgments}

\bibliography{auto_generated}

\providecommand{\href}[2]{#2}\begingroup\raggedright\begin{thebibliography}{10}%
\makeatletter
\providecommand{\hrefCMSnoop }[0]{\@secondoftwo}%
\makeatother
\providecommand{\doi}{\texttt{doi:}\begingroup \urlstyle{tt}\Url}

\bibitem{StandardModel67_1}
\hrefCMSnoop {}{S.~L. Glashow, ``{Partial-symmetries of weak interactions}'',}
  \textit{ Nucl. Phys.} \textbf{ 22} (1961) 579,
\href{http://dx.doi.org/10.1016/0029-5582(61)90469-2}{\doi{10.1016/0029-5582(61)90469-2}}.

\bibitem{Englert:1964et}
\hrefCMSnoop {}{F.~Englert and R.~Brout, ``{Broken symmetry and the mass of
  gauge vector mesons}'',} \textit{ Phys. Rev. Lett.} \textbf{ 13} (1964) 321,
  \href{http://dx.doi.org/10.1103/PhysRevLett.13.321}{\doi{10.1103/PhysRevLett.13.321}}.

\bibitem{Higgs:1964ia}
\hrefCMSnoop {}{P.~W. Higgs, ``{Broken symmetries, massless particles and gauge
  fields}'',} \textit{ Phys. Lett.} \textbf{ 12} (1964) 132,
  \href{http://dx.doi.org/10.1016/0031-9163(64)91136-9}{\doi{10.1016/0031-9163(64)91136-9}}.

\bibitem{Higgs:1964pj}
\hrefCMSnoop {}{P.~W. Higgs, ``{Broken symmetries and the masses of gauge
  bosons}'',} \textit{ Phys. Rev. Lett.} \textbf{ 13} (1964) 508,
  \href{http://dx.doi.org/10.1103/PhysRevLett.13.508}{\doi{10.1103/PhysRevLett.13.508}}.

\bibitem{Guralnik:1964eu}
\hrefCMSnoop {}{G.~S. Guralnik, C.~R. Hagen, and T.~W.~B. Kibble, ``{Global
  conservation laws and massless particles}'',} \textit{ Phys. Rev. Lett.}
  \textbf{ 13} (1964) 585,
  \href{http://dx.doi.org/10.1103/PhysRevLett.13.585}{\doi{10.1103/PhysRevLett.13.585}}.

\bibitem{StandardModel67_2}
\hrefCMSnoop {}{S.~Weinberg, ``{A model of leptons}'',} \textit{ Phys. Rev.
  Lett.} \textbf{ 19} (1967) 1264,
\href{http://dx.doi.org/10.1103/PhysRevLett.19.1264}{\doi{10.1103/PhysRevLett.19.1264}}.

\bibitem{StandardModel67_3}
\hrefCMSnoop {}{A.~Salam, ``Weak and electromagnetic interactions'',} in
  \textit{ Elementary particle physics: relativistic groups and analyticity},
  N.~Svartholm, ed., p.~367.
\newblock Almqvist \& Wiksell, Stockholm, 1968.
\newblock Proceedings of the eighth Nobel symposium.

\bibitem{Aad:2012tfa}
\hrefCMSnoop {}{{ATLAS Collaboration}, ``{Observation of a new particle in the
  search for the Standard Model Higgs boson with the ATLAS detector at the
  LHC}'',} \textit{ Phys. Lett. B} \textbf{ 716} (2012) 1,
  \href{http://dx.doi.org/10.1016/j.physletb.2012.08.020}{\doi{10.1016/j.physletb.2012.08.020}},
\href{http://www.arXiv.org/abs/1207.7214}{\texttt{arXiv:1207.7214}}.

\bibitem{Chatrchyan:2012xdj}
\hrefCMSnoop {}{{CMS Collaboration}, ``{Observation of a new boson at a mass of
  125\GeV with the CMS experiment at the LHC}'',} \textit{ Phys. Lett. B}
  \textbf{ 716} (2012) 30,
  \href{http://dx.doi.org/10.1016/j.physletb.2012.08.021}{\doi{10.1016/j.physletb.2012.08.021}},
\href{http://www.arXiv.org/abs/1207.7235}{\texttt{arXiv:1207.7235}}.

\bibitem{Chatrchyan:2013lba}
\hrefCMSnoop {}{{CMS Collaboration}, ``{Observation of a new boson with mass
  near 125\GeV in $\Pp\Pp$ collisions at $\sqrt{s}=7$ and~8\TeV}'',} \textit{
  JHEP} \textbf{ 06} (2013) 081,
  \href{http://dx.doi.org/10.1007/JHEP06(2013)081}{\doi{10.1007/JHEP06(2013)081}},
\href{http://www.arXiv.org/abs/1303.4571}{\texttt{arXiv:1303.4571}}.

\bibitem{Khachatryan:2014iha}
\hrefCMSnoop {}{{CMS Collaboration}, ``{Constraints on the Higgs boson width
  from off-shell production and decay to $\PZ$-boson pairs}'',} \textit{ Phys.
  Lett. B} \textbf{ 736} (2014) 64,
  \href{http://dx.doi.org/10.1016/j.physletb.2014.06.077}{\doi{10.1016/j.physletb.2014.06.077}},
\href{http://www.arXiv.org/abs/1405.3455}{\texttt{arXiv:1405.3455}}.

\bibitem{Aad:2015xua}
\hrefCMSnoop {}{{ATLAS Collaboration}, ``{Constraints on the off-shell Higgs
  boson signal strength in the high-mass $\ZZ$ and $\WW$ final states with the
  ATLAS detector}'',} \textit{ Eur. Phys. J. C} \textbf{ 75} (2015) 335,
  \href{http://dx.doi.org/10.1140/epjc/s10052-015-3542-2}{\doi{10.1140/epjc/s10052-015-3542-2}},
\href{http://www.arXiv.org/abs/1503.01060}{\texttt{arXiv:1503.01060}}.

\bibitem{Khachatryan:2015mma}
\hrefCMSnoop {}{{CMS Collaboration}, ``{Limits on the Higgs boson lifetime and
  width from its decay to four charged leptons}'',} \textit{ Phys. Rev. D}
  \textbf{ 92} (2015) 072010,
  \href{http://dx.doi.org/10.1103/PhysRevD.92.072010}{\doi{10.1103/PhysRevD.92.072010}},
\href{http://www.arXiv.org/abs/1507.06656}{\texttt{arXiv:1507.06656}}.

\bibitem{Khachatryan:2016ctc}
\hrefCMSnoop {}{{CMS Collaboration}, ``{Search for Higgs boson off-shell
  production in proton-proton collisions at 7 and 8 TeV and derivation of
  constraints on its total decay width}'',} \textit{ JHEP} \textbf{ 09} (2016)
  051,
  \href{http://dx.doi.org/10.1007/JHEP09(2016)051}{\doi{10.1007/JHEP09(2016)051}},
\href{http://www.arXiv.org/abs/1605.02329}{\texttt{arXiv:1605.02329}}.

\bibitem{Aaboud:2018puo}
\hrefCMSnoop {}{{ATLAS Collaboration}, ``{Constraints on off-shell Higgs boson
  production and the Higgs boson total width in $\ZZ\to4\ell$ and
  $\ZZ\to2\ell2\Pgn$ final states with the ATLAS detector}'',} \textit{ Phys.
  Lett. B} \textbf{ 786} (2018) 223,
  \href{http://dx.doi.org/10.1016/j.physletb.2018.09.048}{\doi{10.1016/j.physletb.2018.09.048}},
\href{http://www.arXiv.org/abs/1808.01191}{\texttt{arXiv:1808.01191}}.

\bibitem{Caola:2013yja}
\hrefCMSnoop {}{F.~Caola and K.~Melnikov, ``{Constraining the Higgs boson width
  with \ZZ production at the LHC}'',} \textit{ Phys. Rev. D} \textbf{ 88}
  (2013) 054024,
  \href{http://dx.doi.org/10.1103/PhysRevD.88.054024}{\doi{10.1103/PhysRevD.88.054024}},
\href{http://www.arXiv.org/abs/1307.4935}{\texttt{arXiv:1307.4935}}.

\bibitem{Kauer:2012hd}
\hrefCMSnoop {}{N.~Kauer and G.~Passarino, ``{Inadequacy of zero-width
  approximation for a light Higgs boson signal}'',} \textit{ JHEP} \textbf{ 08}
  (2012) 116,
  \href{http://dx.doi.org/10.1007/JHEP08(2012)116}{\doi{10.1007/JHEP08(2012)116}},
\href{http://www.arXiv.org/abs/1206.4803}{\texttt{arXiv:1206.4803}}.

\bibitem{Campbell:2013una}
\hrefCMSnoop {}{J.~M. Campbell, R.~K. Ellis, and C.~Williams, ``{Bounding the
  Higgs width at the LHC using full analytic results for $\Pg\Pg\to
  \Pe^{-}\Pe^{+} \PGm^{-} \PGm^{+}$}'',} \textit{ JHEP} \textbf{ 04} (2014)
  060,
  \href{http://dx.doi.org/10.1007/JHEP04(2014)060}{\doi{10.1007/JHEP04(2014)060}},
\href{http://www.arXiv.org/abs/1311.3589}{\texttt{arXiv:1311.3589}}.

\bibitem{Khachatryan:2014jba}
\hrefCMSnoop {}{{CMS Collaboration}, ``{Precise determination of the mass of
  the Higgs boson and tests of compatibility of its couplings with the standard
  model predictions using proton collisions at 7 and 8 $\,\text {TeV}$}'',}
  \textit{ Eur. Phys. J. C} \textbf{ 75} (2015) 212,
  \href{http://dx.doi.org/10.1140/epjc/s10052-015-3351-7}{\doi{10.1140/epjc/s10052-015-3351-7}},
\href{http://www.arXiv.org/abs/1412.8662}{\texttt{arXiv:1412.8662}}.

\bibitem{Aad:2014aba}
\hrefCMSnoop {}{{ATLAS Collaboration}, ``{Measurement of the Higgs boson mass
  from the $\PH\rightarrow \gamma\gamma$ and $\PH \rightarrow \PZ\PZ^{*}
  \rightarrow 4\ell$ channels with the ATLAS detector using 25\fbinv of
  $\Pp\Pp$ collision data}'',} \textit{ Phys. Rev. D} \textbf{ 90} (2014)
  052004,
  \href{http://dx.doi.org/10.1103/PhysRevD.90.052004}{\doi{10.1103/PhysRevD.90.052004}},
\href{http://www.arXiv.org/abs/1406.3827}{\texttt{arXiv:1406.3827}}.

\bibitem{Sirunyan:2017exp}
\hrefCMSnoop {}{{CMS Collaboration}, ``{Measurements of properties of the Higgs
  boson decaying into the four-lepton final state in $\Pp\Pp$ collisions at
  $\sqrt{s}=13\TeV$}'',} \textit{ JHEP} \textbf{ 11} (2017) 047,
  \href{http://dx.doi.org/10.1007/JHEP11(2017)047}{\doi{10.1007/JHEP11(2017)047}},
\href{http://www.arXiv.org/abs/1706.09936}{\texttt{arXiv:1706.09936}}.

\bibitem{deFlorian:2016spz}
\hrefCMSnoop {}{D.~de~Florian {et~al.}, ``Handbook of {LHC} {H}iggs cross
  sections: 4. deciphering the nature of the {H}iggs sector'',} CERN Report
  CERN-2017-002-M, 2016.
\newblock
  \href{http://dx.doi.org/10.23731/CYRM-2017-002}{\doi{10.23731/CYRM-2017-002}},
  \href{http://www.arXiv.org/abs/1610.07922}{\texttt{arXiv:1610.07922}}.

\bibitem{Chatrchyan:2012jja}
\hrefCMSnoop {}{{CMS Collaboration}, ``{On the mass and spin-parity of the
  Higgs boson candidate via its decays to $\PZ$ boson pairs}'',} \textit{ Phys.
  Rev. Lett.} \textbf{ 110} (2013) 081803,
  \href{http://dx.doi.org/10.1103/PhysRevLett.110.081803}{\doi{10.1103/PhysRevLett.110.081803}},
\href{http://www.arXiv.org/abs/1212.6639}{\texttt{arXiv:1212.6639}}.

\bibitem{Chatrchyan:2013mxa}
\hrefCMSnoop {}{{CMS Collaboration}, ``{Measurement of the properties of a
  Higgs boson in the four-lepton final state}'',} \textit{ Phys. Rev. D}
  \textbf{ 89} (2014) 092007,
  \href{http://dx.doi.org/10.1103/PhysRevD.89.092007}{\doi{10.1103/PhysRevD.89.092007}},
\href{http://www.arXiv.org/abs/1312.5353}{\texttt{arXiv:1312.5353}}.

\bibitem{Khachatryan:2014kca}
\hrefCMSnoop {}{{CMS Collaboration}, ``{Constraints on the spin-parity and
  anomalous \HVV couplings of the Higgs boson in proton collisions at 7 and
  8\TeV}'',} \textit{ Phys. Rev. D} \textbf{ 92} (2015) 012004,
  \href{http://dx.doi.org/10.1103/PhysRevD.92.012004}{\doi{10.1103/PhysRevD.92.012004}},
\href{http://www.arXiv.org/abs/1411.3441}{\texttt{arXiv:1411.3441}}.

\bibitem{Khachatryan:2016tnr}
\hrefCMSnoop {}{{CMS Collaboration}, ``{Combined search for anomalous
  pseudoscalar \HVV couplings in \VH ($\PH\to\bbar$) production and $\PH\to\VV$
  decay}'',} \textit{ Phys. Lett. B} \textbf{ 759} (2016) 672,
  \href{http://dx.doi.org/10.1016/j.physletb.2016.06.004}{\doi{10.1016/j.physletb.2016.06.004}},
\href{http://www.arXiv.org/abs/1602.04305}{\texttt{arXiv:1602.04305}}.

\bibitem{Sirunyan:2017tqd}
\hrefCMSnoop {}{{CMS Collaboration}, ``{Constraints on anomalous Higgs boson
  couplings using production and decay information in the four-lepton final
  state}'',} \textit{ Phys. Lett. B} \textbf{ 775} (2017) 1,
  \href{http://dx.doi.org/10.1016/j.physletb.2017.10.021}{\doi{10.1016/j.physletb.2017.10.021}},
\href{http://www.arXiv.org/abs/1707.00541}{\texttt{arXiv:1707.00541}}.

\bibitem{Aad:2013xqa}
\hrefCMSnoop {}{{ATLAS Collaboration}, ``{Evidence for the spin-0 nature of the
  Higgs boson using ATLAS data}'',} \textit{ Phys. Lett. B} \textbf{ 726}
  (2013) 120,
  \href{http://dx.doi.org/10.1016/j.physletb.2013.08.026}{\doi{10.1016/j.physletb.2013.08.026}},
\href{http://www.arXiv.org/abs/1307.1432}{\texttt{arXiv:1307.1432}}.

\bibitem{Aad:2015mxa}
\hrefCMSnoop {}{{ATLAS Collaboration}, ``{Study of the spin and parity of the
  Higgs boson in diboson decays with the ATLAS detector}'',} \textit{ Eur.
  Phys. J. C} \textbf{ 75} (2015) 476,
  \href{http://dx.doi.org/10.1140/epjc/s10052-015-3685-1}{\doi{10.1140/epjc/s10052-015-3685-1}},
\href{http://www.arXiv.org/abs/1506.05669}{\texttt{arXiv:1506.05669}}.

\bibitem{Aad:2016nal}
\hrefCMSnoop {}{{ATLAS Collaboration}, ``{Test of CP Invariance in vector-boson
  fusion production of the Higgs boson using the Optimal Observable method in
  the ditau decay channel with the ATLAS detector}'',} \textit{ Eur. Phys. J.
  C} \textbf{ 76} (2016) 658,
  \href{http://dx.doi.org/10.1140/epjc/s10052-016-4499-5}{\doi{10.1140/epjc/s10052-016-4499-5}},
\href{http://www.arXiv.org/abs/1602.04516}{\texttt{arXiv:1602.04516}}.

\bibitem{Aaboud:2017oem}
\hrefCMSnoop {}{{ATLAS Collaboration}, ``{Measurement of inclusive and
  differential cross sections in the $\PH \rightarrow \PZ\PZ^{*} \rightarrow
  4\ell$ decay channel in $\Pp\Pp$ collisions at $\sqrt{s}=13\TeV$ with the
  ATLAS detector}'',} \textit{ JHEP} \textbf{ 10} (2017) 132,
  \href{http://dx.doi.org/10.1007/JHEP10(2017)132}{\doi{10.1007/JHEP10(2017)132}},
\href{http://www.arXiv.org/abs/1708.02810}{\texttt{arXiv:1708.02810}}.

\bibitem{Aaboud:2017vzb}
\hrefCMSnoop {}{{ATLAS Collaboration}, ``{Measurement of the Higgs boson
  coupling properties in the $H\rightarrow ZZ^{*} \rightarrow 4\ell$ decay
  channel at $\sqrt{s}$ = 13 TeV with the ATLAS detector}'',} \textit{ JHEP}
  \textbf{ 03} (2018) 095,
  \href{http://dx.doi.org/10.1007/JHEP03(2018)095}{\doi{10.1007/JHEP03(2018)095}},
\href{http://www.arXiv.org/abs/1712.02304}{\texttt{arXiv:1712.02304}}.

\bibitem{Aaboud:2018xdt}
\hrefCMSnoop {}{{ATLAS Collaboration}, ``{Measurements of Higgs boson
  properties in the diphoton decay channel with 36 fb$^{-1}$ of $\Pp\Pp$
  collision data at $\sqrt{s} = 13$ TeV with the ATLAS detector}'',} \textit{
  Phys. Rev. D} \textbf{ 98} (2018) 052005,
  \href{http://dx.doi.org/10.1103/PhysRevD.98.052005}{\doi{10.1103/PhysRevD.98.052005}},
\href{http://www.arXiv.org/abs/1802.04146}{\texttt{arXiv:1802.04146}}.

\bibitem{Gainer:2014hha}
J.~S. Gainer\hrefCMSnoop {}{ {et~al.}, ``{Beyond geolocating: Constraining
  higher dimensional operators in $\PH \to 4\ell$ with off-shell production and
  more}'',} \textit{ Phys. Rev. D} \textbf{ 91} (2015) 035011,
  \href{http://dx.doi.org/10.1103/PhysRevD.91.035011}{\doi{10.1103/PhysRevD.91.035011}},
\href{http://www.arXiv.org/abs/1403.4951}{\texttt{arXiv:1403.4951}}.

\bibitem{Englert:2014aca}
\hrefCMSnoop {}{C.~Englert and M.~Spannowsky, ``{Limitations and opportunities
  of off-shell coupling measurements}'',} \textit{ Phys. Rev. D} \textbf{ 90}
  (2014) 053003,
  \href{http://dx.doi.org/10.1103/PhysRevD.90.053003}{\doi{10.1103/PhysRevD.90.053003}},
\href{http://www.arXiv.org/abs/1405.0285}{\texttt{arXiv:1405.0285}}.

\bibitem{Ghezzi:2014qpa}
\hrefCMSnoop {}{M.~Ghezzi, G.~Passarino, and S.~Uccirati, ``{Bounding the Higgs
  width using effective field theory}'',} in \textit{ {Proceedings, 12th DESY
  Workshop on Elementary Particle Physics: Loops and Legs in Quantum Field
  Theory (LL2014)}}, p.~072.
\newblock Weimar, Germany, April, 2014.
\newblock \href{http://www.arXiv.org/abs/1405.1925}{\texttt{arXiv:1405.1925}}.
\newblock [PoS(LL2014)072].
\href{http://dx.doi.org/10.22323/1.211.0072}{\doi{10.22323/1.211.0072}}.

\bibitem{Sirunyan:2018qlb}
\hrefCMSnoop {}{{CMS Collaboration}, ``{Search for a new scalar resonance
  decaying to a pair of $\PZ$ bosons in proton-proton collisions at
  $\sqrt{s}=13\TeV$}'',} \textit{ JHEP} \textbf{ 06} (2018) 127,
  \href{http://dx.doi.org/10.1007/JHEP06(2018)127}{\doi{10.1007/JHEP06(2018)127}},
\href{http://www.arXiv.org/abs/1804.01939}{\texttt{arXiv:1804.01939}}.

\bibitem{Nelson:1986ki}
\hrefCMSnoop {}{C.~A. Nelson, ``{Correlation between decay planes in
  Higgs-boson decays into a $\PW$ Pair (into a $\PZ$ Pair)}'',} \textit{ Phys.
  Rev. D} \textbf{ 37} (1988) 1220,
\href{http://dx.doi.org/10.1103/PhysRevD.37.1220}{\doi{10.1103/PhysRevD.37.1220}}.

\bibitem{Soni:1993jc}
\hrefCMSnoop {}{A.~Soni and R.~M. Xu, ``{Probing CP violation via Higgs decays
  to four leptons}'',} \textit{ Phys. Rev. D} \textbf{ 48} (1993) 5259,
  \href{http://dx.doi.org/10.1103/PhysRevD.48.5259}{\doi{10.1103/PhysRevD.48.5259}},
\href{http://www.arXiv.org/abs/hep-ph/9301225}{\texttt{arXiv:hep-ph/9301225}}.

\bibitem{Plehn:2001nj}
\hrefCMSnoop {}{T.~Plehn, D.~L. Rainwater, and D.~Zeppenfeld, ``{Determining
  the structure of Higgs couplings at the LHC}'',} \textit{ Phys. Rev. Lett.}
  \textbf{ 88} (2002) 051801,
  \href{http://dx.doi.org/10.1103/PhysRevLett.88.051801}{\doi{10.1103/PhysRevLett.88.051801}},
\href{http://www.arXiv.org/abs/hep-ph/0105325}{\texttt{arXiv:hep-ph/0105325}}.

\bibitem{Choi:2002jk}
\hrefCMSnoop {}{S.~Y. Choi, D.~J. Miller, M.~M. M{\" u}hlleitner, and P.~M.
  Zerwas, ``{Identifying the Higgs spin and parity in decays to Z pairs}'',}
  \textit{ Phys. Lett. B} \textbf{ 553} (2003) 61,
  \href{http://dx.doi.org/10.1016/S0370-2693(02)03191-X}{\doi{10.1016/S0370-2693(02)03191-X}},
\href{http://www.arXiv.org/abs/hep-ph/0210077}{\texttt{arXiv:hep-ph/0210077}}.

\bibitem{Buszello:2002uu}
\hrefCMSnoop {}{C.~P. Buszello, I.~Fleck, P.~Marquard, and J.~J. van~der Bij,
  ``{Prospective analysis of spin- and CP-sensitive variables in $\PH \to
  \PZ\PZ \to \ell_1^+ \ell_1^- \ell_2^+ \ell_2^-$ at the LHC}'',} \textit{ Eur.
  Phys. J. C} \textbf{ 32} (2004) 209,
  \href{http://dx.doi.org/10.1140/epjc/s2003-01392-0}{\doi{10.1140/epjc/s2003-01392-0}},
\href{http://www.arXiv.org/abs/hep-ph/0212396}{\texttt{arXiv:hep-ph/0212396}}.

\bibitem{Hankele:2006ma}
\hrefCMSnoop {}{V.~Hankele, G.~Klamke, D.~Zeppenfeld, and T.~Figy, ``{Anomalous
  Higgs boson couplings in vector boson fusion at the CERN LHC}'',} \textit{
  Phys. Rev. D} \textbf{ 74} (2006) 095001,
  \href{http://dx.doi.org/10.1103/PhysRevD.74.095001}{\doi{10.1103/PhysRevD.74.095001}},
\href{http://www.arXiv.org/abs/hep-ph/0609075}{\texttt{arXiv:hep-ph/0609075}}.

\bibitem{Accomando:2006ga}
\hrefCMSnoop {}{E.~Accomando {et~al.}, ``{Workshop on CP studies and
  non-standard Higgs physics}'',} (2006).
\href{http://www.arXiv.org/abs/hep-ph/0608079}{\texttt{arXiv:hep-ph/0608079}}.

\bibitem{Godbole:2007cn}
\hrefCMSnoop {}{R.~M. Godbole, D.~J. Miller, and M.~M. M{\" u}hlleitner,
  ``{Aspects of CP violation in the $\PH\PZ\PZ$ coupling at the LHC}'',}
  \textit{ JHEP} \textbf{ 12} (2007) 031,
  \href{http://dx.doi.org/10.1088/1126-6708/2007/12/031}{\doi{10.1088/1126-6708/2007/12/031}},
\href{http://www.arXiv.org/abs/0708.0458}{\texttt{arXiv:0708.0458}}.

\bibitem{Hagiwara:2009wt}
\hrefCMSnoop {}{K.~Hagiwara, Q.~Li, and K.~Mawatari, ``{Jet angular correlation
  in vector-boson fusion processes at hadron colliders}'',} \textit{ JHEP}
  \textbf{ 07} (2009) 101,
  \href{http://dx.doi.org/10.1088/1126-6708/2009/07/101}{\doi{10.1088/1126-6708/2009/07/101}},
\href{http://www.arXiv.org/abs/0905.4314}{\texttt{arXiv:0905.4314}}.

\bibitem{Gao:2010qx}
Y.~Gao\hrefCMSnoop {}{ {et~al.}, ``{Spin determination of single-produced
  resonances at hadron colliders}'',} \textit{ Phys. Rev. D} \textbf{ 81}
  (2010) 075022,
  \href{http://dx.doi.org/10.1103/PhysRevD.81.075022}{\doi{10.1103/PhysRevD.81.075022}},
\href{http://www.arXiv.org/abs/1001.3396}{\texttt{arXiv:1001.3396}}.

\bibitem{DeRujula:2010ys}
A.~De~R{\' u}jula\hrefCMSnoop {}{ {et~al.}, ``{Higgs look-alikes at the
  LHC}'',} \textit{ Phys. Rev. D} \textbf{ 82} (2010) 013003,
  \href{http://dx.doi.org/10.1103/PhysRevD.82.013003}{\doi{10.1103/PhysRevD.82.013003}},
\href{http://www.arXiv.org/abs/1001.5300}{\texttt{arXiv:1001.5300}}.

\bibitem{Christensen:2010pf}
\hrefCMSnoop {}{N.~D. Christensen, T.~Han, and Y.~Li, ``{Testing CP Violation
  in ZZH Interactions at the LHC}'',} \textit{ Phys. Lett. B} \textbf{ 693}
  (2010) 28,
  \href{http://dx.doi.org/10.1016/j.physletb.2010.08.008}{\doi{10.1016/j.physletb.2010.08.008}},
\href{http://www.arXiv.org/abs/1005.5393}{\texttt{arXiv:1005.5393}}.

\bibitem{Bolognesi:2012mm}
S.~Bolognesi\hrefCMSnoop {}{ {et~al.}, ``{Spin and parity of a single-produced
  resonance at the LHC}'',} \textit{ Phys. Rev. D} \textbf{ 86} (2012) 095031,
  \href{http://dx.doi.org/10.1103/PhysRevD.86.095031}{\doi{10.1103/PhysRevD.86.095031}},
\href{http://www.arXiv.org/abs/1208.4018}{\texttt{arXiv:1208.4018}}.

\bibitem{Ellis:2012xd}
\hrefCMSnoop {}{J.~Ellis, D.~S. Hwang, V.~Sanz, and T.~You, ``{A fast track
  towards the `Higgs' spin and parity}'',} \textit{ JHEP} \textbf{ 11} (2012)
  134,
  \href{http://dx.doi.org/10.1007/JHEP11(2012)134}{\doi{10.1007/JHEP11(2012)134}},
\href{http://www.arXiv.org/abs/1208.6002}{\texttt{arXiv:1208.6002}}.

\bibitem{Chen:2012jy}
\hrefCMSnoop {}{Y.~Chen, N.~Tran, and R.~Vega-Morales, ``{Scrutinizing the
  Higgs signal and background in the $2\cPe2\cPm$ golden channel}'',} \textit{
  JHEP} \textbf{ 01} (2013) 182,
  \href{http://dx.doi.org/10.1007/JHEP01(2013)182}{\doi{10.1007/JHEP01(2013)182}},
\href{http://www.arXiv.org/abs/1211.1959}{\texttt{arXiv:1211.1959}}.

\bibitem{Artoisenet:2013puc}
P.~Artoisenet\hrefCMSnoop {}{ {et~al.}, ``{A framework for Higgs
  characterisation}'',} \textit{ JHEP} \textbf{ 11} (2013) 043,
  \href{http://dx.doi.org/10.1007/JHEP11(2013)043}{\doi{10.1007/JHEP11(2013)043}},
\href{http://www.arXiv.org/abs/1306.6464}{\texttt{arXiv:1306.6464}}.

\bibitem{Anderson:2013afp}
I.~Anderson\hrefCMSnoop {}{ {et~al.}, ``{Constraining anomalous \HVV
  interactions at proton and lepton colliders}'',} \textit{ Phys. Rev. D}
  \textbf{ 89} (2014) 035007,
  \href{http://dx.doi.org/10.1103/PhysRevD.89.035007}{\doi{10.1103/PhysRevD.89.035007}},
\href{http://www.arXiv.org/abs/1309.4819}{\texttt{arXiv:1309.4819}}.

\bibitem{Chen:2013waa}
M.~Chen\hrefCMSnoop {}{ {et~al.}, ``{Role of interference in unraveling the \ZZ
  couplings of the newly discovered boson at the LHC}'',} \textit{ Phys. Rev.
  D} \textbf{ 89} (2014) 034002,
  \href{http://dx.doi.org/10.1103/PhysRevD.89.034002}{\doi{10.1103/PhysRevD.89.034002}},
\href{http://www.arXiv.org/abs/1310.1397}{\texttt{arXiv:1310.1397}}.

\bibitem{Dolan:2014upa}
\hrefCMSnoop {}{M.~J. Dolan, P.~Harris, M.~Jankowiak, and M.~Spannowsky,
  ``{Constraining $CP$-violating Higgs sectors at the LHC using gluon
  fusion}'',} \textit{ Phys. Rev. D} \textbf{ 90} (2014) 073008,
  \href{http://dx.doi.org/10.1103/PhysRevD.90.073008}{\doi{10.1103/PhysRevD.90.073008}},
\href{http://www.arXiv.org/abs/1406.3322}{\texttt{arXiv:1406.3322}}.

\bibitem{Gonzalez-Alonso:2014eva}
\hrefCMSnoop {}{M.~Gonzalez-Alonso, A.~Greljo, G.~Isidori, and D.~Marzocca,
  ``{Pseudo-observables in Higgs decays}'',} \textit{ Eur. Phys. J. C} \textbf{
  75} (2015) 128,
  \href{http://dx.doi.org/10.1140/epjc/s10052-015-3345-5}{\doi{10.1140/epjc/s10052-015-3345-5}},
\href{http://www.arXiv.org/abs/1412.6038}{\texttt{arXiv:1412.6038}}.

\bibitem{Greljo:2015sla}
\hrefCMSnoop {}{A.~Greljo, G.~Isidori, J.~M. Lindert, and D.~Marzocca,
  ``{Pseudo-observables in electroweak Higgs production}'',} \textit{ Eur.
  Phys. J. C} \textbf{ 76} (2016) 158,
  \href{http://dx.doi.org/10.1140/epjc/s10052-016-4000-5}{\doi{10.1140/epjc/s10052-016-4000-5}},
\href{http://www.arXiv.org/abs/1512.06135}{\texttt{arXiv:1512.06135}}.

\bibitem{Gritsan:2016hjl}
\hrefCMSnoop {}{A.~V. Gritsan, R.~R{\"o}ntsch, M.~Schulze, and M.~Xiao,
  ``{Constraining anomalous Higgs boson couplings to the heavy flavor fermions
  using matrix element techniques}'',} \textit{ Phys. Rev. D} \textbf{ 94}
  (2016) 055023,
  \href{http://dx.doi.org/10.1103/PhysRevD.94.055023}{\doi{10.1103/PhysRevD.94.055023}},
\href{http://www.arXiv.org/abs/1606.03107}{\texttt{arXiv:1606.03107}}.

\bibitem{Chatrchyan:2008zzk}
\hrefCMSnoop {}{{CMS Collaboration}, ``The {CMS} experiment at the {CERN}
  {LHC}'',} \textit{ JINST} \textbf{ 3} (2008) S08004,
\href{http://dx.doi.org/10.1088/1748-0221/3/08/S08004}{\doi{10.1088/1748-0221/3/08/S08004}}.

\bibitem{Frixione:2007vw}
\hrefCMSnoop {}{S.~Frixione, P.~Nason, and C.~Oleari, ``{Matching NLO QCD
  computations with parton shower simulations: the POWHEG method}'',} \textit{
  JHEP} \textbf{ 11} (2007) 070,
  \href{http://dx.doi.org/10.1088/1126-6708/2007/11/070}{\doi{10.1088/1126-6708/2007/11/070}},
\href{http://www.arXiv.org/abs/0709.2092}{\texttt{arXiv:0709.2092}}.

\bibitem{Bagnaschi:2011tu}
\hrefCMSnoop {}{E.~Bagnaschi, G.~Degrassi, P.~Slavich, and A.~Vicini, ``{Higgs
  production via gluon fusion in the POWHEG approach in the SM and in the
  MSSM}'',} \textit{ JHEP} \textbf{ 02} (2012) 088,
  \href{http://dx.doi.org/10.1007/JHEP02(2012)088}{\doi{10.1007/JHEP02(2012)088}},
\href{http://www.arXiv.org/abs/1111.2854}{\texttt{arXiv:1111.2854}}.

\bibitem{Nason:2009ai}
\hrefCMSnoop {}{P.~Nason and C.~Oleari, ``{NLO Higgs boson production via
  vector-boson fusion matched with shower in POWHEG}'',} \textit{ JHEP}
  \textbf{ 02} (2010) 037,
  \href{http://dx.doi.org/10.1007/JHEP02(2010)037}{\doi{10.1007/JHEP02(2010)037}},
\href{http://www.arXiv.org/abs/0911.5299}{\texttt{arXiv:0911.5299}}.

\bibitem{Luisoni:2013kna}
\hrefCMSnoop {}{G.~Luisoni, P.~Nason, C.~Oleari, and F.~Tramontano,
  ``{$\PH\PW^{\pm}$/$\PH\PZ$ + 0 and 1 jet at NLO with the POWHEG BOX
  interfaced to GoSam and their merging within MiNLO}'',} \textit{ JHEP}
  \textbf{ 10} (2013) 083,
  \href{http://dx.doi.org/10.1007/JHEP10(2013)083}{\doi{10.1007/JHEP10(2013)083}},
\href{http://www.arXiv.org/abs/1306.2542}{\texttt{arXiv:1306.2542}}.

\bibitem{Hartanto:2015uka}
\hrefCMSnoop {}{H.~B. Hartanto, B.~Jager, L.~Reina, and D.~Wackeroth, ``{Higgs
  boson production in association with top quarks in the POWHEG BOX}'',}
  \textit{ Phys. Rev. D} \textbf{ 91} (2015) 094003,
  \href{http://dx.doi.org/10.1103/PhysRevD.91.094003}{\doi{10.1103/PhysRevD.91.094003}},
\href{http://www.arXiv.org/abs/1501.04498}{\texttt{arXiv:1501.04498}}.

\bibitem{Hamilton:2012np}
\hrefCMSnoop {}{K.~Hamilton, P.~Nason, and G.~Zanderighi, ``{MINLO: multi-scale
  improved NLO}'',} \textit{ JHEP} \textbf{ 10} (2012) 155,
  \href{http://dx.doi.org/10.1007/JHEP10(2012)155}{\doi{10.1007/JHEP10(2012)155}},
\href{http://www.arXiv.org/abs/1206.3572}{\texttt{arXiv:1206.3572}}.

\bibitem{MCFM}
\hrefCMSnoop {}{J.~M. Campbell and R.~K. Ellis, ``{MCFM for the Tevatron and
  the LHC}'',} \textit{ Nucl. Phys. Proc. Suppl.} \textbf{ 205-206} (2010) 10,
  \href{http://dx.doi.org/10.1016/j.nuclphysbps.2010.08.011}{\doi{10.1016/j.nuclphysbps.2010.08.011}},
\href{http://www.arXiv.org/abs/1007.3492}{\texttt{arXiv:1007.3492}}.

\bibitem{Campbell:2011bn}
\hrefCMSnoop {}{J.~M. Campbell, R.~K. Ellis, and C.~Williams, ``{Vector boson
  pair production at the LHC}'',} \textit{ JHEP} \textbf{ 07} (2011) 018,
  \href{http://dx.doi.org/10.1007/JHEP07(2011)018}{\doi{10.1007/JHEP07(2011)018}},
\href{http://www.arXiv.org/abs/1105.0020}{\texttt{arXiv:1105.0020}}.

\bibitem{Campbell:2015vwa}
\hrefCMSnoop {}{J.~M. Campbell and R.~K. Ellis, ``{Higgs constraints from
  vector boson fusion and scattering}'',} \textit{ JHEP} \textbf{ 04} (2015)
  030,
  \href{http://dx.doi.org/10.1007/JHEP04(2015)030}{\doi{10.1007/JHEP04(2015)030}},
\href{http://www.arXiv.org/abs/1502.02990}{\texttt{arXiv:1502.02990}}.

\bibitem{Ballestrero:2007}
A.~Ballestrero\hrefCMSnoop {}{ {et~al.}, ``{PHANTOM: a Monte Carlo} event
  generator for six parton final states at high energy colliders'',} \textit{
  Comput. Phys. Commun.} \textbf{ 180} (2009) 401,
  \href{http://dx.doi.org/10.1016/j.cpc.2008.10.005}{\doi{10.1016/j.cpc.2008.10.005}},
  \href{http://www.arXiv.org/abs/0801.3359}{\texttt{arXiv:0801.3359}}.

\bibitem{Catani:2007vq}
\hrefCMSnoop {}{S.~Catani and M.~Grazzini, ``{An NNLO subtraction formalism in
  hadron collisions and its application to Higgs boson production at the
  LHC}'',} \textit{ Phys. Rev. Lett.} \textbf{ 98} (2007) 222002,
  \href{http://dx.doi.org/10.1103/PhysRevLett.98.222002}{\doi{10.1103/PhysRevLett.98.222002}},
\href{http://www.arXiv.org/abs/hep-ph/0703012}{\texttt{arXiv:hep-ph/0703012}}.

\bibitem{Grazzini:2008tf}
\hrefCMSnoop {}{M.~Grazzini, ``{NNLO predictions for the Higgs boson signal in
  the $\PH \to \WW \to\ell\cPgn\ell\cPgn$ and $\PH \to \ZZ \to 4\ell$ decay
  channels}'',} \textit{ JHEP} \textbf{ 02} (2008) 043,
  \href{http://dx.doi.org/10.1088/1126-6708/2008/02/043}{\doi{10.1088/1126-6708/2008/02/043}},
\href{http://www.arXiv.org/abs/0801.3232}{\texttt{arXiv:0801.3232}}.

\bibitem{Grazzini:2013mca}
\hrefCMSnoop {}{M.~Grazzini and H.~Sargsyan, ``{Heavy-quark mass effects in
  Higgs boson production at the LHC}'',} \textit{ JHEP} \textbf{ 09} (2013)
  129,
  \href{http://dx.doi.org/10.1007/JHEP09(2013)129}{\doi{10.1007/JHEP09(2013)129}},
\href{http://www.arXiv.org/abs/1306.4581}{\texttt{arXiv:1306.4581}}.

\bibitem{Caola:2015psa}
\hrefCMSnoop {}{F.~Caola, K.~Melnikov, R.~R{\"o}ntsch, and L.~Tancredi, ``{QCD
  corrections to \ZZ production in gluon fusion at the LHC}'',} \textit{ Phys.
  Rev. D} \textbf{ 92} (2015) 094028,
  \href{http://dx.doi.org/10.1103/PhysRevD.92.094028}{\doi{10.1103/PhysRevD.92.094028}},
\href{http://www.arXiv.org/abs/1509.06734}{\texttt{arXiv:1509.06734}}.

\bibitem{Melnikov:2015laa}
\hrefCMSnoop {}{K.~Melnikov and M.~Dowling, ``{Production of two $\PZ$-bosons
  in gluon fusion in the heavy top quark approximation}'',} \textit{ Phys.
  Lett. B} \textbf{ 744} (2015) 43,
  \href{http://dx.doi.org/10.1016/j.physletb.2015.03.030}{\doi{10.1016/j.physletb.2015.03.030}},
\href{http://www.arXiv.org/abs/1503.01274}{\texttt{arXiv:1503.01274}}.

\bibitem{Campbell:2016ivq}
\hrefCMSnoop {}{J.~M. Campbell, R.~K. Ellis, M.~Czakon, and S.~Kirchner, ``{Two
  loop correction to interference in $\Pg\Pg \to \ZZ$}'',} \textit{ JHEP}
  \textbf{ 08} (2016) 011,
  \href{http://dx.doi.org/10.1007/JHEP08(2016)011}{\doi{10.1007/JHEP08(2016)011}},
\href{http://www.arXiv.org/abs/1605.01380}{\texttt{arXiv:1605.01380}}.

\bibitem{Caola:2016trd}
F.~Caola\hrefCMSnoop {}{ {et~al.}, ``{QCD corrections to vector boson pair
  production in gluon fusion including interference effects with off-shell
  Higgs at the LHC}'',} \textit{ JHEP} \textbf{ 07} (2016) 087,
  \href{http://dx.doi.org/10.1007/JHEP07(2016)087}{\doi{10.1007/JHEP07(2016)087}},
\href{http://www.arXiv.org/abs/1605.04610}{\texttt{arXiv:1605.04610}}.

\bibitem{Grazzini:2015hta}
\hrefCMSnoop {}{M.~Grazzini, S.~Kallweit, and D.~Rathlev, ``{$\PZ\PZ$
  production at the LHC: fiducial cross sections and distributions in NNLO
  QCD}'',} \textit{ Phys. Lett. B} \textbf{ 750} (2015) 407,
  \href{http://dx.doi.org/10.1016/j.physletb.2015.09.055}{\doi{10.1016/j.physletb.2015.09.055}},
\href{http://www.arXiv.org/abs/1507.06257}{\texttt{arXiv:1507.06257}}.

\bibitem{Gieseke:2014gka}
\hrefCMSnoop {}{S.~Gieseke, T.~Kasprzik, and J.~H. Kuehn, ``{Vector-boson pair
  production and electroweak corrections in HERWIG++}'',} \textit{ Eur. Phys.
  J. C} \textbf{ 74} (2014) 2988,
  \href{http://dx.doi.org/10.1140/epjc/s10052-014-2988-y}{\doi{10.1140/epjc/s10052-014-2988-y}},
\href{http://www.arXiv.org/abs/1401.3964}{\texttt{arXiv:1401.3964}}.

\bibitem{Baglio:2013toa}
\hrefCMSnoop {}{J.~Baglio, L.~D. Ninh, and M.~M. Weber, ``{Massive gauge boson
  pair production at the LHC: a next-to-leading order story}'',} \textit{ Phys.
  Rev. D} \textbf{ 88} (2013) 113005,
  \href{http://dx.doi.org/10.1103/PhysRevD.88.113005}{\doi{10.1103/PhysRevD.88.113005}},
  \href{http://www.arXiv.org/abs/1307.4331}{\texttt{arXiv:1307.4331}}.
[Erratum: \DOI{10.1103/PhysRevD.94.099902}].

\bibitem{Ball:2011uy}
\hrefCMSnoop {}{{NNPDF} Collaboration, ``{Unbiased global determination of
  parton distributions and their uncertainties at NNLO and at LO}'',} \textit{
  Nucl. Phys. B} \textbf{ 855} (2012) 153,
  \href{http://dx.doi.org/10.1016/j.nuclphysb.2011.09.024}{\doi{10.1016/j.nuclphysb.2011.09.024}},
\href{http://www.arXiv.org/abs/1107.2652}{\texttt{arXiv:1107.2652}}.

\bibitem{Sjostrand:2014zea}
T.~Sj{\"o}strand\hrefCMSnoop {}{ {et~al.}, ``An introduction to {PYTHIA}
  8.2'',} \textit{ Comput. Phys. Commun.} \textbf{ 191} (2015) 159,
  \href{http://dx.doi.org/10.1016/j.cpc.2015.01.024}{\doi{10.1016/j.cpc.2015.01.024}},
\href{http://www.arXiv.org/abs/1410.3012}{\texttt{arXiv:1410.3012}}.

\bibitem{Agostinelli2003250}
\hrefCMSnoop {}{{GEANT4} Collaboration, ``{GEANT4} -- a simulation toolkit'',}
  \textit{ Nucl. Instrum. Meth. A} \textbf{ 506} (2003) 250,
\href{http://dx.doi.org/10.1016/S0168-9002(03)01368-8}{\doi{10.1016/S0168-9002(03)01368-8}}.

\bibitem{Sirunyan:2017ulk}
\hrefCMSnoop {}{{CMS Collaboration}, ``Particle-flow reconstruction and global
  event description with the cms detector'',} \textit{ JINST} \textbf{ 12}
  (2017) P10003,
  \href{http://dx.doi.org/10.1088/1748-0221/12/10/P10003}{\doi{10.1088/1748-0221/12/10/P10003}},
\href{http://www.arXiv.org/abs/1706.04965}{\texttt{arXiv:1706.04965}}.

\bibitem{Cacciari:2008gp}
\hrefCMSnoop {}{M.~Cacciari, G.~P. Salam, and G.~Soyez, ``The anti-$\kt$ jet
  clustering algorithm'',} \textit{ JHEP} \textbf{ 04} (2008) 063,
  \href{http://dx.doi.org/10.1088/1126-6708/2008/04/063}{\doi{10.1088/1126-6708/2008/04/063}},
  \href{http://www.arXiv.org/abs/0802.1189}{\texttt{arXiv:0802.1189}}.

\bibitem{Cacciari:2011ma}
\hrefCMSnoop {}{M.~Cacciari, G.~P. Salam, and G.~Soyez, ``{FastJet user
  manual}'',} \textit{ Eur. Phys. J. C} \textbf{ 72} (2012) 1896,
  \href{http://dx.doi.org/10.1140/epjc/s10052-012-1896-2}{\doi{10.1140/epjc/s10052-012-1896-2}},
\href{http://www.arXiv.org/abs/1111.6097}{\texttt{arXiv:1111.6097}}.

\bibitem{Chatrchyan:2012jua}
\hrefCMSnoop {}{{CMS Collaboration}, ``{Identification of $\PQb$-quark jets
  with the CMS experiment}'',} \textit{ JINST} \textbf{ 8} (2013) P04013,
  \href{http://dx.doi.org/10.1088/1748-0221/8/04/P04013}{\doi{10.1088/1748-0221/8/04/P04013}},
\href{http://www.arXiv.org/abs/1211.4462}{\texttt{arXiv:1211.4462}}.

\bibitem{Sirunyan:2017ezt}
\hrefCMSnoop {}{{CMS Collaboration}, ``{Identification of heavy-flavour jets
  with the CMS detector in $\Pp\Pp$ collisions at 13 TeV}'',} \textit{ JINST}
  \textbf{ 13} (2018) P05011,
  \href{http://dx.doi.org/10.1088/1748-0221/13/05/P05011}{\doi{10.1088/1748-0221/13/05/P05011}},
\href{http://www.arXiv.org/abs/1712.07158}{\texttt{arXiv:1712.07158}}.

\bibitem{Barlow:1990vc}
\hrefCMSnoop {}{R.~J. Barlow, ``{Extended maximum likelihood}'',} \textit{
  Nucl. Instrum. Meth. A} \textbf{ 297} (1990) 496,
\href{http://dx.doi.org/10.1016/0168-9002(90)91334-8}{\doi{10.1016/0168-9002(90)91334-8}}.

\bibitem{Oreglia:1980cs}
\href {http://www.slac.stanford.edu/cgi-wrap/getdoc/slac-r-236.pdf}{M.~Oreglia,
  ``{A study of the reactions $\psi^\prime \to \gamma \gamma \psi$}''}.
\newblock PhD thesis, Stanford University, 1980.
\newblock {SLAC} Report {SLAC-R-236}.

\bibitem{Verkerke:2003ir}
\hrefCMSnoop {}{W.~Verkerke and D.~P. Kirkby, ``{The RooFit toolkit for data
  modeling}'',} in \textit{ 13$^\text{th}$ International Conference for
  Computing in High-Energy and Nuclear Physics (CHEP03)}.
\newblock 2003.
\newblock
  \href{http://www.arXiv.org/abs/physics/0306116}{\texttt{arXiv:physics/0306116}}.
\newblock
{CHEP-2003-MOLT007}.

\bibitem{Brun:1997pa}
\hrefCMSnoop {}{R.~Brun and F.~Rademakers, ``{ROOT: An object oriented data
  analysis framework}'',} \textit{ Nucl. Instrum. Meth. A} \textbf{ 389} (1997)
  81,
\href{http://dx.doi.org/10.1016/S0168-9002(97)00048-X}{\doi{10.1016/S0168-9002(97)00048-X}}.

\bibitem{Wilks:1938dza}
\hrefCMSnoop {}{S.~S. Wilks, ``{The large-sample distribution of the likelihood
  ratio for testing composite hypotheses}'',} \textit{ Annals Math. Statist.}
  \textbf{ 9} (1938) 60,
\href{http://dx.doi.org/10.1214/aoms/1177732360}{\doi{10.1214/aoms/1177732360}}.

\bibitem{CMS-PAS-LUM-17-001}
\href {https://cds.cern.ch/record/2257069}{{CMS Collaboration}, ``{CMS
  luminosity measurements for the 2016 data taking period}'',} {CMS Physics
  Analysis Summary} {CMS-PAS-LUM-17-001}, 2017.

\bibitem{CMS-PAS-LUM-17-004}
\href {https://cds.cern.ch/record/2621960}{{CMS Collaboration}, ``{CMS
  luminosity measurement for the 2017 data taking period at $\sqrt{s} =
  13\TeV$}'',} {CMS Physics Analysis Summary} {CMS-PAS-LUM-17-004}, 2018.

\end{thebibliography}\endgroup
\cleardoublepage \appendix\section{The CMS Collaboration \label{app:collab}}\begin{sloppypar}\hyphenpenalty=5000\widowpenalty=500\clubpenalty=5000\vskip\cmsinstskip
\textbf{Yerevan Physics Institute, Yerevan, Armenia}\\*[0pt]
A.M.~Sirunyan, A.~Tumasyan
\vskip\cmsinstskip
\textbf{Institut f\"{u}r Hochenergiephysik, Wien, Austria}\\*[0pt]
W.~Adam, F.~Ambrogi, E.~Asilar, T.~Bergauer, J.~Brandstetter, M.~Dragicevic, J.~Er\"{o}, A.~Escalante~Del~Valle, M.~Flechl, R.~Fr\"{u}hwirth\cmsAuthorMark{1}, V.M.~Ghete, J.~Hrubec, M.~Jeitler\cmsAuthorMark{1}, N.~Krammer, I.~Kr\"{a}tschmer, D.~Liko, T.~Madlener, I.~Mikulec, N.~Rad, H.~Rohringer, J.~Schieck\cmsAuthorMark{1}, R.~Sch\"{o}fbeck, M.~Spanring, D.~Spitzbart, W.~Waltenberger, J.~Wittmann, C.-E.~Wulz\cmsAuthorMark{1}, M.~Zarucki
% ...
\vskip\cmsinstskip
% ...
1:  Also at Vienna University of Technology, Vienna, Austria\\
% ...
\end{sloppypar}
\end{document}